\documentclass[12pt]{iopart}
\usepackage{iopams}  
\usepackage{afterpage}
\usepackage{graphicx}
\usepackage{amssymb}

\begin{document}

\title[Phase lapses in two-level quantum dots]{Phase lapses in
transmission through interacting two-level quantum dots}

\author{C Karrasch$^1$, T Hecht$^2$, A Weichselbaum$^2$, 
J von Delft$^2$, Y Oreg$^3$, and V Meden$^1$}
\address{$^1$ Institut f\"ur Theoretische Physik, Universit\"at
  G\"ottingen, 37077 G\"ottingen, Germany}
\ead{meden@theorie.physik.uni-goettingen.de}
 
\address{$^2$ Physics Department, Arnold
  Sommerfeld Center for Theoretical Physics and Cen\-ter for
  NanoScience, Ludwig-Maximilians-Universit\"at, 80333 Munich,
  Germany} 

\address{$^3$ Department of Condensed
  Matter Physics, The Weizmann Institute of Science, Rehovot 76100,
  Israel} 

\begin{abstract}
  We investigate the appearance of $\pi$ lapses in the transmission
  phase $\theta$ of a two-level quantum dot with Coulomb interaction
  $U$.  Using the numerical and functional renormalization group
  methods we study the entire parameter space for spin-polarized as
  well as spin-degenerate dots, modeled by spinless or spinful 
  electrons, respectively.  We investigate the effect of finite
  temperatures $T$. For small $T$ and sufficiently small
  single-particle spacings $\delta$ of the dot levels we find $\pi$
  phase lapses between two transmission peaks in an overwhelming part
  of the parameter space of the level-lead couplings. For large
  $\delta$ the appearance or not of a phase lapse between resonances
  depends on the relative sign of the level-lead couplings in analogy
  to the $U=0$ case. We show that this generic scenario is the
  same for spin-polarized and spin-degenerate dots.  We emphasize that
  in contrast to dots with more levels, for a two-level dot with small
  $\delta$ and generic dot-lead couplings (that is up to cases with
  special symmetry) the ``universal'' phase lapse behavior is already 
  established at $U=0$. The most important effect of the Coulomb interaction
  is to increase the separation of the transmission resonances. The
  relation of the appearance of phase lapses to the inversion of the
  population of the dot levels is discussed.  For the spin-polarized
  case and low temperatures we compare our results to recent
  mean-field studies.  For small $\delta$ correlations are found to
  strongly alter the mean-field picture.
  
\end{abstract}


\maketitle

\section{Introduction}

The local Coulomb interaction $U>0$ of electrons occupying quantum
dots leads to a variety of effects. Many of them can conveniently be
studied in transport through the dot within the linear regime.
Theoretically as well as experimentally well-investigated examples are
the Coulomb blockade (CB) peaks of the transmission (conductance)
\cite{CBref} as well as the plateaus of width $U$ of the transmission
(conductance) induced by the Kondo effect \cite{Kondo}.  Additional
features of interacting multi-level dots that have recently attracted
considerable theoretical attention are the population inversions 
of the dot levels \cite{Imry,Koenig,Sindel1,Berkovits}, the phase
lapses of the transmission phase $\theta$ or, equivalently, the
zeros of the transmission amplitude $t$ (transmission zeros) 
\cite{Imry,Hackenbroich,Gefen02,Silva,Gefen,Gefenagain} and additional
correlation-induced resonances  of $|t|$ \cite{VF}. They appear
in certain parts of the parameter space when the level occupancies and
the transmission amplitude are investigated as functions of the level
positions, which can be tuned via a nearby plunger gate
  voltage.  Such effects were hitherto mostly studied in a minimal
model involving only two levels.  A very important step towards a
unified understanding of population inversions, phase lapses and correlation-induced resonancess in spin-polarized two-level
dots was recently taken by a multi-stage mapping of the problem on a
generalized Kondo model and a subsequent renormalization group 
and Bethe ansatz analysis of the effective Hamiltonian \cite{Slava}.

Theoretical studies of phase lapses (transmission zeros) are of primary interest in
connection with a series of linear response transmission measurements 
by the Weizman group \cite{ex1,ex2,ex3} on Aharonov-Bohm rings
containing a quantum dot in one arm. Under suitable conditions 
both the phase $\theta$ and magnitude $|t|$
of the transmission amplitude $t = |t|e^{i \theta}$ of
the dot can be extracted from the Aharonov-Bohm oscillations of the
current through the ring \cite{Aharony}.  
When this is done as function of a plunger
gate voltage $V_g$ that linearly shifts the dot's single-particle
energy levels downward, $\varepsilon_j = \varepsilon_j^0 - V_g$
($j=1,2, \dots$ is a level index), a series of well-separated
CB peaks of rather similar width and height was 
observed in $|t(V_g)|$, across which $\theta (V_g)$ continuously increased
by $\pi$, as expected for Breit-Wigner-like resonances.  In each CB
valley between any two successive peaks, $\theta$ always jumped
sharply downward by $\pi$.  This phase lapse behavior was
found to be ``universal'', occurring in a large succession of
valleys for every {\it many}-electron dot studied in Refs.~\cite{ex1,ex2,ex3}.
This universality is puzzling, since naively the behavior of $\theta
(V_g)$ is expected to be ``mesoscopic'', i.e.\ to show a phase lapse in
some CB valleys and none in others, depending on the dot's shape, the
parity of its orbital wavefunctions, etc. Only recently \cite{ex3}, 
also the {\it few}-electron regime was probed experimentally: 
as $V_g$ was increased to 
successively fill up the dot with electrons, starting from electron 
number $N_{\rm e} = 0$, $\theta (V_g)$ was observed to behave 
mesoscopically in the \emph{few}-electron regime,
whereas the above-mentioned universal phase lapse behavior emerged only in the
\emph{many}-electron regime $(N_{\rm e} \gtrsim 15)$. 
 
It was suggested in Ref.\ \cite{ex3} that 
a generic difference between the few- and many-electron dots 
may be that for the latter, transport might simultaneously occur 
through several partially filled single-particle levels in parallel.  
A possible reason could be that the mean (non-interacting) level 
spacing $\delta$ of the topmost filled levels decreases as the number of 
electrons increases, while the charging energy $U$ still implies 
well separated transmission resonances \cite{Felix}.  
This scenario forms the basis of a recent
systematic study by us of the interplay of level spacing, level width
and charging energy on the phase lapses for up to four interacting levels and
spin-polarized electrons \cite{PL1}.  We showed that the universal phase lapse
and transmission zero behavior appearing at small $\delta$ can be understood as 
resulting from a Fano-type interference effect \cite{Fano}
  involving transport through two or more effective dot levels, whose
  position and width have been renormalized by the Coulomb
interaction and coupling to the leads.  The importance of several
overlapping levels for phase lapses had earlier been pointed out by Silvestrov
and Imry in a rather specific model of a single wide and several
narrow levels with strong interaction \cite{Imry}.

Here we supplement our earlier study \cite{PL1} by discussing the relation
between phase lapses and population inversions and by investigating the
role of finite temperatures $T>0$ as well as spin, focusing on $N=2$
levels. 
When spin is included, the Kondo effect plays a role for an
odd average occupation of the dot, but the phase lapse scenario is 
unaffected by this. Experimentally the behavior of the phase in
the presence of the Kondo effect was investigated in 
Refs.\ \cite{Yang} and  \cite{Yang2}.
As in Ref.\ \cite{PL1} we are concerned with the {\it generic} 
behavior and thus investigate the entire parameter space, going beyond
subspaces of higher symmetry [such as left-right (l-r) or 1-2 symmetry  
of the coupling between the left and right leads and the two
levels]. For low temperatures and
sufficiently small single-particle spacings $\delta$ of the dot levels, 
we find $\pi$ phase lapses between two transmission peaks in an overwhelmingly 
large part of the parameter space of the level-lead couplings. 
We point out that the two level case is special compared to models
with $N>2$, as for generic level-lead couplings a transmission zero
and phase lapse occurs between the two transmission peaks 
{\it even at} U=0. The effect of the
interaction is merely to increase the separation of the transmission peaks.    
For large $\delta$ the appearance or not of phase lapses between transmission
peaks depends on the  relative sign of the level-lead  couplings in 
analogy to the noninteracting case \cite{Silva}.   

For spin-polarized dots we in addition compare our $T=0$ results with
the ones of recent mean-field studies \cite{Gefen,Gefenagain}.  In these works
level-lead couplings beyond the subspaces 
with increased symmetry were studied, and a remarkably
more complex behavior was found once the symmetries were
broken.
The importance of considering such generic parameter sets
was independently pointed out in Ref.\ \cite{VF}.  We here elucidate how the
phase lapse behavior is affected by quantum fluctuations, which are expected to
be strong in low-dimensional systems. We find that upon including 
  quantum fluctuations, the
  part of parameter space exhibiting universal $\pi$ phase lapses between well
  separated CB peaks becomes 
  larger than suggested by the mean-field study.  In particular, we
  do not recover certain peculiar features of the mean-field results
  of Refs.\ \cite{Gefen} and \cite{Gefenagain}, 
namely the occurrence, in certain regimes of
  parameter space, of a phase lapse of less than $\pi$ (instead of precisely
  $\pi$), accompanied by the disappearance of the corresponding transmission zero
  \cite{Gefen,Berkovits}. These features thus turn out to be artifacts of the
  mean-field approximation, which misses the rather simple scenario
  for the phase lapse behavior of a two-level dot at small $\delta$: for
  generic level-lead couplings a phase lapse and transmission zero
  between two transmission peaks is already present at $U=0$;
  increasing the Coulomb interaction the peaks become well separated
  while the phase lapse and transmission zero remain in the valley between them.

In the model of a single wide and several narrow 
levels \cite{Imry} a relation between phase lapses and population inversions 
was discussed. Therefore, in phase lapse studies quite often also the 
level occupancies $n_j$, $j=1,2$, are investigated. 
We emphasize that the generic
appearance of a phase lapse and transmission zero even at $U=0$ renders the two-level model
unsuitable for establishing a general relation between phase lapses and population inversions,
as the latter only appear at sufficiently large $U$.   
Furthermore, we show that discontinuities of 
the $n_j$ as a function of $V_g$ are an artifact  of the mean-field 
solution (see Refs.\ \cite{Gefen} and \cite{Berkovits}). 
Within our approaches discontinuities are only found for l-r 
symmetric level-lead  couplings with a relative plus sign of the 
underlying hopping matrix elements and degenerate levels, 
a case which was earlier identified as being nongeneric \cite{VF,Slava},
because the transmission shows only a single peak.   

This paper is organized as follows. In Sec.\ \ref{model} we introduce
our model for the spin-polarized and spin-degenerate two-level dot. We
discuss the relation between the measured magnetic flux dependence of
the interferometer's linear conductance and the magnitude and 
phase of the dot's transmission amplitude. 
The latter can be computed from the one-particle Green 
function of the dot. We present a brief account of our techniques 
to obtain the latter, the numerical (NRG) \cite{Krishna-murthy} 
and functional renormalization 
group (fRG) methods. For an introduction to the use of 
the fRG to quantum dots see Refs.\ \cite{TCV} and
\cite{Christophdiplom}. 
In contrast to the implementation of the NRG used in Ref.\ \cite{PL1},
where we were restricted to l-r symmetric dots at $T=0$, 
we have now adopted the full density matrix (FDM) NRG method of
\cite{Weichsel}, which enables us to investigate dots with 
arbitrary level-lead state 
overlap matrix  elements $t_j^l$ (with $l=L,R$) as well as to study 
finite temperatures. In Secs.\ \ref{resultsnon}-\ref{resultsspin} 
we present our results of the $V_g$ dependence of $|t|$ and 
$\theta$. First we briefly 
discuss the noninteracting  two-level dot with generic level-lead 
couplings and point out that the phase lapse scenario differs from the one 
for more than two levels.  We then investigate  interacting, 
spin-polarized dots, study the relation between phase lapses and population inversions and 
compare to the mean-field results for the phase lapses. The issue of 
continuous versus discontinuous $V_g$ dependence of the level 
occupancies $n_1$ and $n_2$ is commented on. 
Next we study the role of finite temperatures. 
Finally, we consider spin-degenerate levels at small $T$ which implies 
the appearance of Kondo physics  at odd average dot filling. 
Using NRG and fRG we show that the spin does not alter the 
universal phase lapse scenario. Our findings are summarized in 
Sec.\ \ref{summary}.

\section{The model and methods}
\label{model}

In this section we introduce our model for the two-level dot. 
We argue that it is the energy dependent (effective) transmission 
amplitude $\tilde t(\omega)$ which one has to
compute if one is interested in comparing to the measurements
of Refs.\ \cite{ex1,ex2,ex3} of the magnitude of the transmission 
amplitude and its phase.
The amplitude $\tilde t (\omega)$ can be determined 
from the matrix elements of the dot's interacting 
one-particle Green function. We furthermore discuss aspects of the 
NRG and the fRG specific to our problem. 

\subsection{Two-level setup and transmission amplitude}

Our Hamiltonian consists of three parts
\begin{eqnarray}
\label{generalham}
H= H_{\rm lead} + H_{\rm dot} + H_{\rm lead-dot} \;. 
\end{eqnarray}
The two semi-infinite leads are modeled as noninteracting one-dimensional 
tight-binding chains and for simplicity are assumed to be equal
\begin{eqnarray}
\label{leadham}
H_{\rm lead} = -\tau \sum_{l=L,R} \sum_{\sigma} 
\sum_{m=0}^{\infty} (c^\dag_{m,\sigma,l}c_{m+1,\sigma,l} +
\mbox{H.c.}) \; .
\end{eqnarray}
The hopping strength in the leads is $\tau$.
We use standard second quantized notation with $l=L,R$ indicating the left 
and right leads, where the quantum numbers $m$ and $\sigma$ 
label Wannier states and spin, respectively.
The dot is described by
\begin{eqnarray}
\label{dotham}
H_{\rm dot} = \sum_{\sigma} 
\sum_{j=1,2} \varepsilon_{j} d^\dag_{j,\sigma} d_{j,\sigma} 
+ {1 \over 2} U \sum_{\sigma,\sigma'} \sum_{j,j'} 
\left(d^\dag_{j,\sigma} d_{j,\sigma} 
- {\textstyle \frac{1}{2}} \right) \left(d^\dag_{j',\sigma'} 
d_{j',\sigma'}-{\textstyle \frac{1}{2}}\right) ,
\end{eqnarray}
where the term with $j=j'$ and $\sigma =\sigma'$ is excluded from the
sum in the interacting part.  
We define $ \varepsilon_{1/2} = \mp \delta/2 - V_g$. 
In experimental systems the inter- and
intra-level Coulomb repulsion can be expected to be comparable in size
and to avoid a proliferation of parameters we assumed them to be
equal. This assumption is not essential; by relaxing it we have
  checked that our results are robust against inter-level variations
  of the interaction strengths.  Finally, the coupling between dot
and lead states is given by
\begin{eqnarray}
\label{dotleadham}
H_{\rm lead-dot} = -  \sum_{l=L,R} \sum_{\sigma} \sum_{j=1,2} (t_j^l 
c^\dag_{0,\sigma,l} d_{j,\sigma} + {\rm H.c.})
\end{eqnarray}  
with real overlap matrix elements $t_j^l$. For later purposes we
define $s=\mbox{sign}\,(t_1^Lt_1^Rt_2^Lt_2^R)$ and 
$\gamma=\{\Gamma_1^L, \Gamma_1^R, \Gamma_2^L, \Gamma_2^R\}/\Gamma$. 
 
For simplicity, part of our studies will be performed on a 
model of spinless electrons, for which the spin index will be 
dropped. The resulting model may be regarded as a spin-polarized 
version of the spinful model obtained if the latter is put in a very 
large magnetic field. 

The experimental two-path interferometer is characterized by wide base
regions and very narrow point contacts towards the emitter and
collector. Electrons can thus only pass once through either the
upper or lower arm (one of them containing the dot) before reaching
the collector. It is furthermore reasonable to assume that the voltage
drops at the point contacts and the dot remains in equilibrium 
\cite{Aharony}.   
Under these conditions the magnetic flux $\phi$ dependent 
part of the conductance of the Aharonov-Bohm
interferometer $G_{\rm AB}$ is given by \cite{Aharony,Gerland,ex3} 
\begin{eqnarray}
\label{G_AB}
G_{\rm AB} \propto - \int_{-\infty}^\infty d \omega f'(\omega) \,
|t_{\rm dot}(\omega)| \, |t_{\rm ref}| \, \cos{\left[ 2 \pi
    \phi/\phi_0 + \theta_{\rm ref} + \theta_{\rm dot}(\omega)\right]} 
\end{eqnarray}
with the flux quantum $\phi_0$, the derivative $f'$ of the Fermi
function as well as the transmission amplitudes $t_{\rm dot} = |t_{\rm
  dot}| e^{i  \theta_{\rm dot}}$ of the arm containing the dot and 
$t_{\rm ref} = |t_{\rm ref}| e^{i \theta_{\rm ref}}$ of the reference arm.
The transmission $t_{\rm dot}$ is the product of the transmission $\tilde t$ 
through the dot and the transmission  $t_{\rm rest}$ through the rest of the
interferometer arm containing the dot. It is reasonable 
to assume that $t_{\rm rest}$ as well as $t_{\rm ref}$ are only weakly 
energy and gate voltage dependent and thus the $V_g$-dependence of 
the Aharonov-Bohm oscillations of the  measured linear conductance 
of the interferometer is dominated by the $V_g$ dependence of the 
magnitude and phase of the transmission amplitude through the dot. 
As usual \cite{Silva}, 
we compute the energy-averaged 
transmission phase $\theta$ and magnitude $|t|$ of the dot for a fixed
spin direction as the phase and absolute value of 
\begin{eqnarray}
\label{transamp}
t(V_g) = - \int_{-\infty}^\infty d \omega f'(\omega) \tilde t(\omega) \; ,
\end{eqnarray}   
where the frequency integral represents an energy 
  average, weighted by the derivative of the Fermi function,
  as appropriate for a finite electron temperature in the leads.
In the limit $T \to 0$, $- f'$ reduces to a $\delta$-function and
$t(V_g)$ is equal to $\tilde t(\mu)$. We here take the chemical
potential $\mu=0$.

Using scattering theory $\tilde t(\omega)$ (for fixed spin direction) 
can be related to the spin-independent matrix elements 
(in the $j=1,2$ indices of the Wannier states) of the dot's 
one-particle retarded Green function ${\mathcal G}$
\begin{eqnarray}
\label{transformula}     
\tilde t(\omega) & = & 2 \left(\sqrt{\Gamma_1^L \Gamma_1^R} \,
   {\mathcal G}_{1,1}(\omega+i0) + \sqrt{\Gamma_2^L \Gamma_1^R} \,
   {\mathcal G}_{1,2}(\omega+i0) \right. \nonumber \\ 
&& \left. +  s \sqrt{\Gamma_1^L \Gamma_2^R }\,
   {\mathcal G}_{2,1}(\omega+i0) + s \sqrt{\Gamma_2^L \Gamma_2^R} \,
   {\mathcal G}_{2,2}(\omega+i0) \right) \; ,
\end{eqnarray}   
with (after taking the wide band limit; see below) 
\begin{eqnarray}
\label{Gammadef}
\Gamma_j^l =  \pi |t_j^l|^2 \rho_{\rm lead}(0) \geq 0 \; ,
\end{eqnarray}
where $\rho_{\rm lead}(\omega)$ 
denotes the local density of states at the end of each semi-infinite
lead. Without loss of generality we have assumed that $t_1^l \geq 0$,
$t_2^L \geq 0$ and $t_2^R = s |t_2^R|$.   
The spin-independent dot occupancies $n_j$ (per spin direction), that we
will also investigate, follow from the Green 
function ${\mathcal G}_{j,j}$ by integrating over frequency (or can be
computed directly when using NRG).  
For l-r symmetry of the level-lead couplings and at temperature $T=0$, 
$\tilde{t}(0)$ can also be expressed in terms of the spin
independent occupancies \cite{PL1}
\begin{equation}
  \tilde{t}(0)=\sin{([n_e-n_o]\pi)} e^{i(n_e+n_o)/\pi} \; ,
  \label{t_LR}
\end{equation}
where $n_e=n_1+n_2,\,n_o=0$ for $s=+$ and  $n_e=n_1,\,n_o=n_2$ for $s=-$,
respectively.
Here we will compute ${\mathcal G}$ in two ways, using both a 
truncated, that is approximate, fRG scheme, and a numerically exact 
method, the NRG.

\subsection{The fRG approach}

The truncated fRG is an approximation scheme to obtain the self-energy
$\Sigma$ (and thus the one-particle Green function) and higher order 
vertex functions for many-body 
problems \cite{SalmhoferHonerkamp,Ralf,lecturenotes}. 
As a first step in the application of this approach to quantum dots
one integrates out the noninteracting leads within the 
functional integral representation of our  
many-body problem \cite{NegeleOrland}. 
The leads provide a frequency dependent one-particle potential on the dot
levels.  On the imaginary frequency axis it is given by 
\begin{eqnarray}
\label{leadpotdef}
V^{\rm lead}_{j,\sigma;j',\sigma'}(i\omega) = 
\sum_{l} t_j^l t_{j'}^l  
{\rm g}_{\rm lead} (i\omega)  \, \delta_{\sigma,\sigma'} \;,
\end{eqnarray}
where ${\rm g}_{\rm lead}(i \omega)$ denotes the spin-independent 
Green function of the isolated semi-infinite leads taken at the 
last lattice site
\begin{eqnarray}
\label{gleaddef}
{\rm g}_{\rm lead}(i \omega) = \frac{i\omega+\mu}{2 \; \tau^2} \left( 1 - 
 \sqrt{1 - \frac{4 \; \tau^2}{(i\omega+\mu)^2}} \, \right) \; .
\end{eqnarray}
As we are not interested in band
effects we take the wide band limit. The potential then reduces to 
\begin{eqnarray}
\label{leadpotdefwideband}
V^{\rm lead}_{j,\sigma;j',\sigma'}(i\omega) = - i
\sum_{l} \sqrt{\Gamma_j^l \Gamma_{j'}^l} \, \mbox{sign}\,(\omega) \, 
\delta_{\sigma,\sigma'} \;.
\end{eqnarray}
After this step, instead of dealing with an infinite system we only have to 
consider the dot of two interacting levels. 

In the computation of the interacting one-particle Green function 
projected onto the dot system the sum of the dot Hamiltonian with 
$U = 0$ and $V_{j,\sigma;j',\sigma'}^{\rm lead}(i\omega)$ can be 
interpreted as a frequency 
dependent ``single-particle Hamiltonian'' and in the following  
will be denoted by $h_0(i\omega)$. For the spin-polarized case it is a 
$2 \times 2$ matrix in the quantum number $j=1,2$.  
Including spin,  because of the additional quantum number 
$\sigma=\uparrow,\downarrow$, $h_0(i \omega)$ is a $4 \times 4$ matrix 
which is block-diagonal in $\sigma$ (spin conservation).  
As we are here not interested in the role of a magnetic field lifting 
the spin-degeneracy of each level the $\sigma=\uparrow$ and
$\sigma=\downarrow$ blocks are equivalent. The resolvent ${\mathcal
  G}_0(z)=\left[ z - h_0(z) \right]^{-1}$ obtained from $h_0(z)$ 
is equivalent to the  noninteracting propagator of our 
two-level many-body problem projected on the dot levels. 
In the generating functional of the one-particle irreducible vertex 
functions we replace ${\mathcal G}_0(i \omega)$ by
\begin{equation}
\label{cutoffproc}
{\mathcal G}_0^{\Lambda}(i \omega) = \Theta(|\omega|-\Lambda) 
{\mathcal G}_0(i \omega) =  \Theta(|\omega|-\Lambda)  \left[ i\omega -
h_0(i \omega) \right]^{-1}
\end{equation}
with $\Lambda$ being an infrared cutoff running from $\infty$ down to $0$.  
Taking the derivative with respect to 
$\Lambda$ one can derive an exact, infinite hierarchy of coupled differential
equations for vertex functions, such as the self-energy and the
one-particle irreducible two-particle interaction. In 
particular, the flow of the self-energy $\Sigma^\Lambda$ (one-particle
vertex) is determined by $\Sigma^\Lambda$ and 
the two-particle vertex $\gamma^{\Lambda}$, while the flow of 
$\gamma^{\Lambda}$ is determined by $\Sigma^\Lambda$, $\gamma^{\Lambda}$, and 
the flowing three-particle vertex.
The latter could be computed from a flow equation involving
the four-particle vertex, and so on.
At the end of the fRG flow $\Sigma^{\Lambda=0}$ is the self-energy $\Sigma$ 
of the original, cutoff-free problem we are interested 
in \cite{SalmhoferHonerkamp,Ralf} from which the Green function
${\mathcal G}$ can be computed using the Dyson equation. 
A detailed derivation of the fRG flow equations for a general quantum
many-body problem that only requires a basic knowledge of the
functional integral approach to many-particle
physics \cite{NegeleOrland} and the application of the method for a
simple toy problem is presented in Ref.\ \cite{lecturenotes}. For an
overview of the application to quantum dots see Refs.\ \cite{TCV} and 
\cite{Christophdiplom}.

We here truncate the infinite hierarchy of flow equations by only
keeping the self-energy and the frequency-independent part of the
two-particle vertex. Higher order terms can be neglected if the 
bare two-particle interaction is not too large.
By comparison to NRG data this approximation 
scheme was earlier shown to provide excellent results for a variety of
dot systems \cite{VF,TCV,PL1}. 
For further comparison see Fig.\ \ref{fig7} below. 
The present scheme leads to a 
frequency-independent self-energy (see below). As finite frequency 
effects (inelastic processes) become important
at temperatures $T>0$, but these are not accurately treated by the 
level of approximation used here, in the present paper we shall 
show fRG results only for $T=0$.
The truncation leads to the coupled differential flow equations 
\begin{equation}
 \frac{\partial}{\partial\Lambda} \Sigma^{\Lambda}_{k',k} =
 - \frac{1}{2\pi} \sum_{\omega = \pm \Lambda} \sum_{l,l'} \,
 e^{i\omega 0^+} \, {\mathcal G}^{\Lambda}_{l,l'}(i\omega) \,
 \gamma^{\Lambda}_{k',l';k,l} 
\label{finalflowsigma}
\end{equation}
and 
\begin{eqnarray}
&& \frac{\partial}{\partial\Lambda} \gamma^{\Lambda}_{k',l';k,l} =   
\frac{1}{2\pi} \, 
 \sum_{\omega = \pm \Lambda} \, \sum_{m,m'} \, \sum_{n,n'}
 \Big\{ \frac{1}{2} \, {\mathcal G}^{\Lambda}_{m,m'}(i\omega) \, 
   {\mathcal G}^{\Lambda}_{n,n'}(-i\omega) 
 \gamma^{\Lambda}_{k',l';m,n} \, \gamma^{\Lambda}_{m',n';k,l} 
\nonumber \\
&& +  {\mathcal G}^{\Lambda}_{m,m'}(i\omega) \, 
 {\mathcal G}^{\Lambda}_{n,n'}(i\omega)  
 \left[ - \gamma^{\Lambda}_{k',n';k,l} \, \gamma^{\Lambda}_{m',l';n,l} 
        + \gamma^{\Lambda}_{l',n';k,m} \, \gamma^{\Lambda}_{m',k';n,l}
 \right] \Big\}
\label{finalflowgamma}
\end{eqnarray}
where $k$, $l$, etc.\ are multi-indices representing the 
quantum numbers $j,\sigma$ and
\begin{equation}
\label{Glambdadef}
  {\mathcal G}^{\Lambda}(i\omega) = 
 \left[  {\mathcal G}_0^{-1}(i\omega) - \Sigma^{\Lambda} \right]^{-1} \; .
\end{equation}  
In the model with spin-degenerate levels 
each index $k$, $l$ etc.\ can take four different values 
$j=1,2$ and $\sigma=\uparrow, \downarrow$ which gives 16 equations for 
$\Sigma^\Lambda$ and 256 for the two-particle vertex. For a 
spin-polarized two-level dot the multi-indices take two values and
one obtains 4 equations for $\Sigma^\Lambda$ and 16 for the 
two-particle vertex. The number of
independent equations can be significantly reduced (see below)
taking into account the antisymmetry of the two-particle vertex 
and the spin symmetry (for spin-degenerate levels) 
both being preserved by Eqs.\ (\ref{finalflowsigma})
and (\ref{finalflowgamma}). 
The initial conditions at $\Lambda = \Lambda_0 \to \infty$ are given by
$\Sigma_{1,1'}^{\Lambda_0}=0$ while
$\gamma^{\Lambda_0}_{1',2';1,2}$ is given by the bare antisymmetrized
two-body interaction.
In the spin-polarized case the only nonzero components of the 
two-particle vertex at $\Lambda=\Lambda_0\to\infty$ are
\begin{equation}
\label{ini1}
\gamma^{\Lambda_0}_{1,2;1,2}=\gamma^{\Lambda_0}_{2,1;2,1}=
U \hspace{0.7cm} 
\textnormal{and}\hspace{0.7cm} 
\gamma^{\Lambda_0}_{1,2;2,1}=\gamma^{\Lambda_0}_{2,1;1,2}=-U \, .
\end{equation}
In the model including spin the initial conditions take the form
\begin{eqnarray}
\label{ini2}
\gamma^{\Lambda_0}_{1\uparrow,1\downarrow;1\uparrow,1\downarrow}&=U &
\; , \;\;\;\; 
\gamma^{\Lambda_0}_{1\uparrow,2\uparrow;1\uparrow,2\uparrow}=U  \; , \;\;\;\; 
\gamma^{\Lambda_0}_{1\uparrow,2\downarrow;1\uparrow,2\downarrow}=U \nonumber\\
\gamma^{\Lambda_0}_{2\uparrow,2\downarrow;2\uparrow,2\downarrow}&=U & \; , \;\;\;\; 
\gamma^{\Lambda_0}_{1\downarrow,2\downarrow;1\downarrow,2\downarrow}=U \; , \;\;\;\; 
\gamma^{\Lambda_0}_{1\downarrow,2\uparrow;1\downarrow,2\uparrow}=U.
\end{eqnarray}
All other components which do not arise out of these by 
permutations
($\gamma^{\Lambda_0}_{1,2;1',2'}=\gamma^{\Lambda_0}_{1',2';1,2}$ 
and $\gamma^{\Lambda_0}_{1,2;1',2'}=-\gamma^{\Lambda_0}_{1,2;2',1'}$) are zero.
The self-energy matrix and thus the one-particle Green function is  
completely independent of the spin direction and in the following we
suppress the spin indices. 

As already mentioned the present approximation leads to a
frequency-independent self-energy. This allows for a simple
single-particle interpretation of its matrix elements. The sum of the
$\Sigma_{j,j}^\Lambda$ and the bare level position correspond to the
flowing effective level positions, $\varepsilon_j^{\Lambda} =
\varepsilon_j + \Sigma_{j,j}^\Lambda$, while
$t^\Lambda=-\Sigma_{1,2}^\Lambda = -\Sigma_{2,1}^\Lambda$ is a hopping
between the levels 1 and 2 generated in the fRG flow. The fRG
formalism then reduces to a set of coupled differential flow equations
for $\varepsilon_j^{\Lambda}$, $t^\Lambda$ and a few (one in the
spin-polarized case and seven for spin-degenerate levels) independent
components of the two-particle vertex. These flow equations
can easily be integrated numerically using standard 
routines.  It is important to note that
although we start out with intra- and inter-level Coulomb interactions of
equal strengths they generically become different during the fRG flow
(because of the different $\Gamma_j^l$). Furthermore, additional
interaction terms which are initially zero will be generated in the
flow.  The set of equations significantly simplifies if the flow of
the vertex is neglected while the results remain qualitatively the
same. Within this additional approximation and for a spin-polarized
dot the flow equations for $\varepsilon_j^{\Lambda}$ and $t^\Lambda$
are explicitly given in Ref.\ \cite{VF}.  In certain limiting cases it
is even possible to analytically solve the differential equations
\cite{VF,TCV}. However, in the present work, the flow of the vertex 
is retained which clearly improves the quality of 
approximation \cite{TCV}.

At the end of the fRG flow, the full Green function takes  the
form 
$[{\mathcal G}(i\omega)]^{-1}_{j,j'} = i \omega \delta_{j,j'} -
h_{j,j'}(i\omega)$  with an effective,
{\it noninteracting} (but $V_g$-, $U$- and $\omega$-dependent) ``Hamiltonian'' 
\begin{eqnarray}
\label{effham}
h_{j,j'}(i \omega) = h_{0;j,j'}(i \omega) -  \Sigma_{j,j'} \; .  
\end{eqnarray}
In a last step we have to perform the analytic continuation to 
the real frequency axis $i\omega \to \omega +i 0$. This is 
straightforward, as the only frequency dependence of $h(i \omega)$ 
is the trivial one of the lead contribution Eq.\ (\ref{leadpotdefwideband}). 
Then $\tilde t(\omega)$ can be computed using 
Eq.\ (\ref{transformula}). 

\subsection{The NRG approach}

The numerical renormalization group was invented by K.~G.~Wilson in 1974
as a nonperturbative renormalization scheme for the Kondo model
\cite{KondoModel}. It was later extended to the fermionic
 \cite{Krishna-murthy, Krishna2} Anderson model
which describes a localized electronic state
coupled to a fermionic bath.
The NRG allows thermodynamic and dynamic properties of such strongly
correlated systems to be calculated at zero and finite temperature
\cite{Costi,BullaS,Bulla2,Hofstetter,Verstraete,Weichsel}.

The key idea of NRG is to discretize the conduction band of the bath
logarithmically, leading to a tight-binding chain
for which  the hopping matrix elements between the successive sites
fall off exponentially with $\Lambda_{\rm NRG}^{-n/2}$,
where $\Lambda_{\rm NRG}>1$ is the discretization parameter, typically
$1<\Lambda_{\rm NRG}<3$, and $n$ is the site index. This energy scale separation
ensures that the problem can be solved iteratively by adding one site at a
time and diagonalizing the enlarged system at each step, thereby
resolving successively smaller and smaller energy scales. Thus, by
choosing the length $N$ of the chain so large
that the corresponding energy scale $\sim\Lambda_{\rm NRG}^{-N/2}$
is smaller than all other energies in the problem,
all relevant energy scales can be resolved and treated
properly. Since  the dimension of the Hilbert space of
the chain increases exponentially with the length of
the chain, a truncation scheme has to be adopted,
according to which only the lowest $N_{\rm kept}$
eigenstates of the chain are retained at each iteration.
Recently, it was shown that by also keeping track of
discarded states a complete, but approximate, basis of states can
be constructed \cite{Anders}. This can be used
to calculate spectral functions which rigorously
satisfy relevant sum rules \cite{Weichsel}.

In order to obtain the transmission through the dot
$\tilde{t}(\omega)$ Eq.\ (\ref{transformula})
we follow \cite{Weichsel} and \cite{BullaS} to compute the imaginary
part of the local Green functions at temperature $T$,
using the Lehmann representation
\begin{eqnarray}
 \mbox{Im} \, \mathcal{G}_{j,j\prime}(\omega)=
  &-\pi&
  \frac{e^{-\omega_n/T}}{Z}\,
  \langle n| d_{j,\sigma} |m\rangle
  \,
  \langle m| d_{j\prime,\sigma}^\dag |n\rangle
  \,
  \mathcal{\delta}(\omega-[\omega_m-\omega_n])
  \nonumber
  \\*
  &-\pi&
  \frac{e^{-\omega_n/T}}{Z}\,
    \langle n| d_{j\prime,\sigma}^\dag |m\rangle
  \,
  \langle m| d_{j,\sigma} |n\rangle
  \,
  \mathcal{\delta}(\omega+[\omega_m-\omega_n]),
\end{eqnarray}
with $Z=\sum_n e^{- \omega_n/T}$, the many-body eigenstates $\left| n
\right>$ and eigenenergies $\omega_n$.
Since these are causal functions, the real part can be accessed by
performing a Kramers-Kronig transformation \cite{KK}.

In the next three sections we present our results. In 
Sec.\ \ref{resultsnon} for the $U=0$
case. In Sec.\ \ref{resultsgeneric} we present the generic 
phase lapse scenario for interacting spin-polarized dots, 
compare to the mean-field results and investigate the role of 
finite temperatures. Finally, in Sec.\ \ref{resultsspin} we study the
spinful two-level dot. 

\section{Results: Noninteracting dots}
\label{resultsnon}

The large number of parameters makes it essential to analyze 
the transmission for the noninteracting case before considering 
the effect of two-particle interactions. We focus on $T=0$.
A  closed expression for $|t(V_g)|$ and $\theta(V_g)$ (for a fixed
spin direction) at $U=0$ 
can be obtained from Eqs.~(\ref{transamp}) and 
(\ref{transformula}) by replacing $ {\mathcal G}(0+i0)$ 
by ${\mathcal G}_0(0+i0)$, 
\begin{eqnarray}
\label{U0conduct}
\mbox{} \hspace{-1.5cm} |t(V_g)| & = &\frac{ 2 \left[\Gamma_1^L \Gamma_1^R
  \varepsilon^2_{2}+ \Gamma_2^L \Gamma_2^R
  \varepsilon^2_{1} + 2 s \sqrt{\Gamma_1^L \Gamma_1^R 
\Gamma_2^L \Gamma_2^R}  \varepsilon_{1}
\varepsilon_{2}\right]^{1/2}}{\left[ \left( \Gamma_1^L \Gamma_2^R + \Gamma_2^L
\Gamma_1^R  - 2 s \sqrt{\Gamma_1^L \Gamma_1^R 
\Gamma_2^L \Gamma_2^R}- \varepsilon_{1}
\varepsilon_{2}\right)^2 +\left(  \varepsilon_{1} \Gamma_2 +
\varepsilon_{2} \Gamma_1 \right)^2\right]^{1/2} } \; , \\
\label{U0phase}
\mbox{} \hspace{-1.5cm} \theta(V_g) & = & \arctan{ \left[
\frac{\varepsilon_1\Gamma_2+\varepsilon_2\Gamma_1}{\varepsilon_1\varepsilon_
2-\left(\sqrt{\Gamma_1^L\Gamma_2^R}-s\sqrt{\Gamma_1^R\Gamma_2^L}\right)^2}
\right]} \, \mbox{mod} \, \pi \; ,
\end{eqnarray}
with $\Gamma_j = \sum_l \Gamma_j^l$. For later use we define
\begin{eqnarray}
\label{Gammatotdef}
\Gamma=\sum_{j,l} \Gamma_j^l \; .
\end{eqnarray}
For a fixed set of $\Gamma_j^l$ the $\delta$ dependence of 
$|t(V_g)|$ and $\theta(V_g)$ is shown in the first columns of Figs.\
\ref{fig1} and \ref{fig2} for $U/\Gamma=0.2$. The results are
qualitatively the same as those obtained for $U=0$.  
For generic level-lead couplings  $\Gamma_j^l$ 
the gate voltage dependence of
Eq.~(\ref{U0conduct}) in the limit of small and large $\delta/\Gamma$
is dominated by two peaks (of height $\leq 1$) and a 
transmission zero. Associated with the transmission zero is a $\pi$
phase lapse at the same gate voltage.  
The transmission zero (and phase lapse) follows from perfect destructive interference at a 
particular $V_g$ and in the limit of a strong
asymmetry in the coupling of the two levels to the leads, 
$\Gamma_1 \ll \Gamma_2$ or vice versa, can be understood as  
resulting from a Fano anti-resonance \cite{Fano}. Apparently the
Fano anti-resonance with vanishing transmission 
is robust if one goes away from this limit towards more symmetric 
level-lead couplings. Across each of the transmission resonances 
$\theta$ increases roughly by $\pi$ as expected for a Breit-Wigner resonance. 
Further details of $|t(V_g)|$ and $\theta(V_g)$ 
depend on $s$. For $s=+$ the transmission zero (and phase lapse) is located between 
the two conductance peaks for all $\delta$. For $\delta \to 0$ the 
resonance peak positions depend on the asymmetry of the $\Gamma_j^l$ and 
the separation of the peaks is small if the $\Gamma_j^l$ are close to
l-r symmetry, that is close to 
$\Gamma_j^L=\Gamma_j^R$. For l-r symmetric dots 
and $\delta=0$ the transmission zero (and phase lapse) disappears (not shown in the figures). 
This is an example of a 
submanifold in parameter space with nongeneric behavior. A complete 
account of such cases (which also remain nongeneric for $U>0$) 
is given in Refs.\ \cite{VF} and \cite{Slava}. 
As they require fine tuning these parameter sets are presumably 
irrelevant in connection with the
experiments and we will here only briefly mention results obtained in
such cases. 

For $s=-$ and fixed $\Gamma_j^l$
the position of the transmission zeros and phase lapses with respect to the CB peaks 
is different for small or large $\delta/\Gamma$ (see Fig.\ \ref{fig2}). 
At small $\delta/\Gamma$ it is located between 
the two conductance peaks, whereas
for large $\delta/\Gamma$  it lies on one of the outer sides of 
these peaks \cite{Silva}. In the crossover regime
between these limiting cases the height of one of the peaks decreases, 
while the
other becomes broader and splits up into two resonances separated by a
minimum with non-vanishing conductance [see Fig.\ \ref{fig2} (g) and (j)]. 
The crossover scale $\delta_c$ depends on the choice of 
$\Gamma_j^l$. 
For large $\delta/\Gamma$, $|t|$ has three local 
maxima, although the height of one of the maxima 
is significantly smaller than the height of the other two (not shown
in Fig.\ \ref{fig2}). 
For fixed, asymmetric $\Gamma_j^l$ and $\delta \to 0$ the separation 
of the two conductance peaks for $s=-$ is significantly larger 
than for $s=+$ [compare Figs.\ \ref{fig1} (a) and
\ref{fig2} (a)]. 

It is important to note that for small $\delta/\Gamma$ 
essential features of the universal phase lapse regime established in the
experiments are already found at $U=0$: regardless of the sign $s$ 
for generic $\Gamma_j^l$ (that is with the exception of a few cases
with increased
symmetry) two transmission resonances are separated by a transmission zero and $\pi$
phase lapse. At $U=0$ the peak separation is too small and the shape of 
the $V_g$ dependence of the transmission and phase close to the peaks 
is qualitatively different from those observed experimentally 
(namely Lorentzian-like for the magnitude of the transmission, s-shaped for 
the phase). As we show
next the latter problems do not arise for sufficiently large
interaction $U$, which in particular leads to an increased 
separation $U+\delta$ of the transmission peaks.

\section{Results: Spin-polarized dots}
\label{resultsgeneric}

\subsection{The generic phase lapse scenario}

\begin{figure}[t]	
	\centering
	      \vspace{-0.3cm}\includegraphics[width=0.352\textwidth,height=3.75cm,clip]{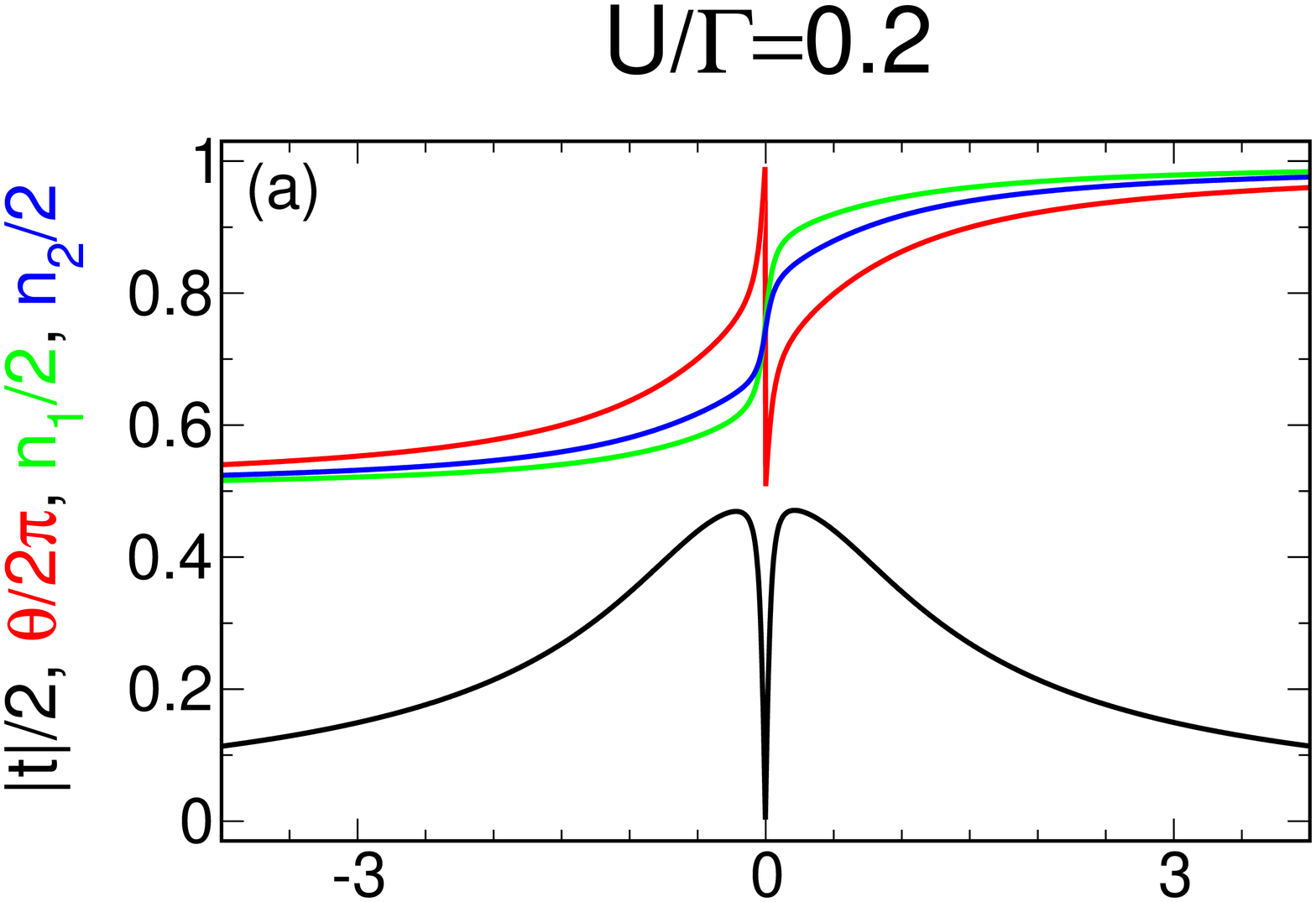}
	      \includegraphics[width=0.292\textwidth,height=3.75cm,clip]{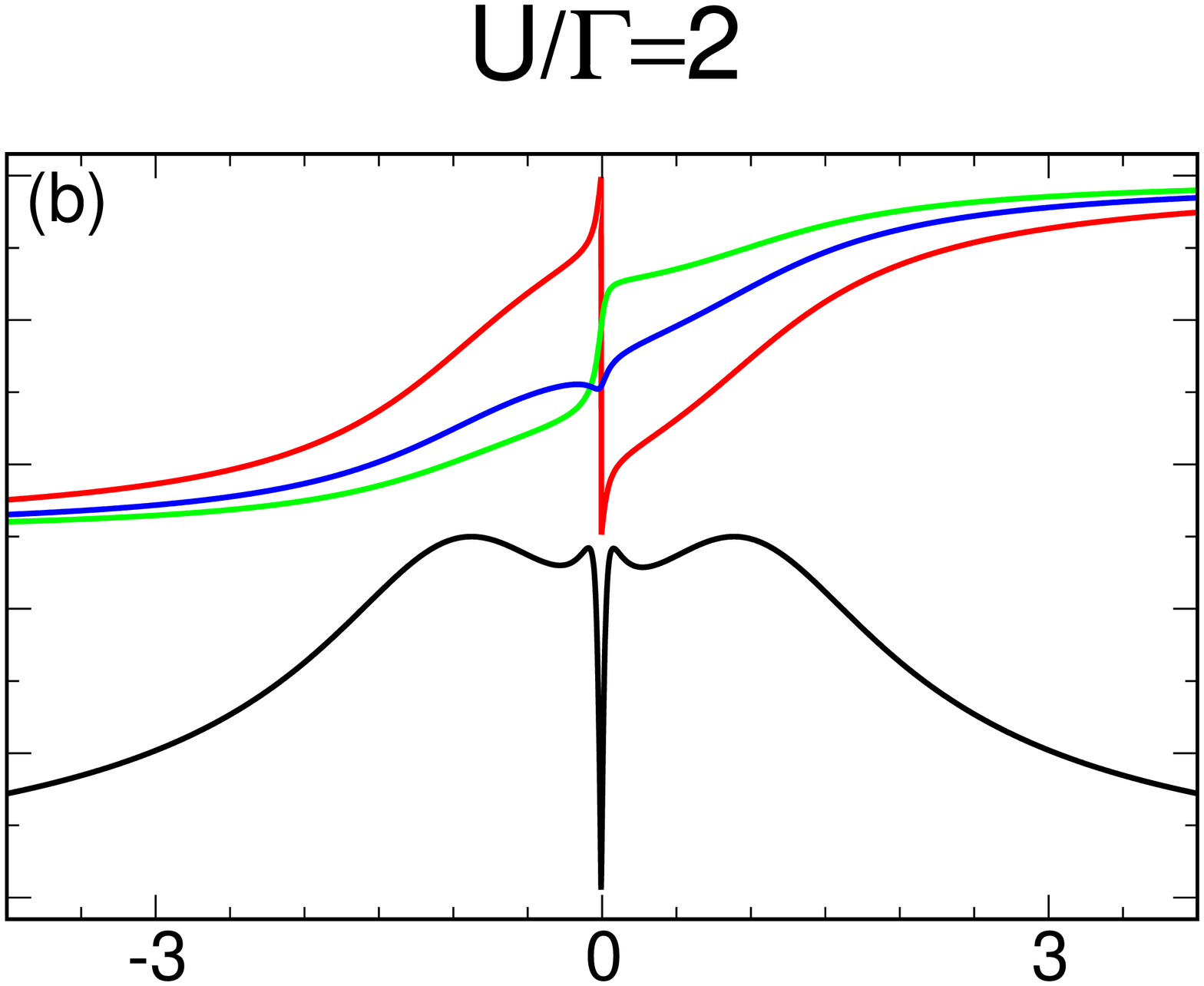}
        \includegraphics[width=0.33\textwidth,height=3.75cm,clip]{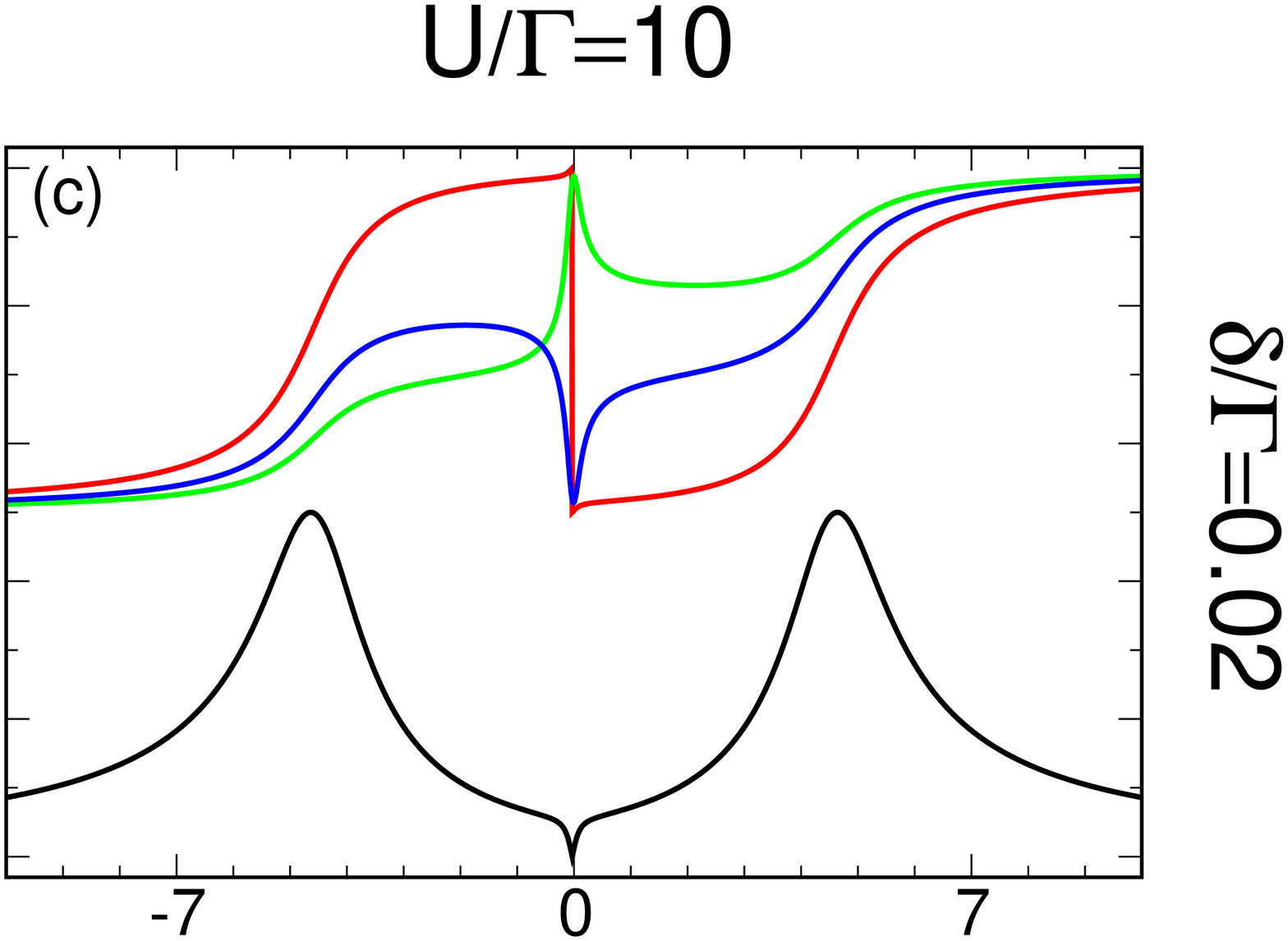}\vspace{0.3cm}
	      \includegraphics[width=0.352\textwidth,height=3.1cm,clip]{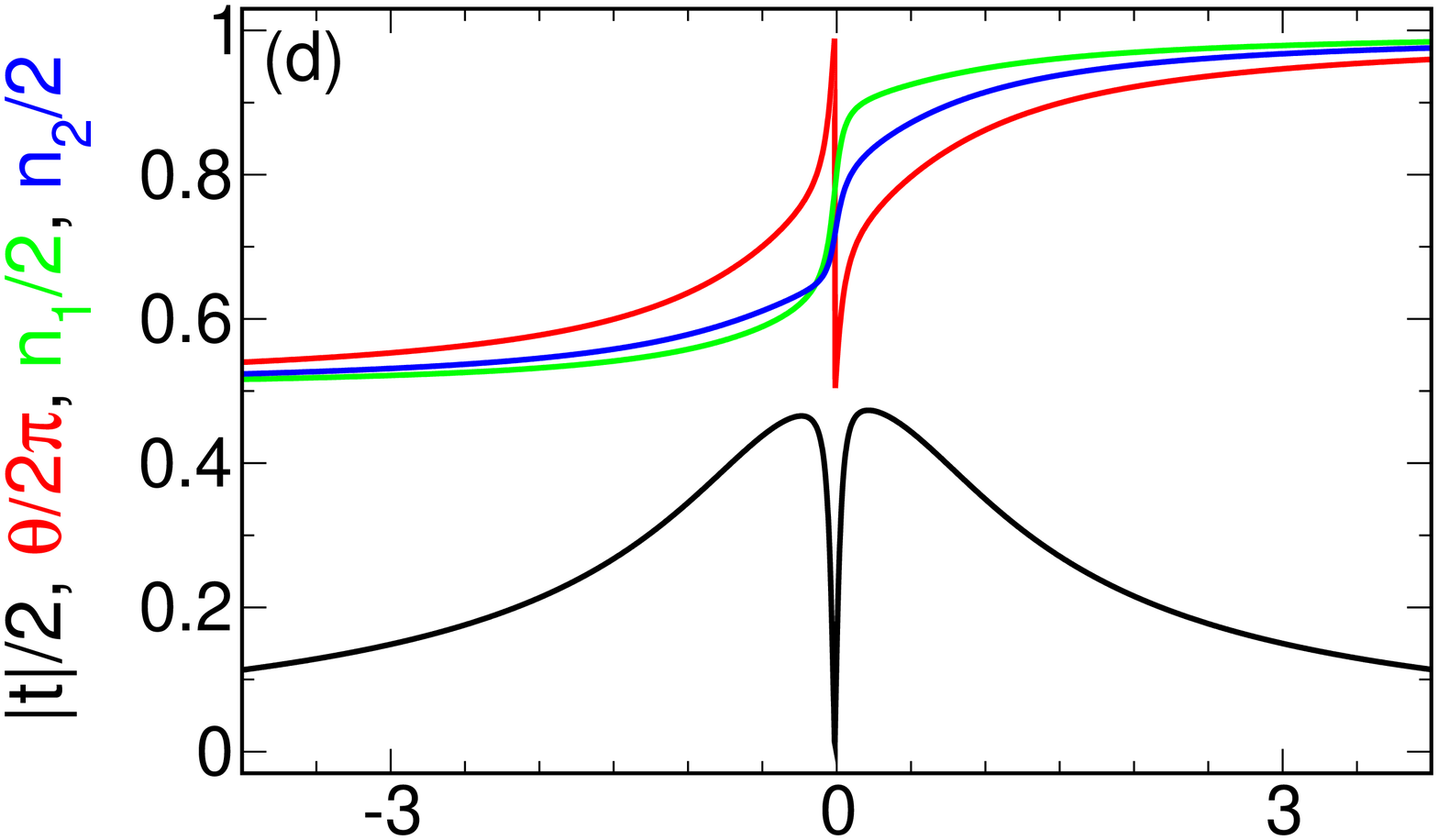}\hspace{0.001\textwidth}
	      \includegraphics[width=0.292\textwidth,height=3.1cm,clip]{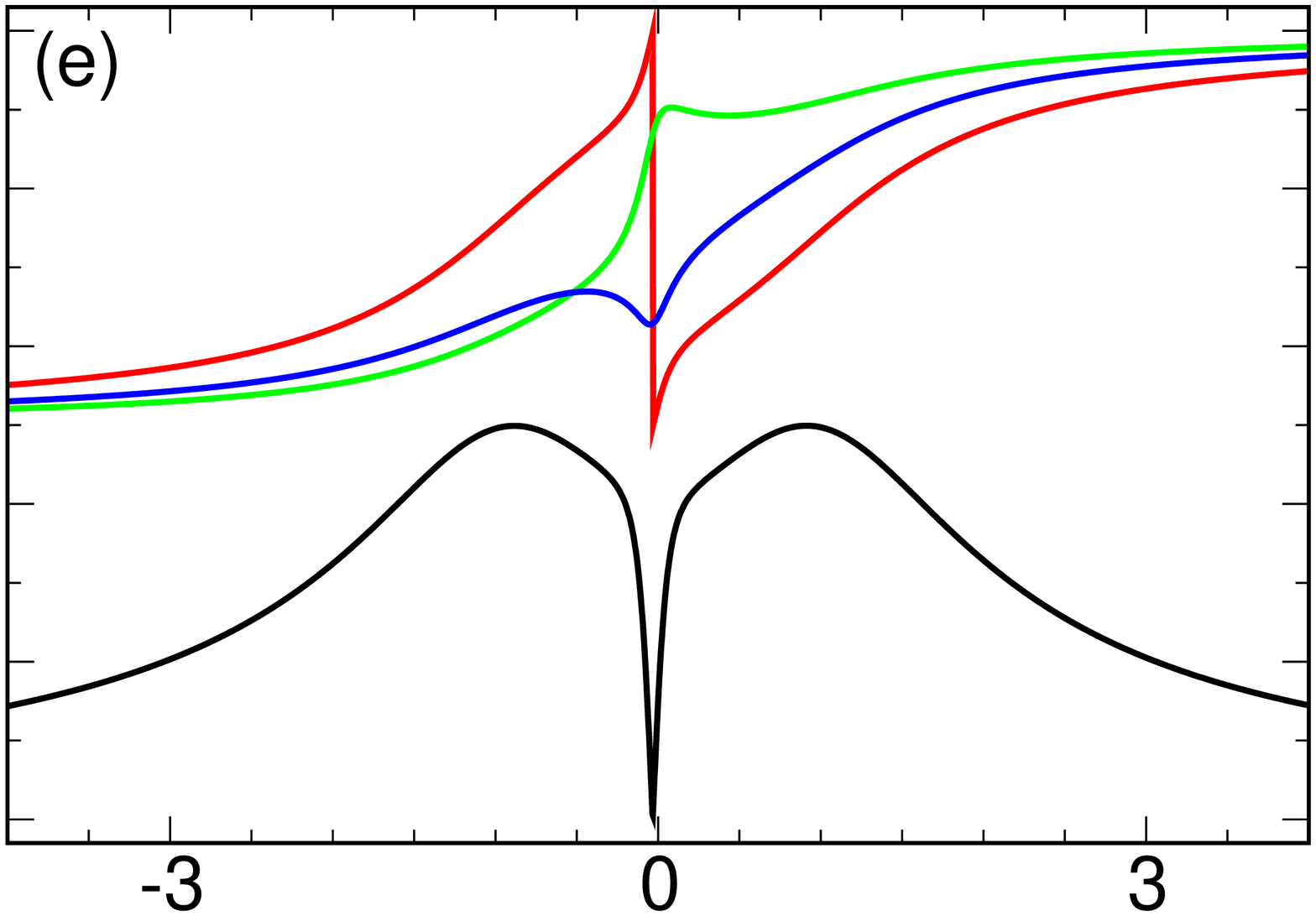}\hspace{0.001\textwidth}
        \includegraphics[width=0.33\textwidth,height=3.1cm,clip]{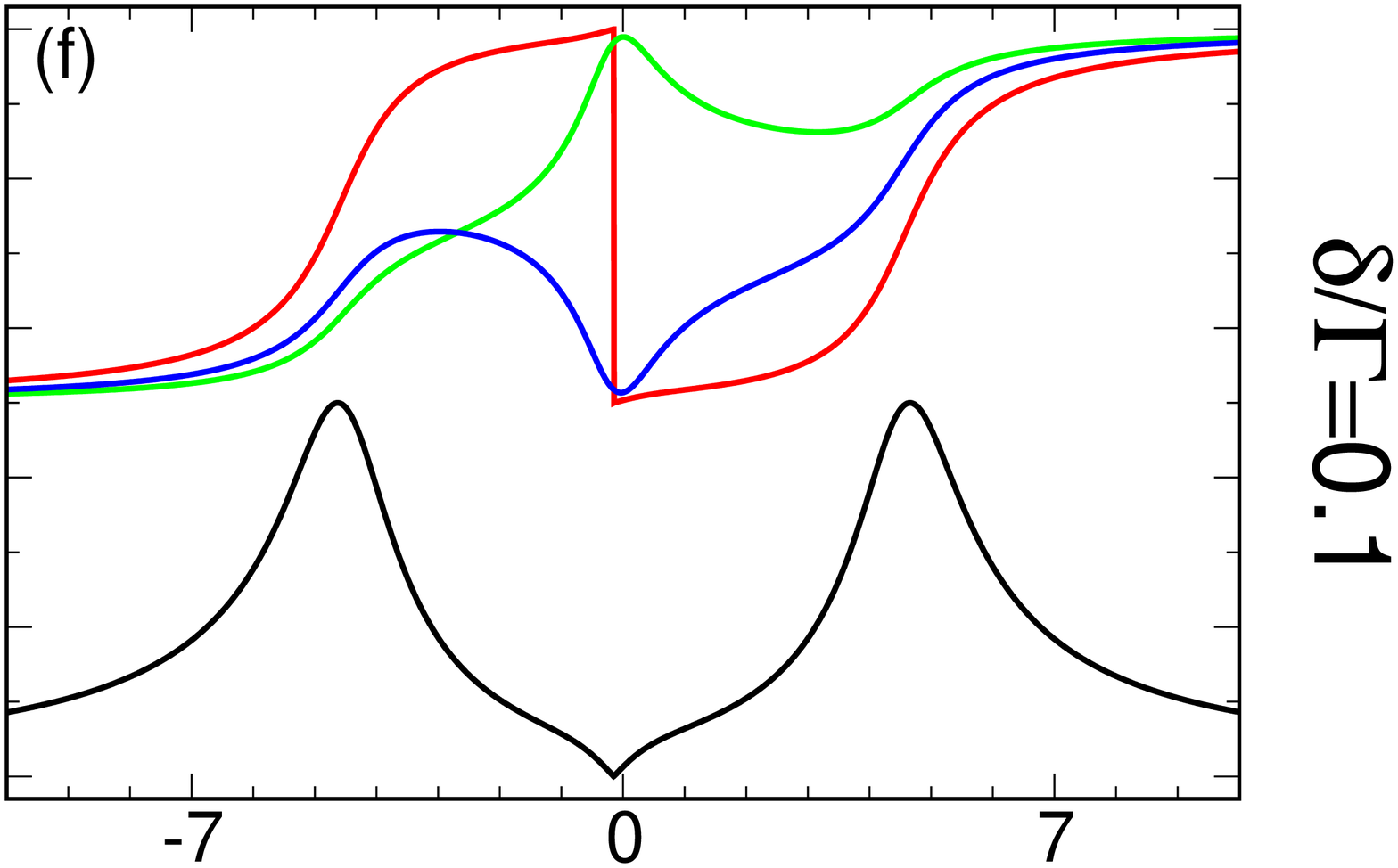}\vspace{0.3cm}
	      \includegraphics[width=0.352\textwidth,height=3.1cm,clip]{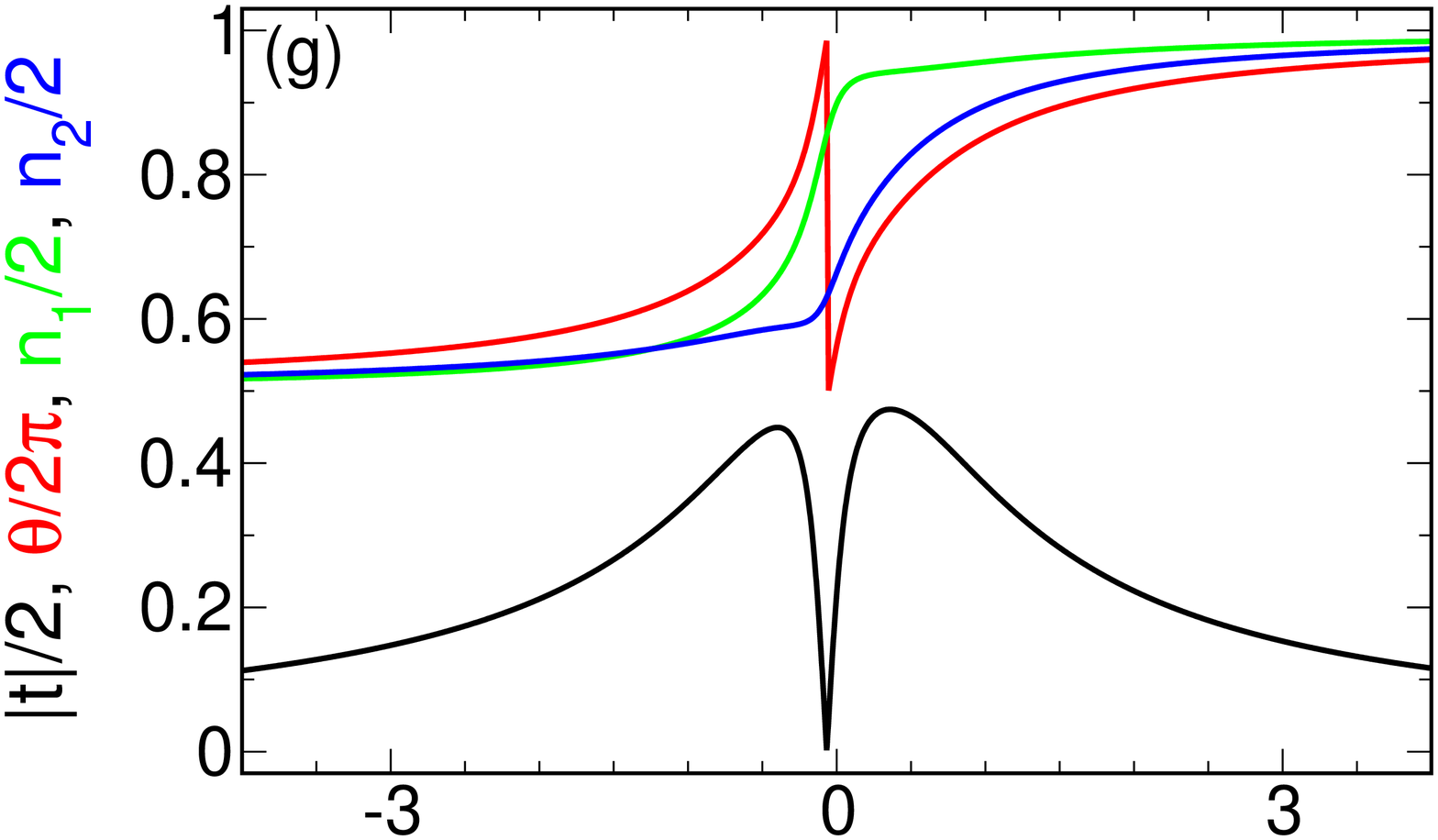}\hspace{0.001\textwidth}
	      \includegraphics[width=0.292\textwidth,height=3.1cm,clip]{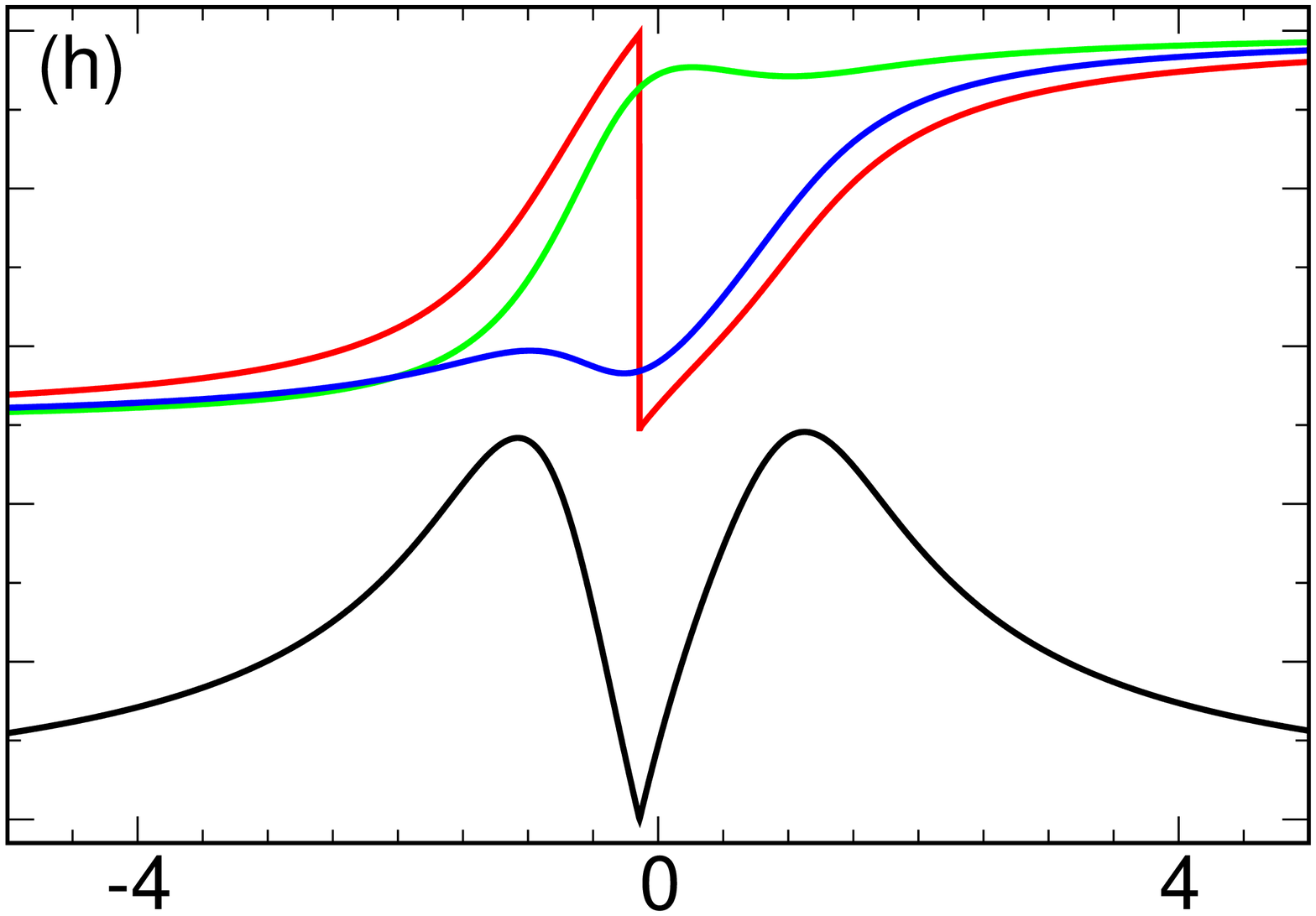}\hspace{0.001\textwidth}
        \includegraphics[width=0.333\textwidth,height=3.1cm,clip]{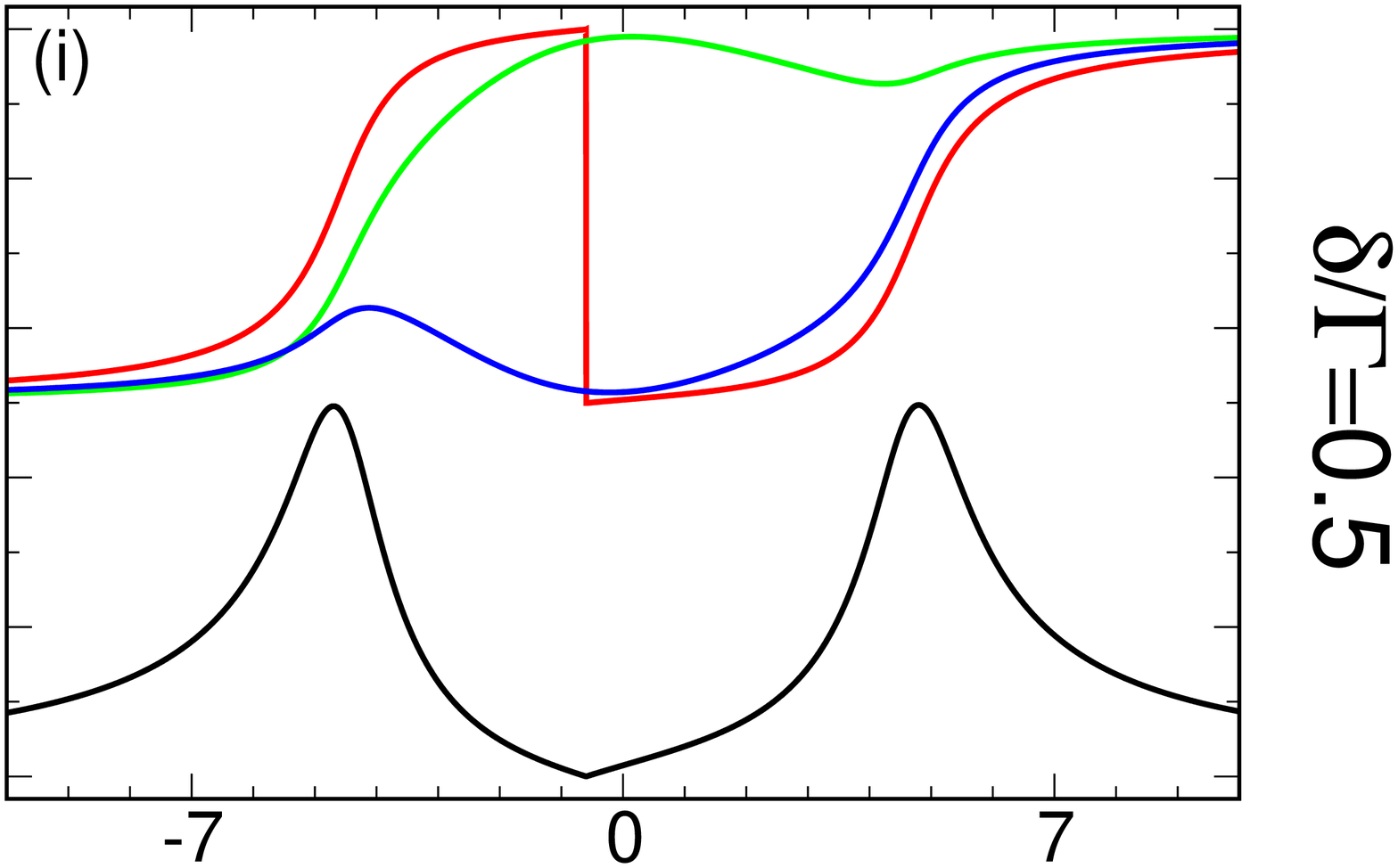}\vspace{0.3cm}
	      \includegraphics[width=0.352\textwidth,height=3.1cm,clip]{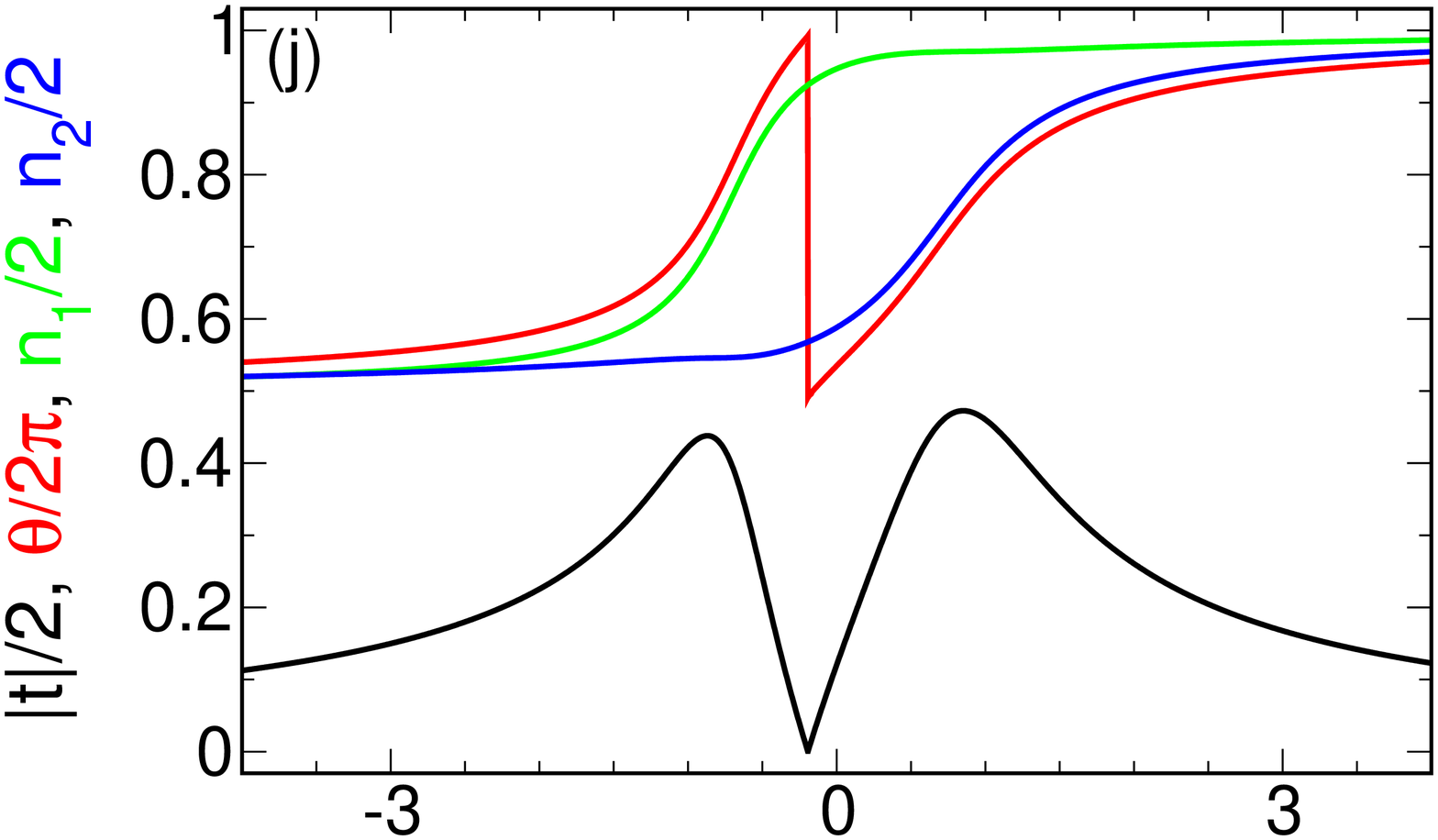}\hspace{0.001\textwidth}
	      \includegraphics[width=0.292\textwidth,height=3.1cm,clip]{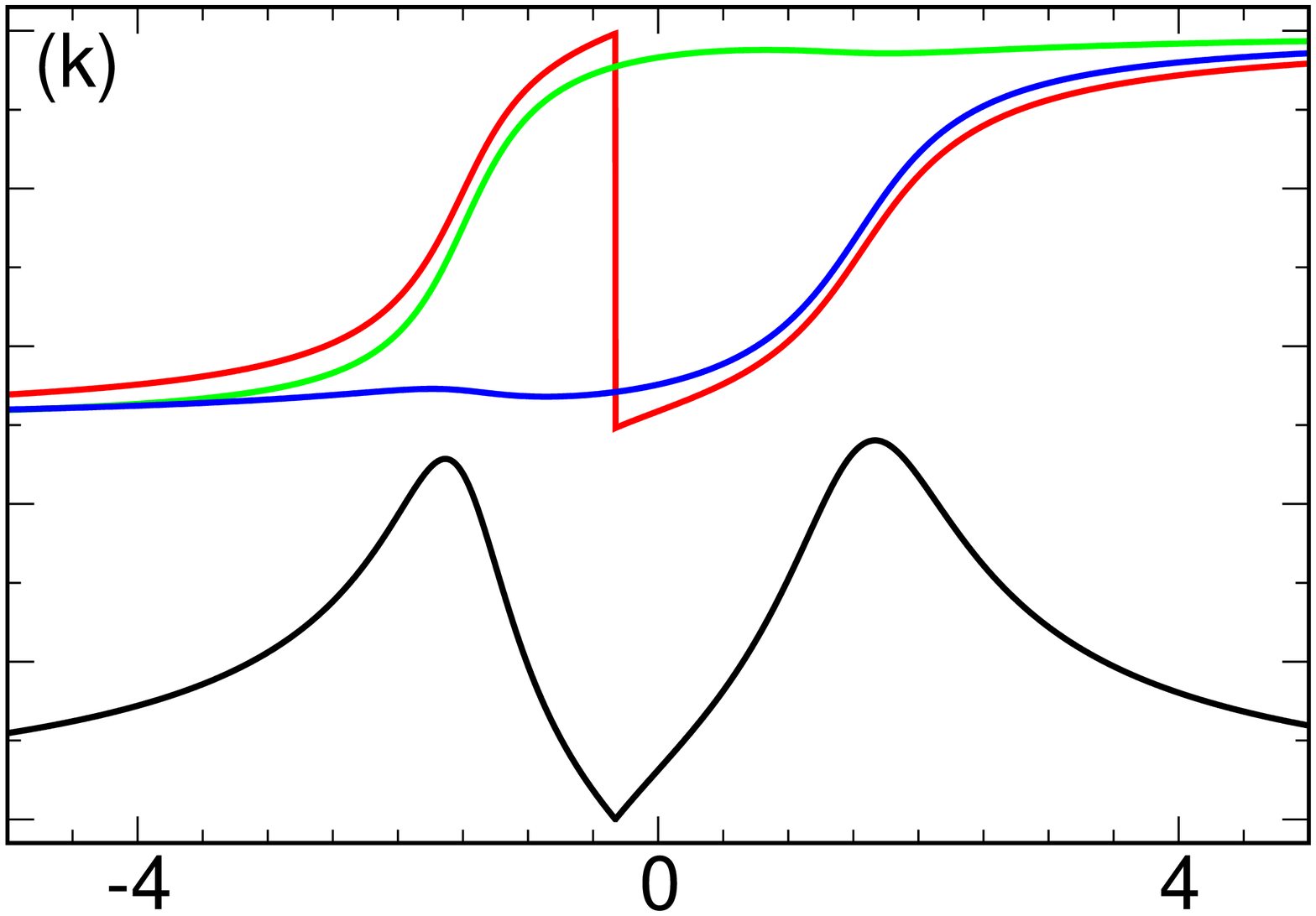}\hspace{0.001\textwidth}
        \includegraphics[width=0.33\textwidth,height=3.1cm,clip]{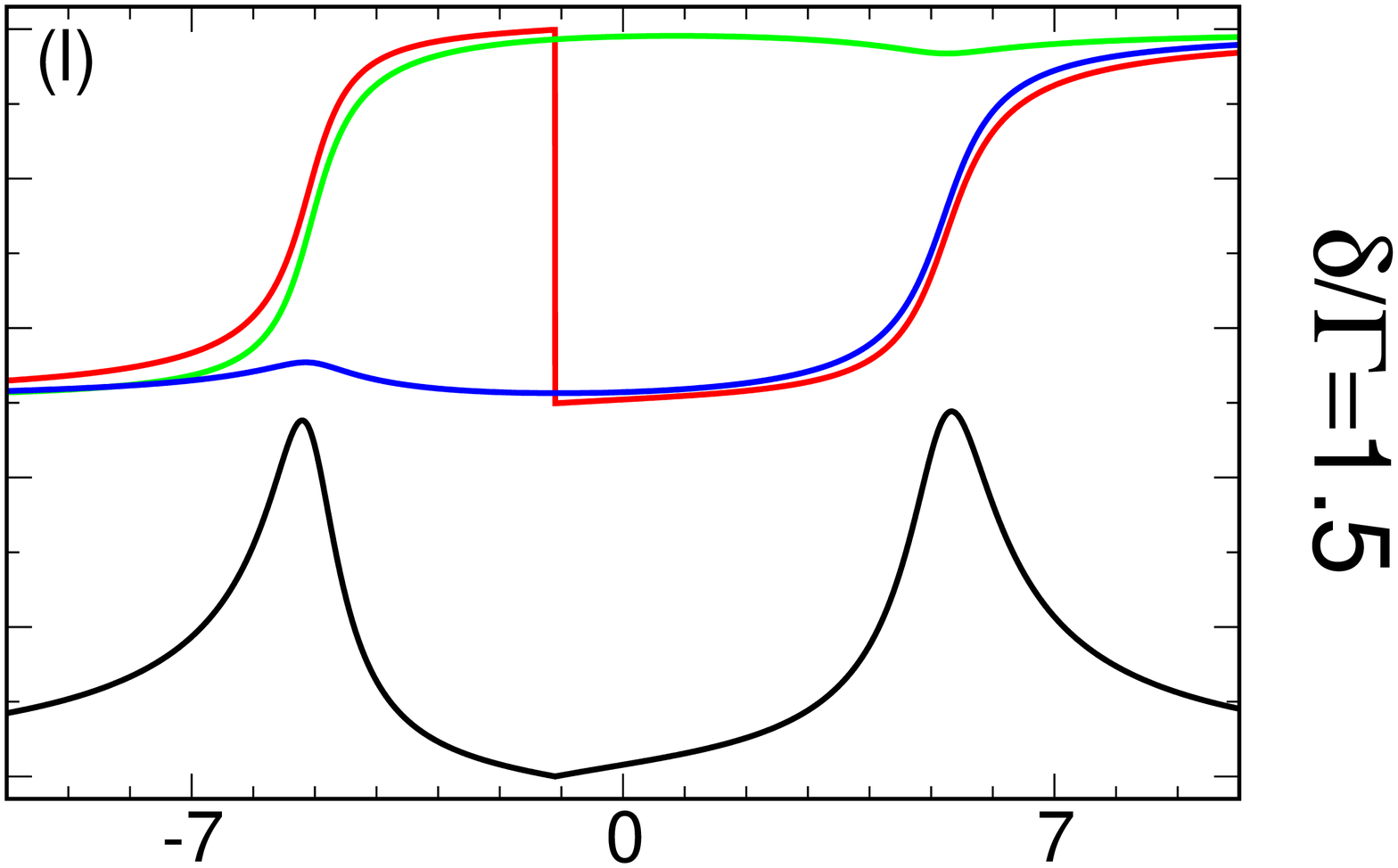}\vspace{0.3cm}
	      \includegraphics[width=0.352\textwidth,height=3.65cm,clip]{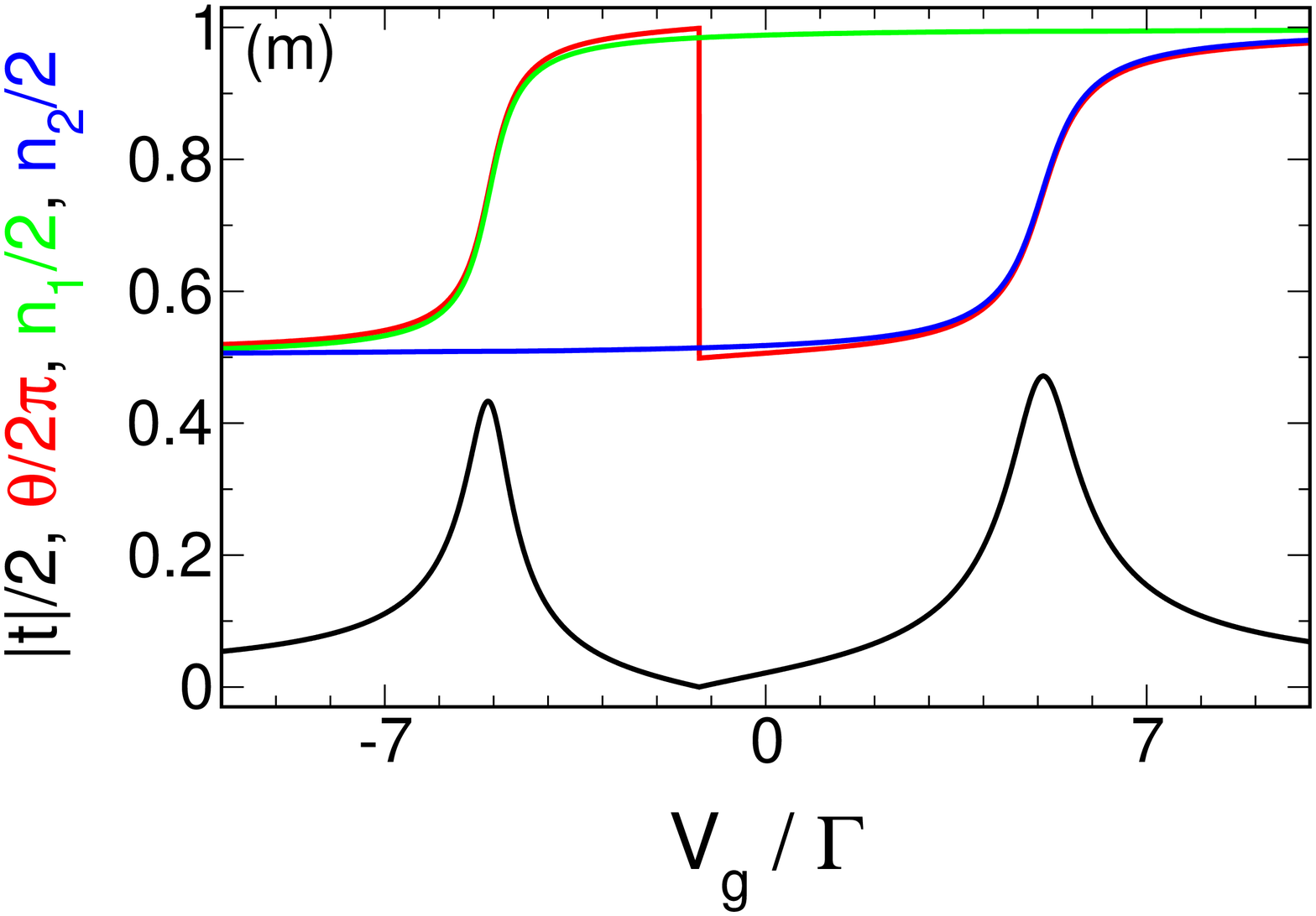}\hspace{0.001\textwidth}
	      \includegraphics[width=0.292\textwidth,height=3.65cm,clip]{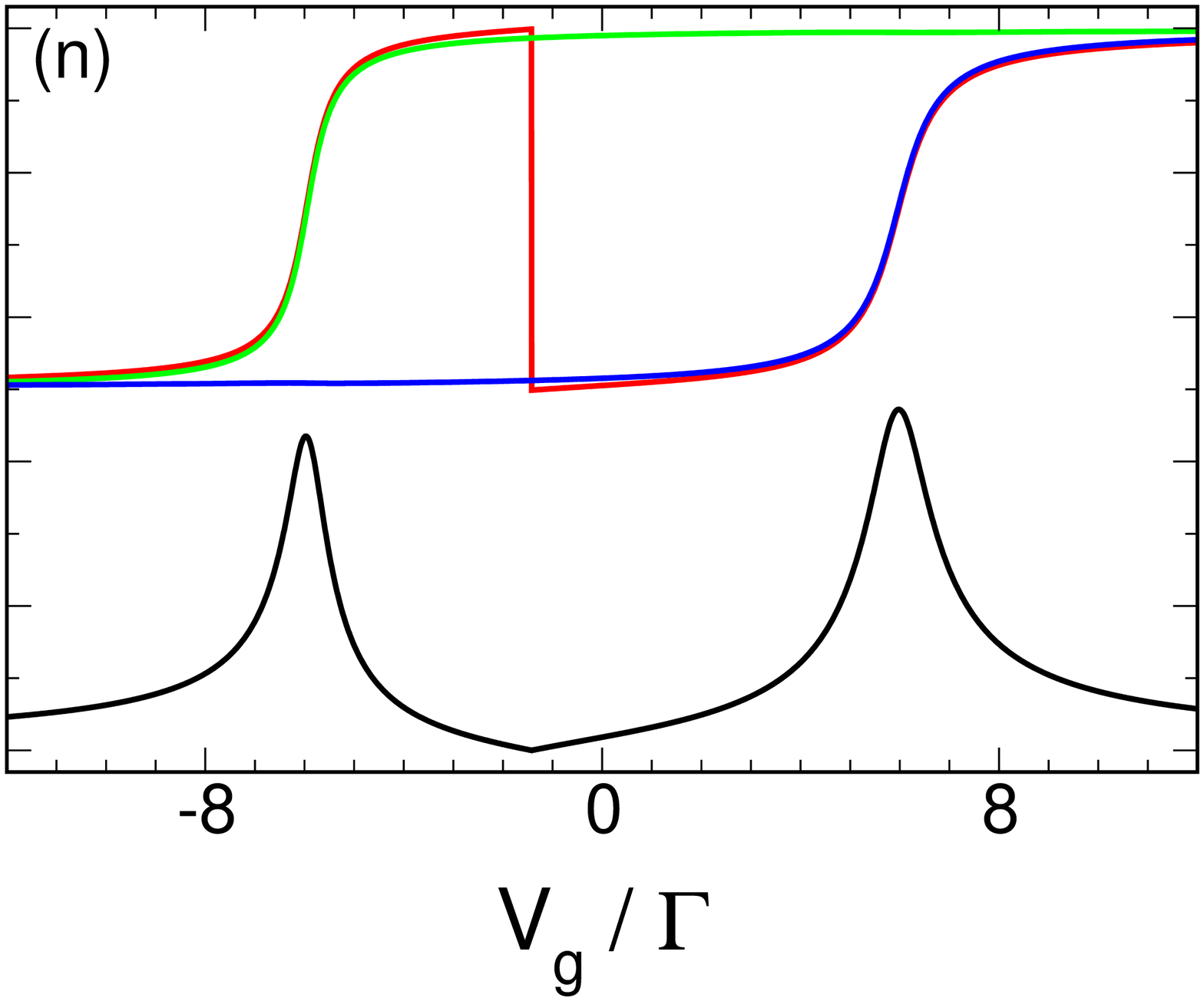}\hspace{0.001\textwidth}
        \includegraphics[width=0.33\textwidth,height=3.65cm,clip]{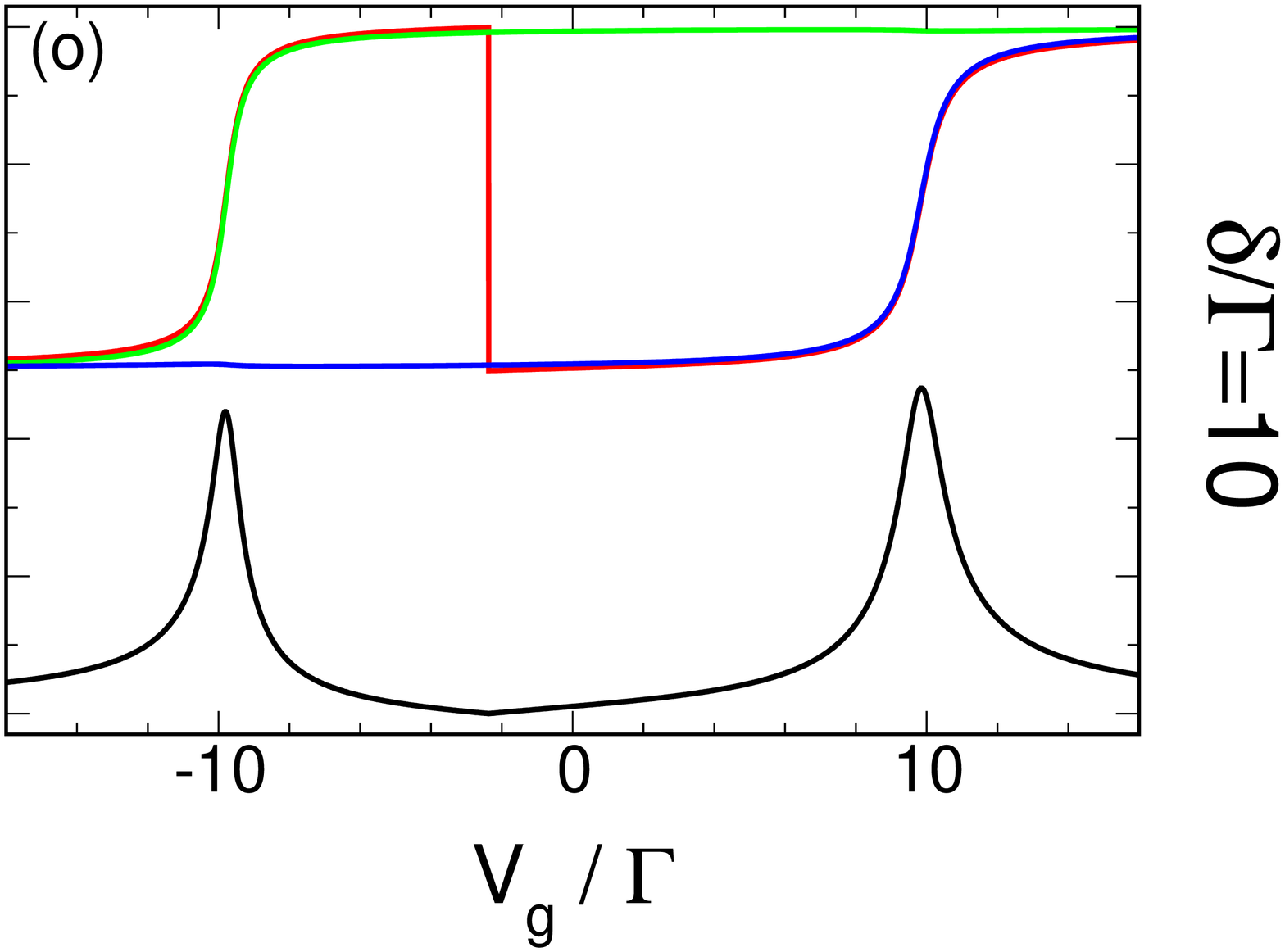}
        \caption{Systematic account of the energy
          scales $U/\Gamma$ and $\delta/\Gamma$ 
that govern the gate voltage $V_g$ dependence of the magnitude of 
the transmission $|t|$ (black), the transmission phase 
$\theta$ (red) and the level occupancies (green and blue) of 
a spin-polarized two-level dot at $T=0$. The parameters are 
$\gamma=\{0.1, 0.3, 0.4, 0.2\}$ and $s=+$. For better 
visibility $n_{1/2}$ were shifted by $1$. The depicted behavior 
is the generic one and in particular qualitatively independent of the actual 
choice of $\gamma$ (up to certain cases of increased symmetry; for
examples see the text). The behavior at $U/\Gamma=0.2$ is qualitatively
the same as the one at $U=0$. The results were obtained using
the truncated fRG.} 
\label{fig1}
\end{figure}

\begin{figure}[t]	
	\centering
	      \vspace{-0.3cm}\includegraphics[width=0.352\textwidth,height=3.75cm,clip]{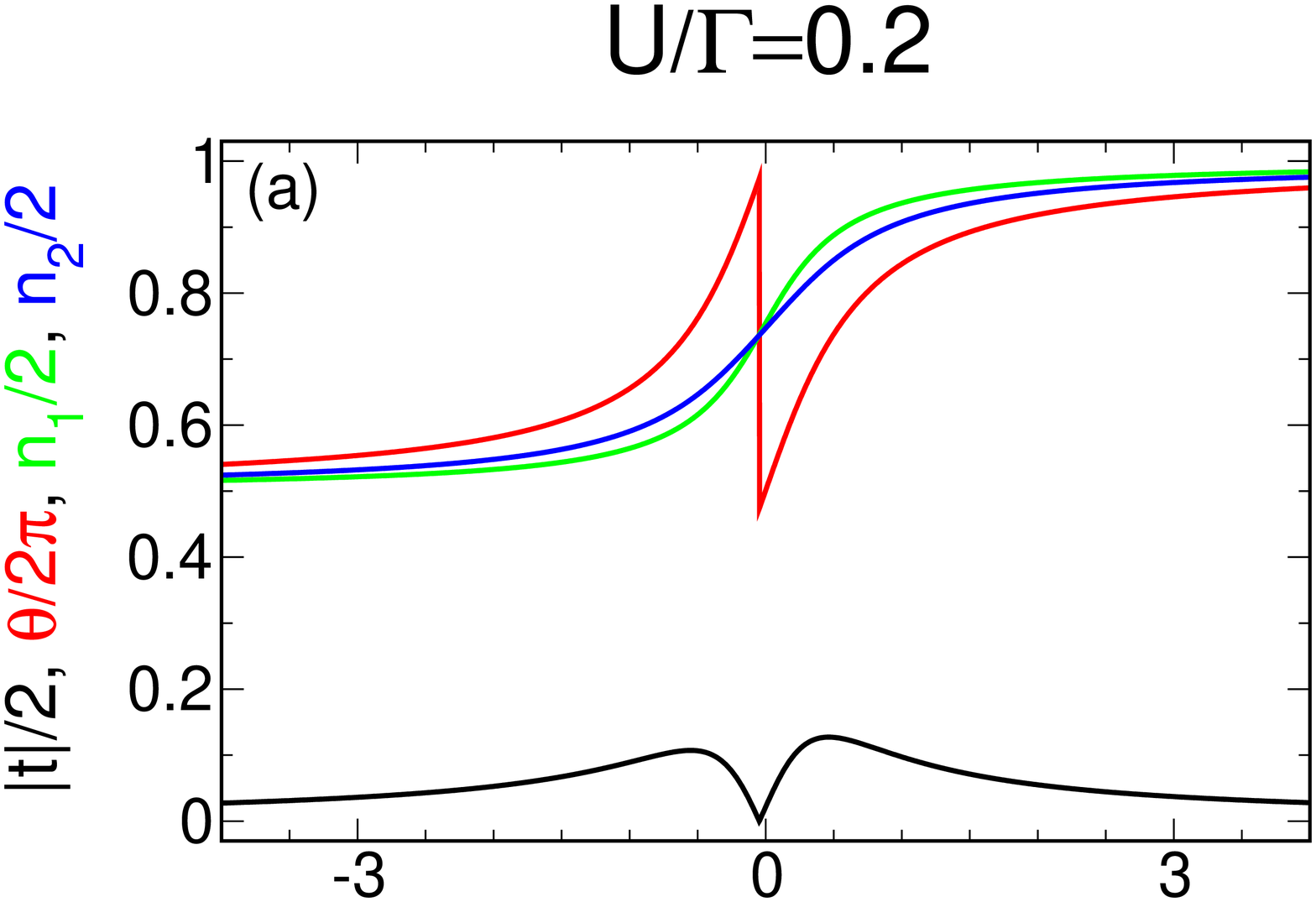}
	      \includegraphics[width=0.292\textwidth,height=3.75cm,clip]{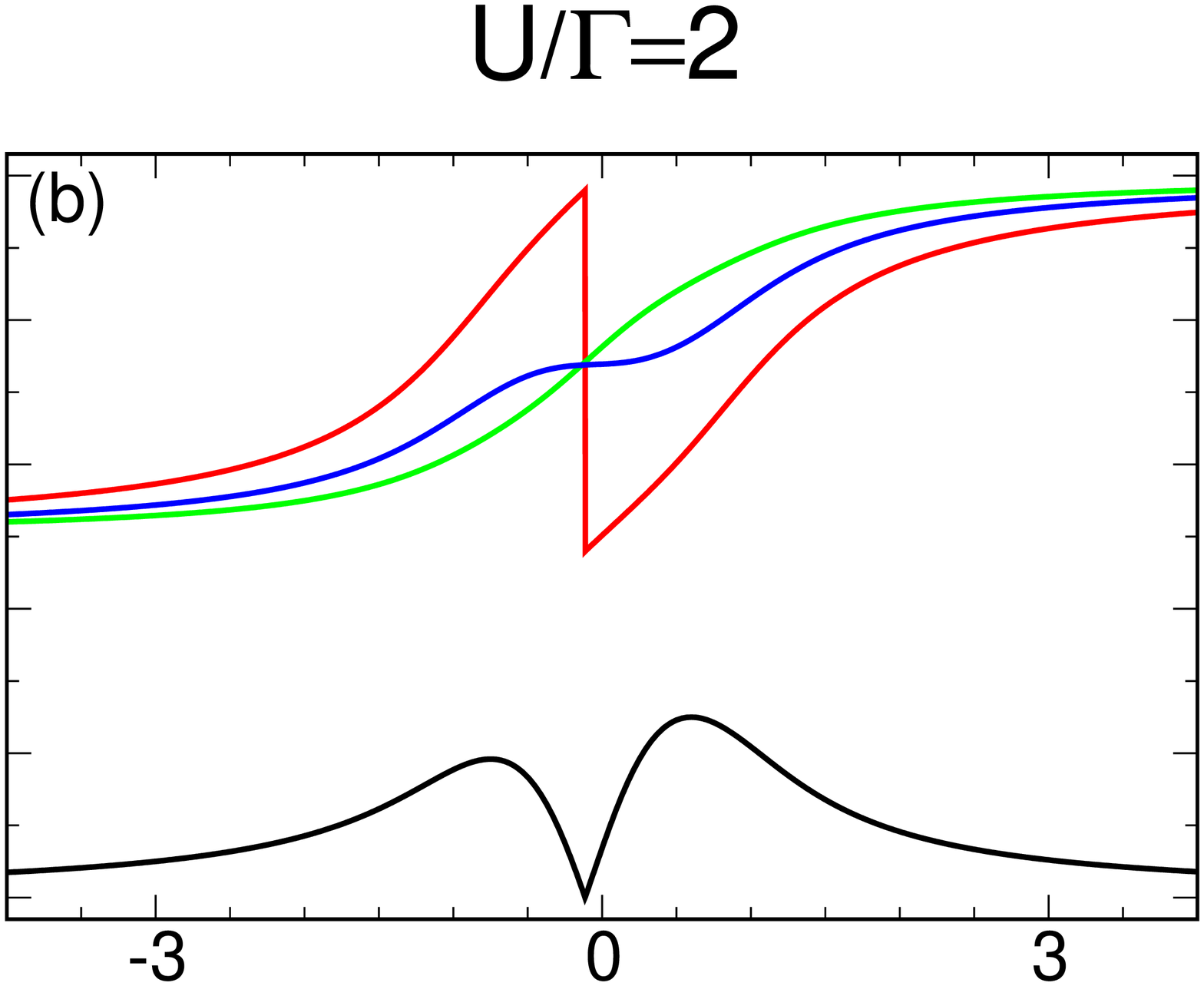}
        \includegraphics[width=0.33\textwidth,height=3.75cm,clip]{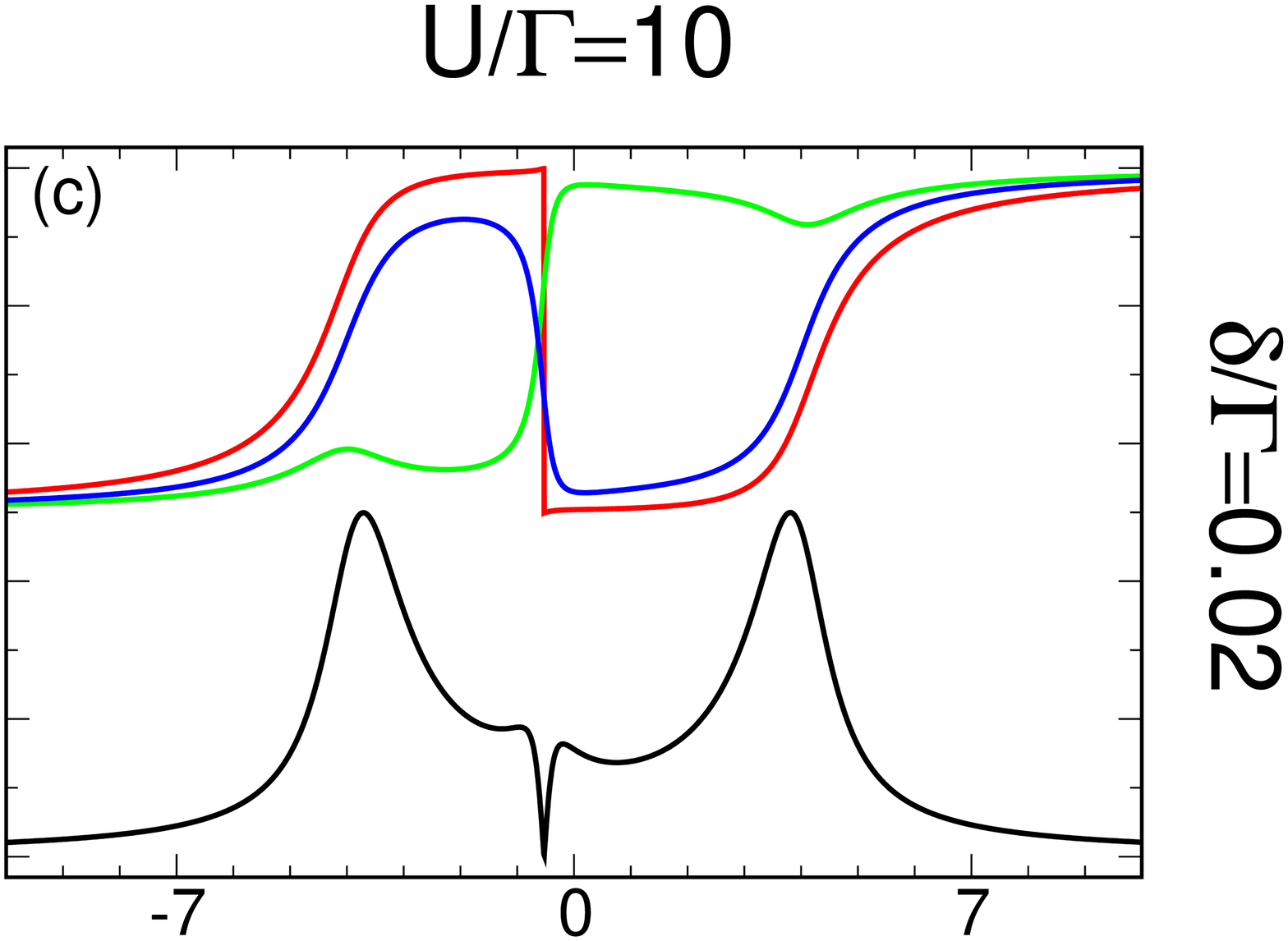}\vspace{0.3cm}
	      \includegraphics[width=0.352\textwidth,height=3.1cm,clip]{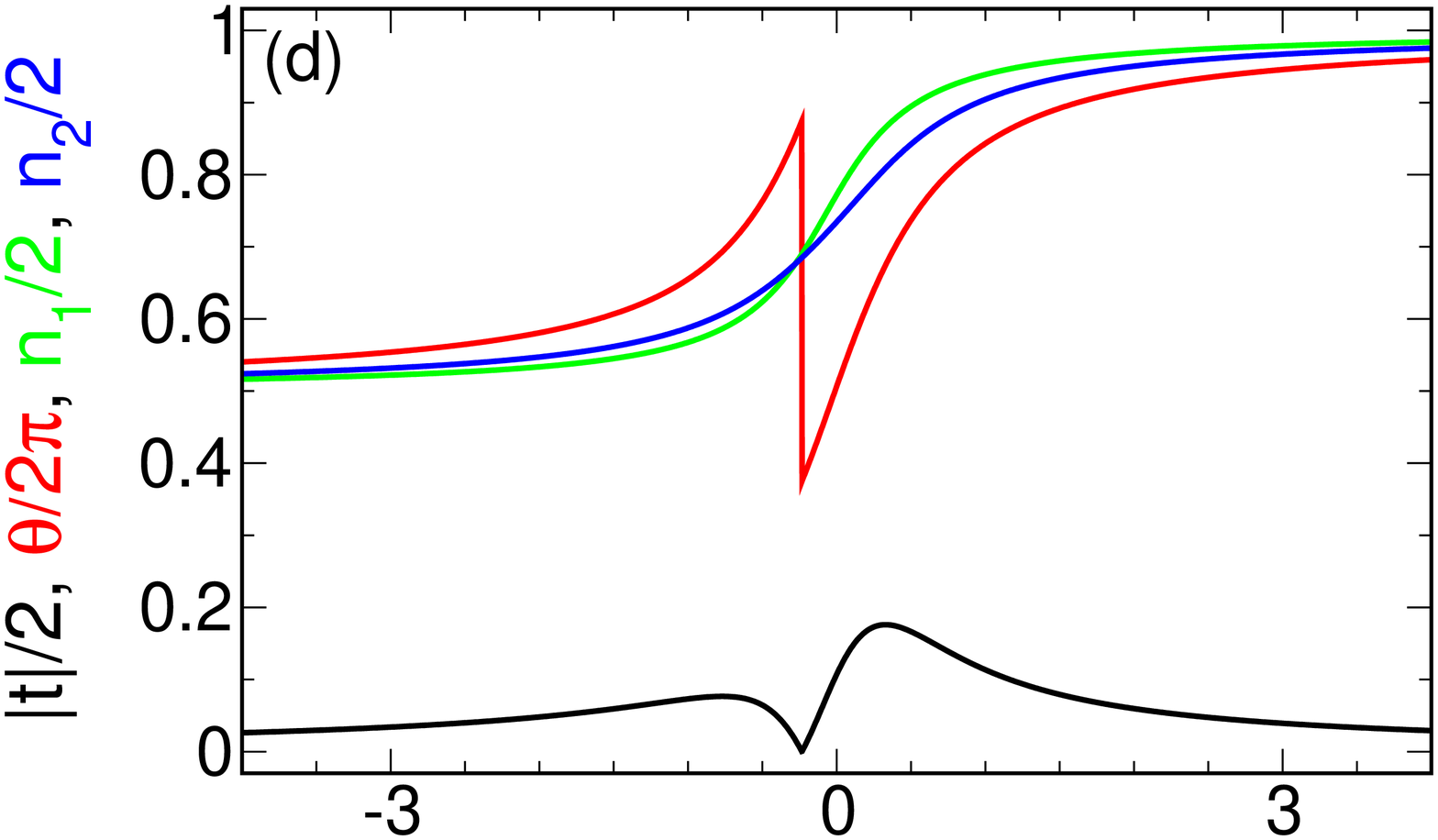}\hspace{0.001\textwidth}
	      \includegraphics[width=0.292\textwidth,height=3.1cm,clip]{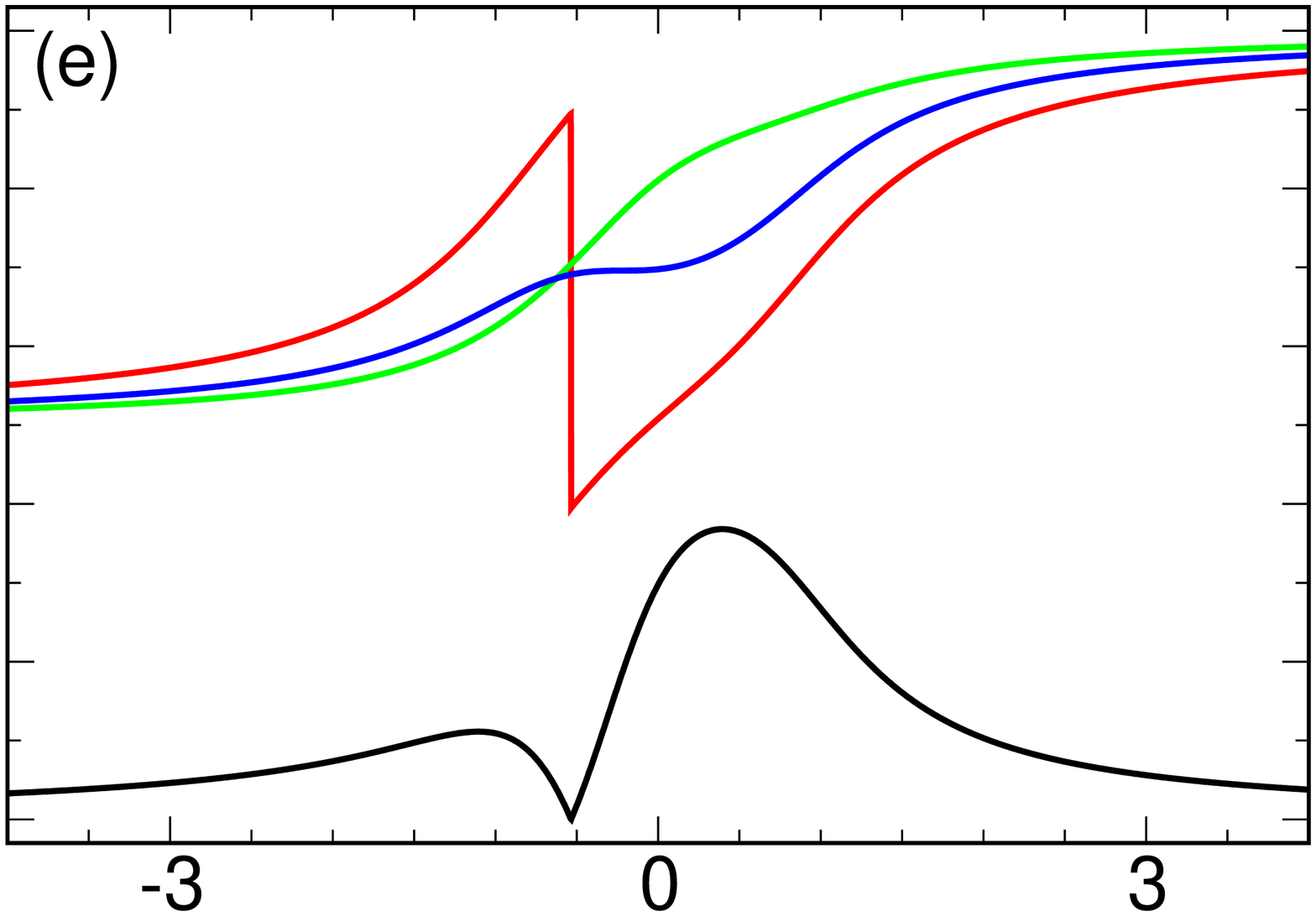}\hspace{0.001\textwidth}
        \includegraphics[width=0.33\textwidth,height=3.1cm,clip]{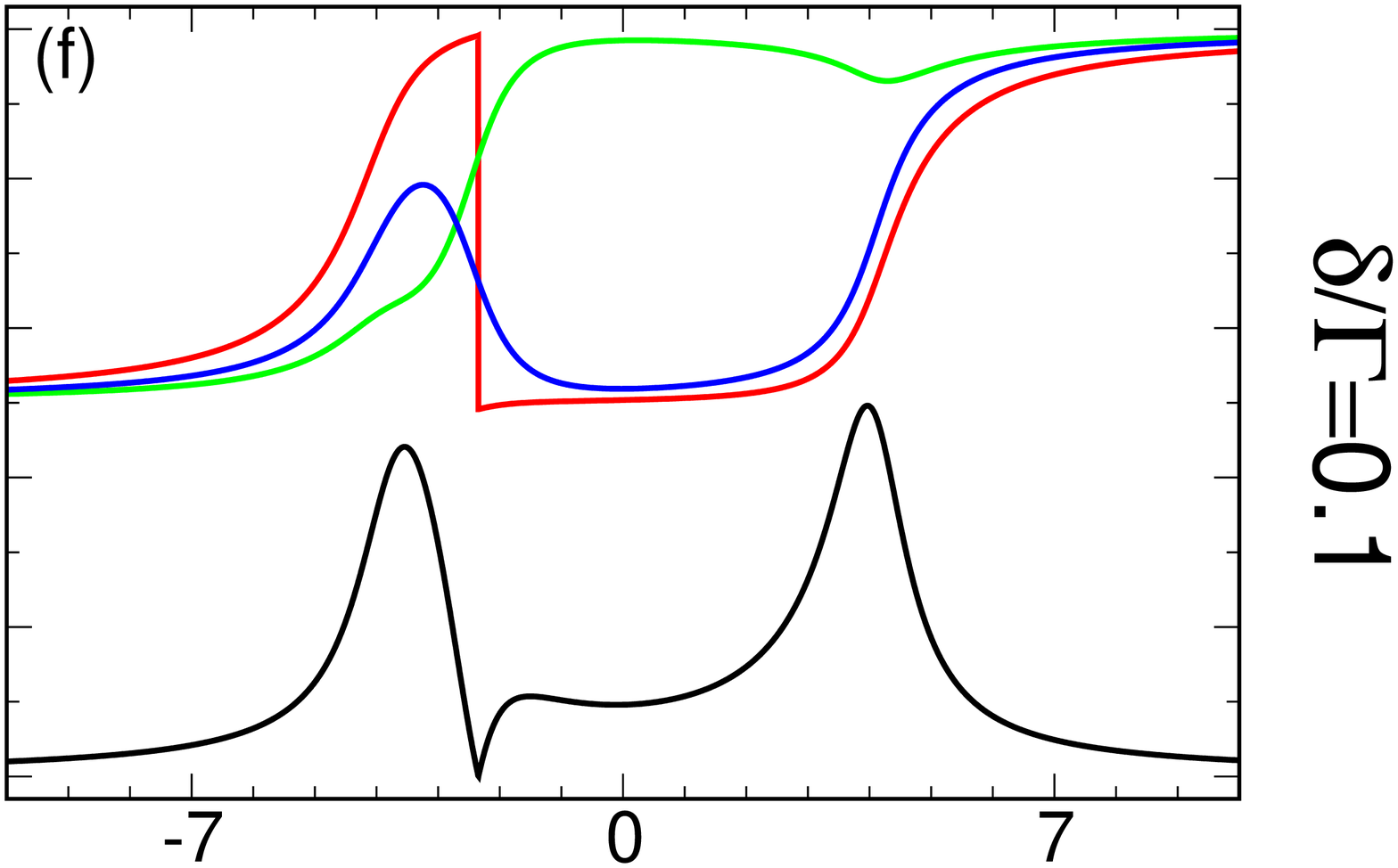}\vspace{0.3cm}
	      \includegraphics[width=0.352\textwidth,height=3.1cm,clip]{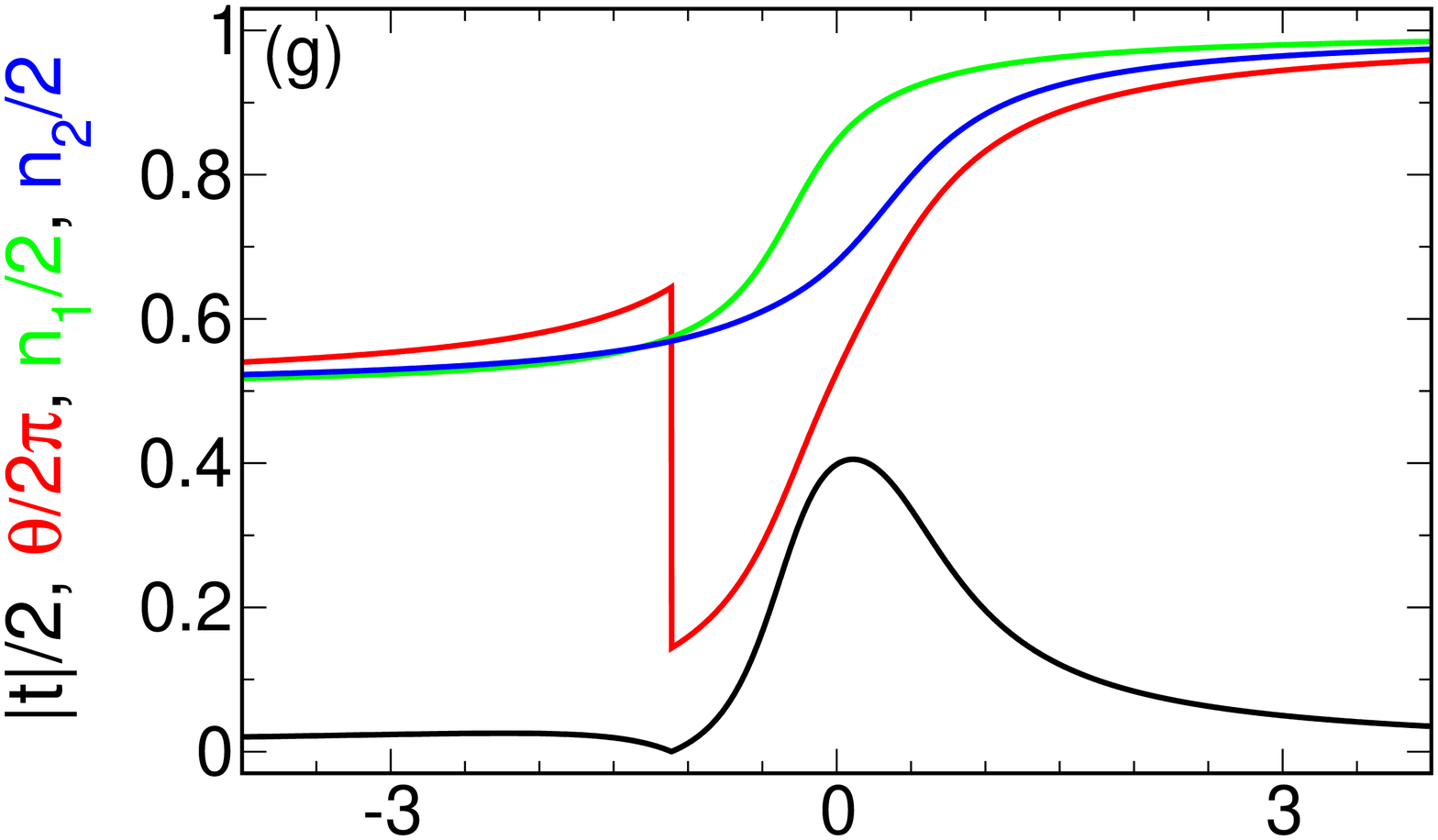}\hspace{0.001\textwidth}
	      \includegraphics[width=0.292\textwidth,height=3.1cm,clip]{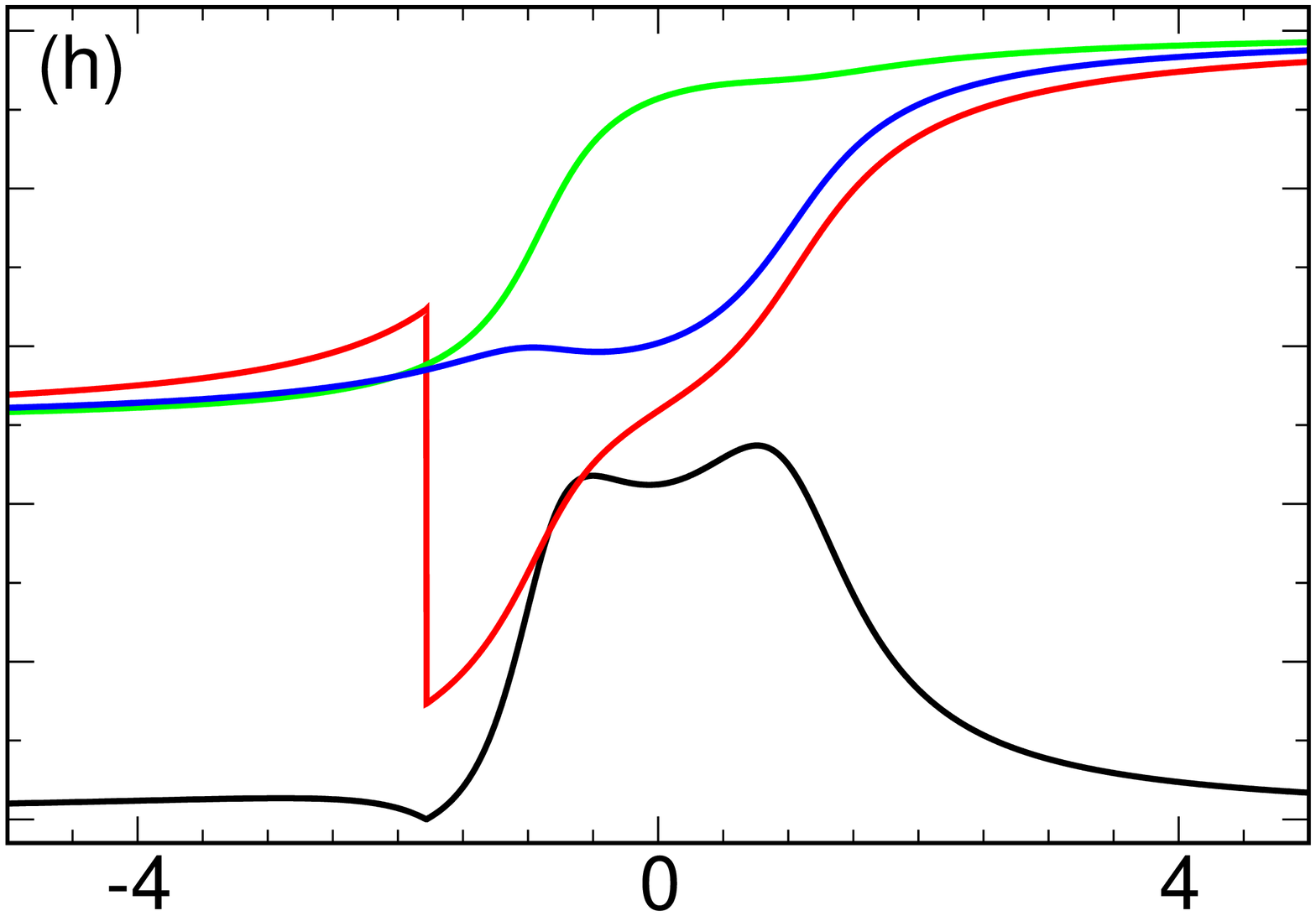}\hspace{0.001\textwidth}
        \includegraphics[width=0.333\textwidth,height=3.1cm,clip]{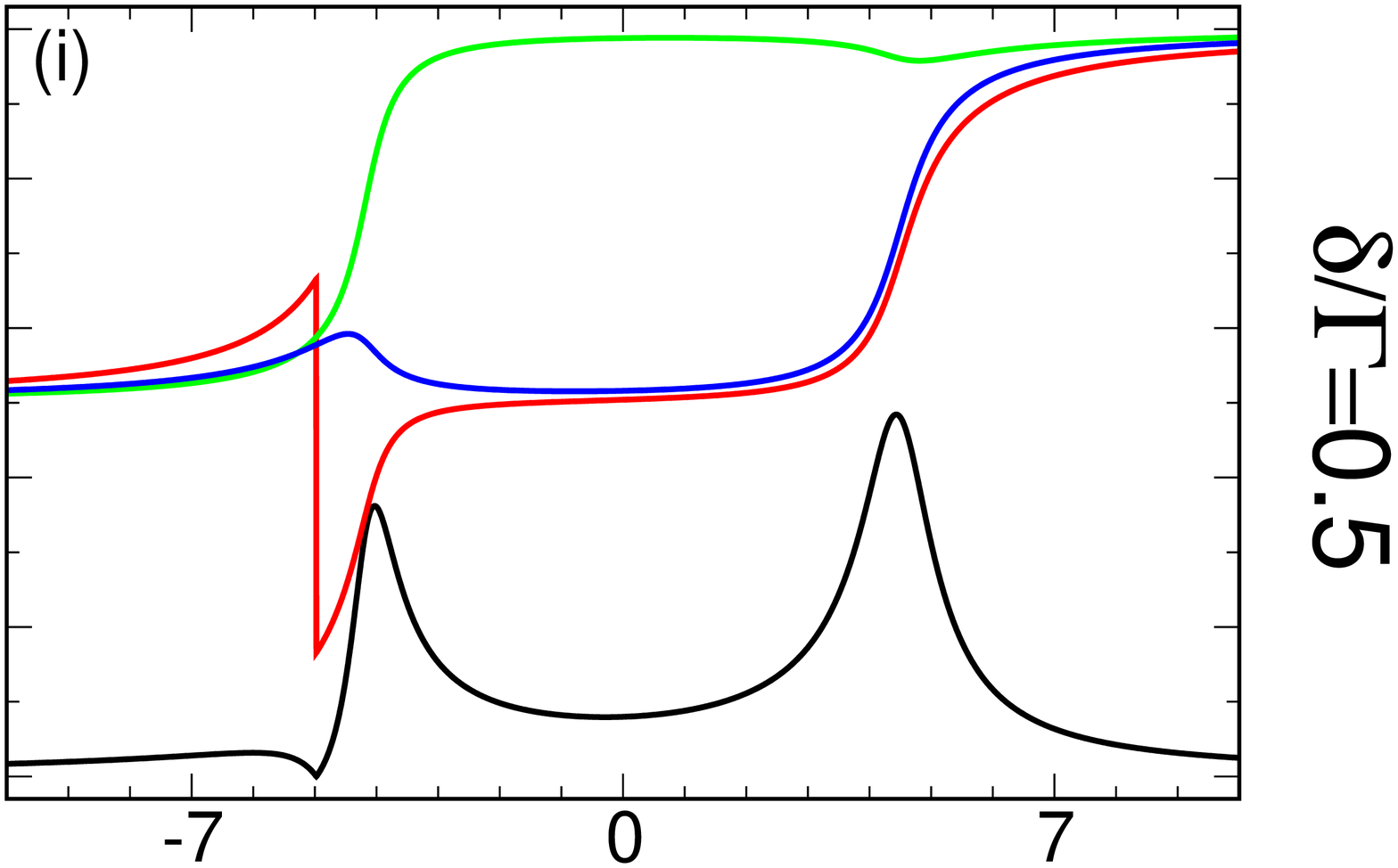}\vspace{0.3cm}
	      \includegraphics[width=0.352\textwidth,height=3.1cm,clip]{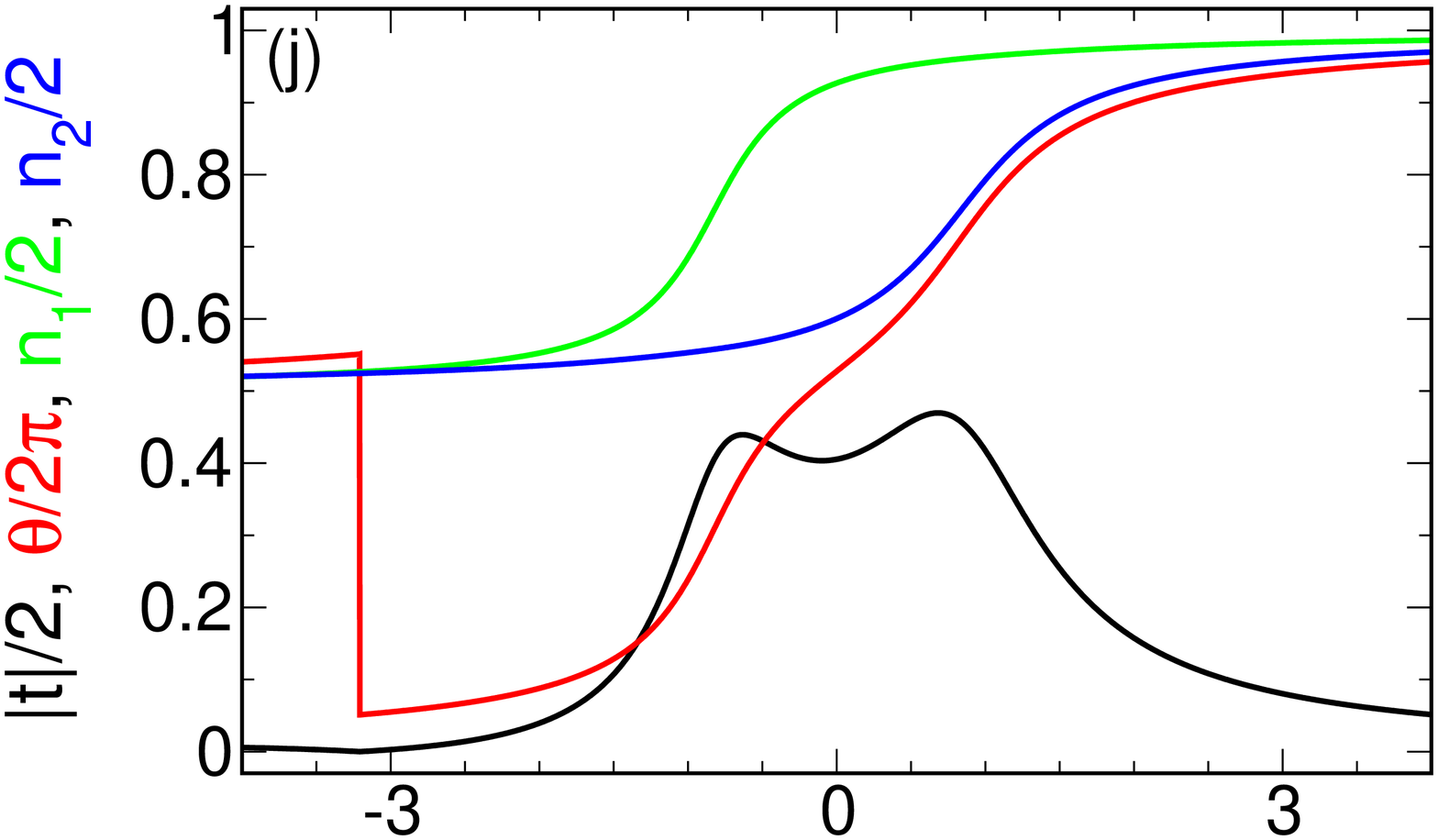}\hspace{0.001\textwidth}
	      \includegraphics[width=0.292\textwidth,height=3.1cm,clip]{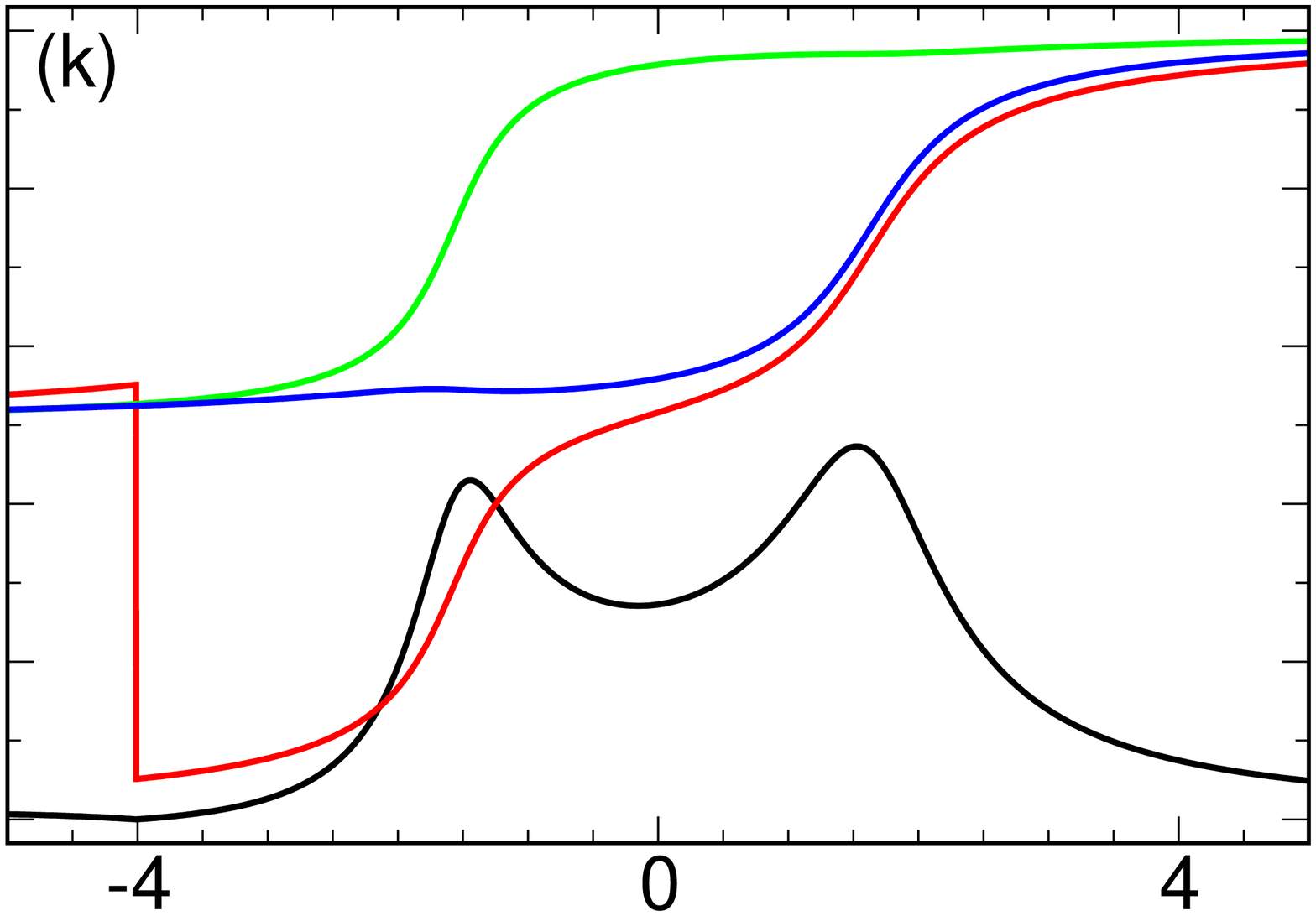}\hspace{0.001\textwidth}
        \includegraphics[width=0.33\textwidth,height=3.1cm,clip]{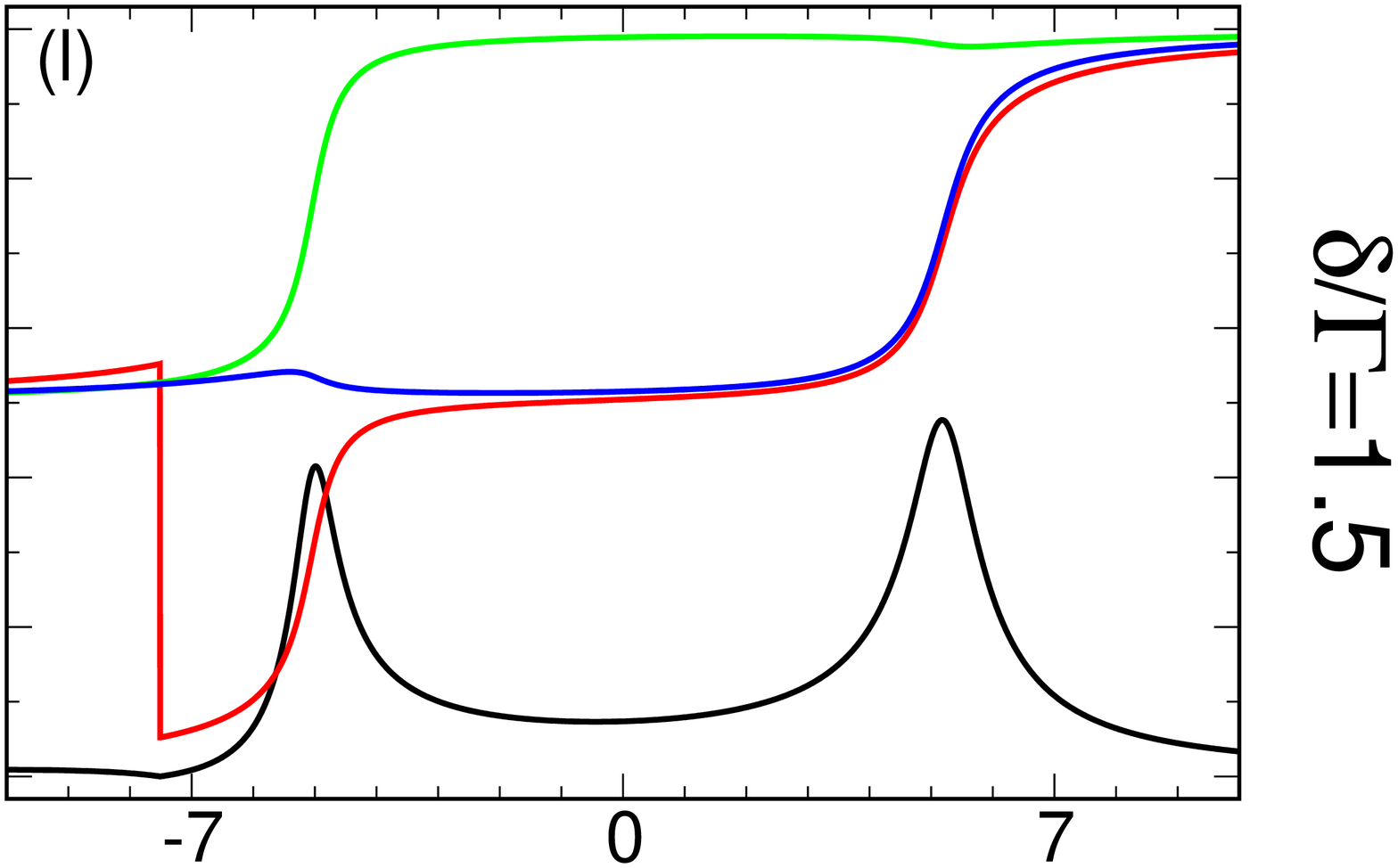}\vspace{0.3cm}
	      \includegraphics[width=0.352\textwidth,height=3.65cm,clip]{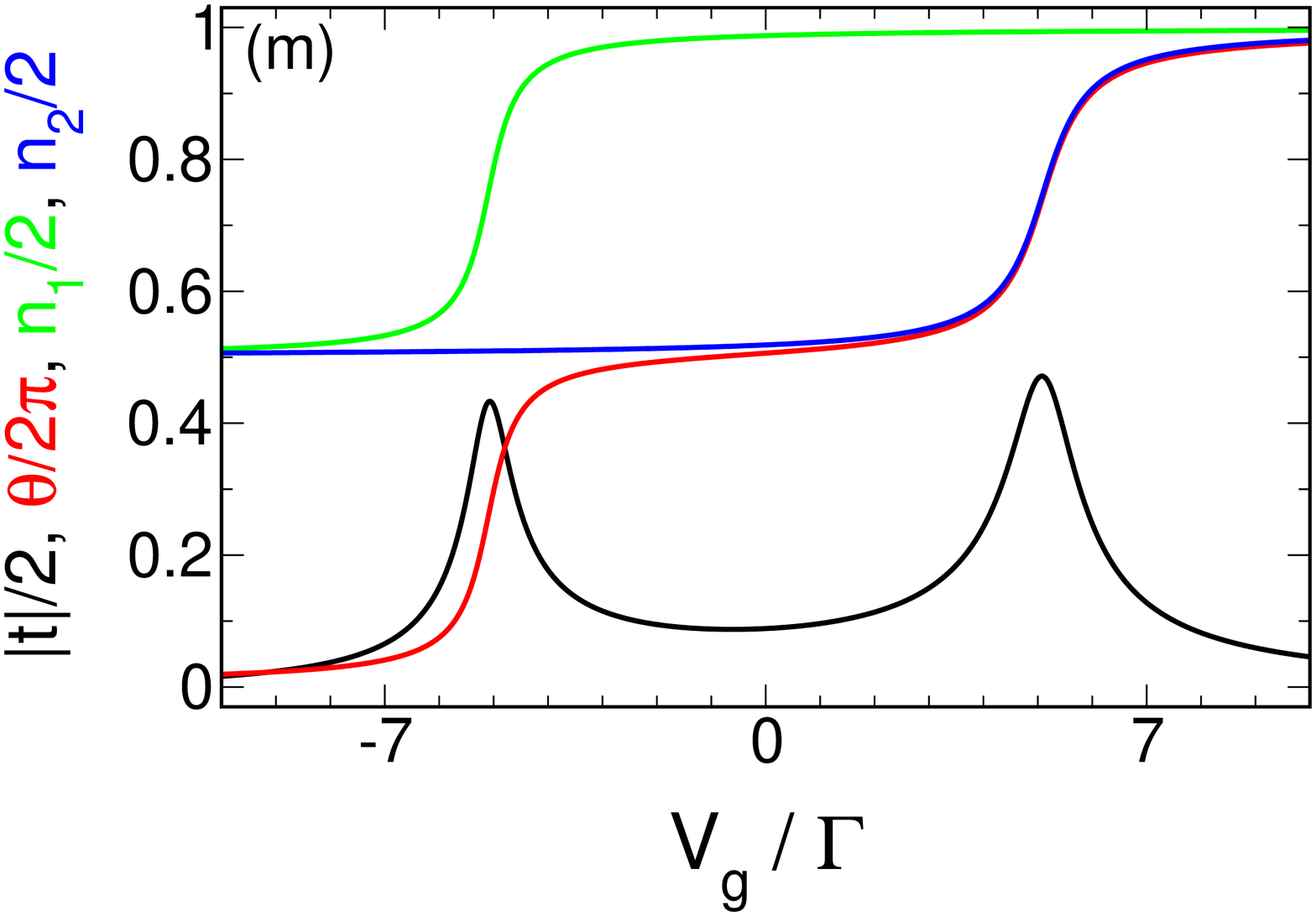}\hspace{0.001\textwidth}
	      \includegraphics[width=0.292\textwidth,height=3.65cm,clip]{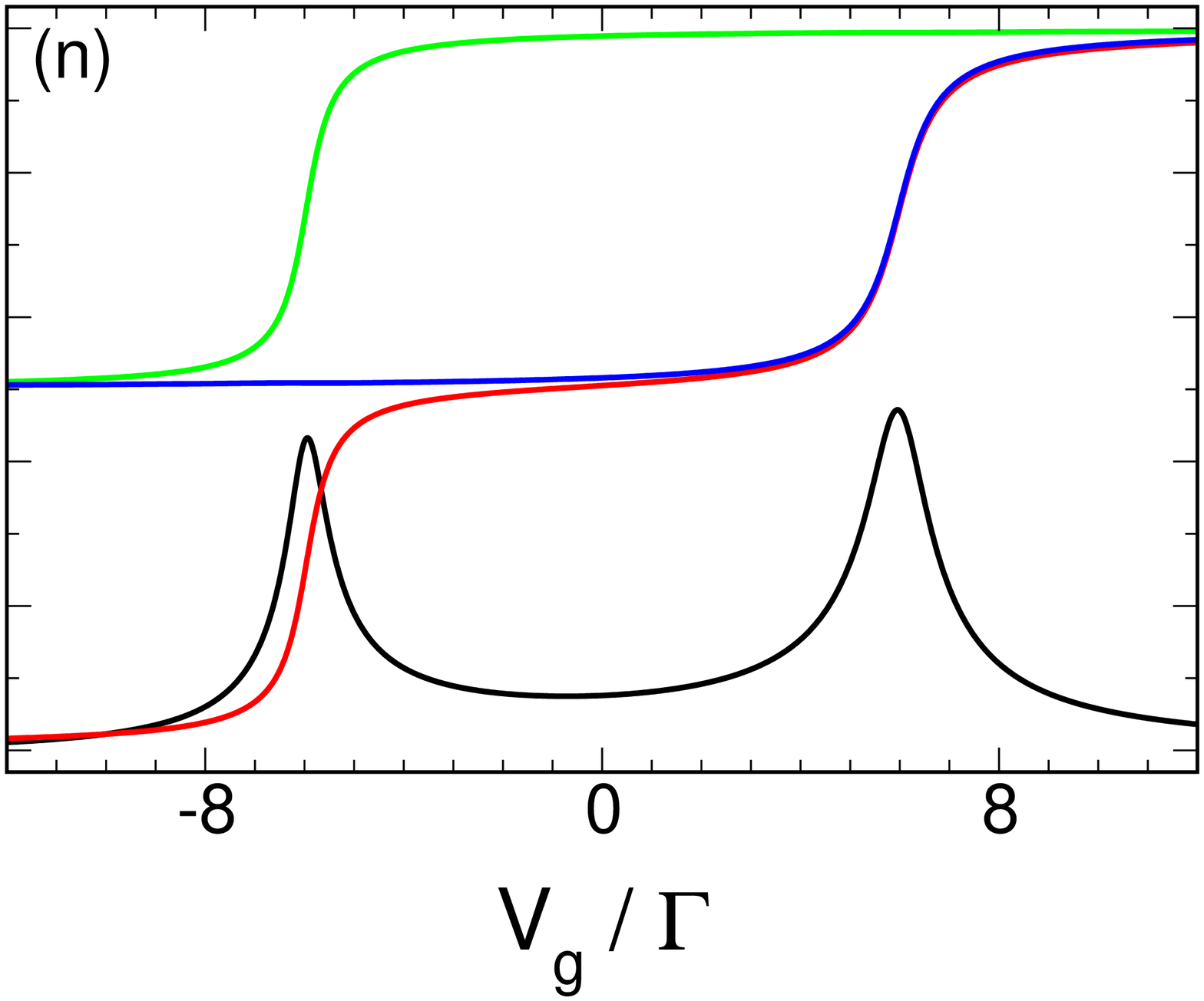}\hspace{0.001\textwidth}
        \includegraphics[width=0.33\textwidth,height=3.65cm,clip]{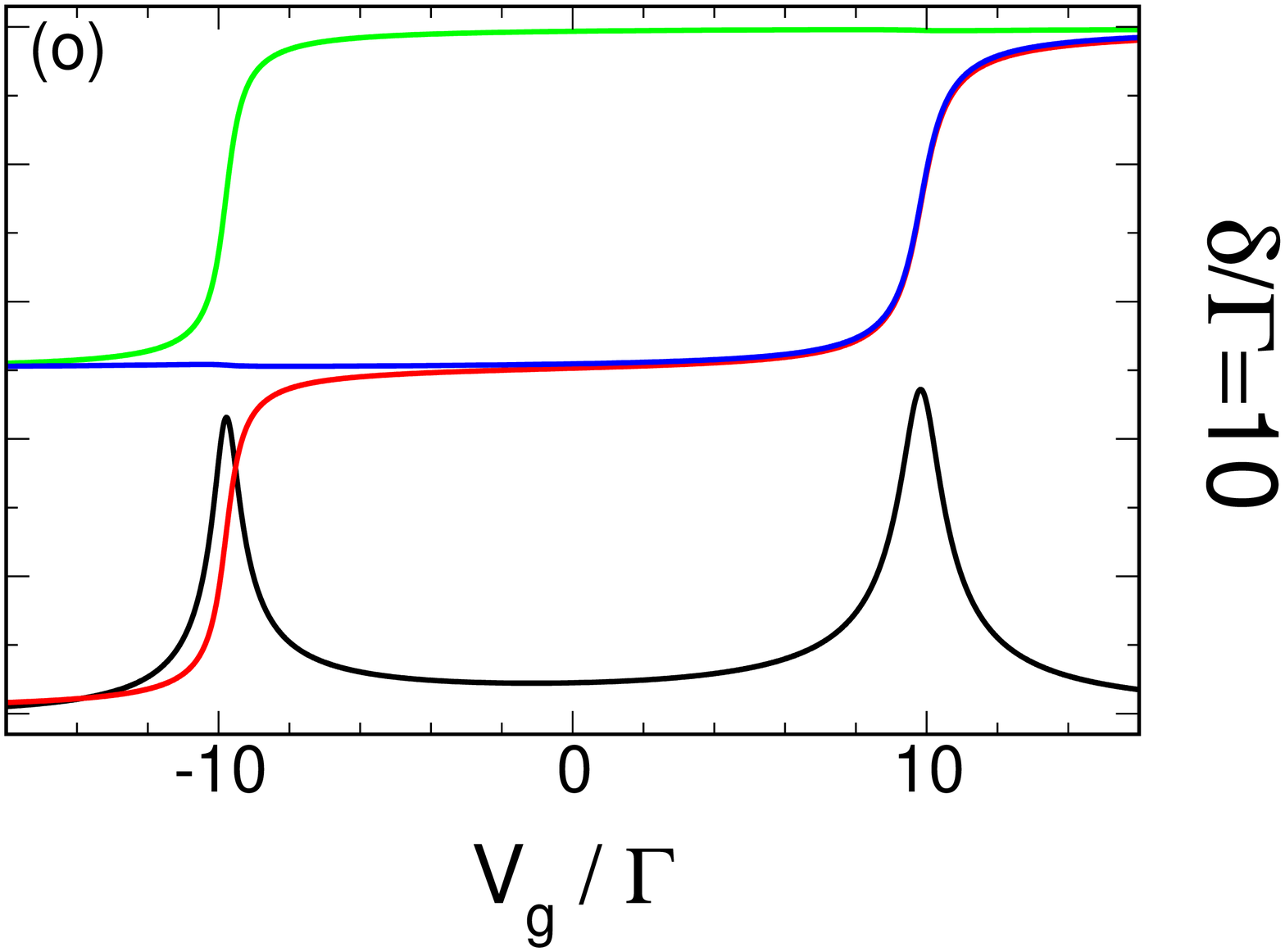}
        \caption{The same as in Fig.\ \ref{fig1}, but for $s=-$.} 
\label{fig2}
\end{figure}

In Ref.\ \cite{PL1} it was shown that fRG and NRG results 
for $|t(V_g)|$ and $\theta(V_g)$ agree quantitatively up to 
fairly large $U$. For a generic set of couplings $\gamma$
we present fRG data for $\theta(V_g)$ and  $|t(V_g)|$ together 
with the occupancies of the levels $n_j$ for
different $U$ and $\delta$ in Figs. \ref{fig1} ($s=+$) and 
\ref{fig2} ($s=-$). Increasing $U/\Gamma$ the
separation of the transmission peaks in the limit of small 
and large $\delta/\Gamma$ increases and is eventually 
given by $U+\delta$. Even though this charging effect appears to be
straightforward it is important to note that in particular the
groundstate at small $\delta/\Gamma$ is highly correlated.
This becomes explicit from the mapping of the present problem on
a generalized single impurity Anderson and Kondo model as 
discussed in Ref.\ \cite{Slava}. An indication of strong 
correlation effects are the correlation-induced resonances of the transmission 
found in Ref.\ \cite{VF}, which we briefly mention below.  
With increasing $U$, even at small $\delta/\Gamma$ 
the gate voltage dependence of $\theta(V_g)$ across the 
transmission resonances becomes s-shaped and the resonances more
Lorentzian-like [see third columns of Figs.\ \ref{fig1} and \ref{fig2}].    
Obviously, for $s=+$ the transmission zero and phase lapse remain between the two 
transmission peaks for all $\delta$ and $U$ (see Fig.\ \ref{fig1}). 
For $s=-$ this only holds for sufficiently small level spacings as,
similar to the $U=0$ case, with increasing $\delta/\Gamma$ 
a crossover sets in to a regime in which the transmission zero and phase lapse 
are no longer between the peaks. Analogously to the $U=0$ case, 
the crossover scale $\delta_c$ depends on the 
particular choice of $\Gamma_j^l$. 
As can be seen from the second row of Fig.\ \ref{fig2} (the CB peaks
at large $U$ have still almost equal height), with 
increasing $U/\Gamma$, $\delta_c$ is pushed towards larger values. 
The Coulomb interaction thus stabilizes the parameter regime 
of universal phase lapses.  This shows that the effect 
of the Coulomb interaction leading to universal $\pi$ phase lapses between 
separated CB peaks in a two-level dot is rather straightforward:
for small $\delta/\Gamma$ the phase lapse and transmission zero are already present at 
$U=0$, and the effect of finite $U$ is simply that the CB peaks 
become well separated because of charging effects. They also lead 
to a Lorentzian-like lineshape of the peaks and an s-like variation
of $\theta$ across them.---The present scenario  
has to be contrasted to the one obtained for
$N>2$ levels discussed in Ref.\ \cite{PL1}. The generic appearance 
of $N-1$ transmission zeros and phase lapses separating the transmission peaks  at small 
$\delta/\Gamma$ and $U=0$ is specific to the case with $N=2$ levels. 
For  $N>2$ the number of transmission zeros and phase lapses at $U=0$ strongly depends on 
the parameters  and the mechanism leading to
universal $\pi$ phase lapses at sufficiently large $U$ (at small $\delta/\Gamma$)
is much more involved \cite{PL1}. 
This shows that although important insights can be gained from 
studying two-levels to  achieve a {\it complete} understanding of 
the phase lapse scenario it is  essential to study dots with Coulomb 
interaction and more than two levels \cite{PL1}. 

The lineshape of $|t|$ shows characteristic differences in the limits
of small and large $\delta/\Gamma$.
In the universal regime at small $\delta/\Gamma$ and for 
sufficiently large $U/\Gamma$ the two CB peaks have equal width 
of order $\Gamma$ (not $\Gamma_j$) and equal height which is consistent 
with the expectation that at each peak the transport occurs 
through both bare levels simultaneously. A similar behavior is 
observed in the experiments. In the mesoscopic regime 
($\delta/\Gamma\gg1$) the width of 
the $j$'th peak is given by $\Gamma_j$ and the relative height $h_j$ by 
$\Gamma_j^L/\Gamma_j^R$, independent of the value of $U$. This effectively
noninteracting lineshape can be understood from the gate voltage 
dependence of the  effective level positions
$\varepsilon_j^{\Lambda=0}$ at the end
of the fRG flow. When one level is charged the effective level
position of the other level is pushed downwards by $U$. 
Besides this, the gate voltage dependence of the 
$\varepsilon_j^{\Lambda=0}$ remains linear, leading to two 
transmission peaks at gate voltages $\varepsilon_j^{\Lambda=0}(V_g)=0$ 
with separation $U+\delta$, but with the same width and height as for 
$U=0$. The 
hopping between the two effective levels generated in the fRG flow 
is small and can be neglected.  

Apart from the CB peaks, for
sufficiently large $U/\Gamma$ the transmission shows additional 
features at small $V_g/\Gamma$ [see Figs.\ \ref{fig1} (b), (c), 
(e) and \ref{fig2} (c)].  
These are the correlation-induced resonances mentioned in the introduction, which
have been found to be most pronounced at $\delta=0$ and in this case
occur for interactions larger than a critical $U_c$ which depends on
the $\Gamma_j^l$ and $s$ \cite{VF,Slava}. Their appearance indicates 
that the groundstate at small $\delta/\Gamma$ is strongly correlated
(as mentioned in Ref.\ \cite{VF} the correlation-induced resonances are not captured by a mean-field
analysis; see below).
Associated with  the correlation-induced resonances is a
sharp increase of $\theta$ [see  Figs.\ \ref{fig1} (b) and (e)].  
At large $U/\Gamma$ the correlation-induced resonances are exponentially (in $U/\Gamma$) 
sharp features that vanish quickly with increasing $T$ (see below),
which might be one of the reasons why up to now they have not been
observed in experiments. The correlation-induced resonances are not directly linked to the
universal phase lapse scenario. 

For increasing $U/\Gamma$ at fixed $\delta/\Gamma$ and decreasing
$\delta/\Gamma$ at fixed $U/\Gamma$ we observe an increased tendency
towards population inversion of the $n_j$. We define that a population inversion occurs if (1)
$n_1(V_g^{\rm PI})=n_2(V_g^{\rm PI})$ at a certain $V_g^{\rm PI}$ and
(2) one $n_j$ has positive and one negative slope at $V_g^{\rm PI}$ so
that the filling of one level causes a tendency for the other to
empty.  For large $U/\Gamma$ and small $\delta/\Gamma$ it is mainly
the more strongly coupled level (in Figs. \ref{fig1} and \ref{fig2},
this is the level 2 shown in blue) whose
population increases across both CB peaks while it is depopulated in
between. This behavior is reminiscent of the one discussed in the
model with a broad and several narrow levels \cite{Imry},  where
a relation between population inversion and phase lapse behavior was proposed. Remarkably, for
sufficiently large $U/\Gamma$, we find population inversion even for small asymmetries
$\Gamma_{2}/\Gamma_1$ (which is only 1.5 in the example of Figs.\
\ref{fig1} and \ref{fig2}).  We emphasize that despite this
resemblance to the observation of Ref.\ \cite{Imry}, the $N=2$ model is
not appropriate to establish a general relation between the appearance
of population inversions and $\pi$ phase lapses at small $\delta/\Gamma$ \cite{Gefen,Berkovits}.
While the latter are already present at $U=0$, the former only develop
with increasing $U$ [compare Figs.\ \ref{fig1} (a), (d) or \ref{fig2}
(a), (d) to Figs.\ \ref{fig1} (c), (f) or \ref{fig2} (c),(f)]. Note 
that the gate voltage $V_g^{\rm PI}$ at
which the population inversion occurs is generically \emph{not} 
identical to the position of the phase lapse and transmission zero [see Figs.\ \ref{fig1} (b), 
(c), (f) and \ref{fig2} (c), (f)] \cite{Slava}. However, for l-r symmetric 
$\Gamma_j^l$ Eq.\ (\ref{t_LR}) ensures that if a population inversion occurs its position
is identical to the one of the phase lapse and transmission zero. 

As can be seen in Figs.\ \ref{fig1} (c), (f) and \ref{fig2} (c), (f)
for small $\delta/\Gamma$ and large $U/\Gamma$ the $n_j$ show a rather
strong gate voltage dependence between the CB peaks. Nevertheless the 
total dot occupancy $n_1+n_2$ is only weakly $V_g$ dependent and close
to 1 within the entire CB valley. This is reminiscent of the 
plateau-like occupancy in the local moment regime of the single impurity 
Anderson model showing the Kondo effect. As discussed in
Ref.\ \cite{Slava} a relation to this model can indeed be established.   

We note in passing that with the exception of the nongeneric case of 
l-r symmetric $\Gamma_j^l$, $s=+$ and $\delta=0$, the $n_j$ are 
continuous functions of $V_g$.

\subsection{Comparison to mean-field theory}

In Refs.\ \cite{Gefen} and \cite{Gefenagain} 
Golosov and Gefen (GG) analyze the phase lapse scenario
of the spin-polarized interacting two-level dot within the mean-field
approximation. However, they anticipated themselves that quantum 
fluctuations not captured in the mean-field approach could be 
 important. Examples of this had been
  pointed out already in Refs.\ \cite{Sindel1} and \cite{VF}. 
  Thus, GG emphasized that
  the effects of such fluctuations on their results need to be studied
  in subsequent work.  
  The present subsection is devoted to this task.

\begin{figure}[t]
        \centering
        \includegraphics[width=0.455\textwidth,height=3.8cm,clip]{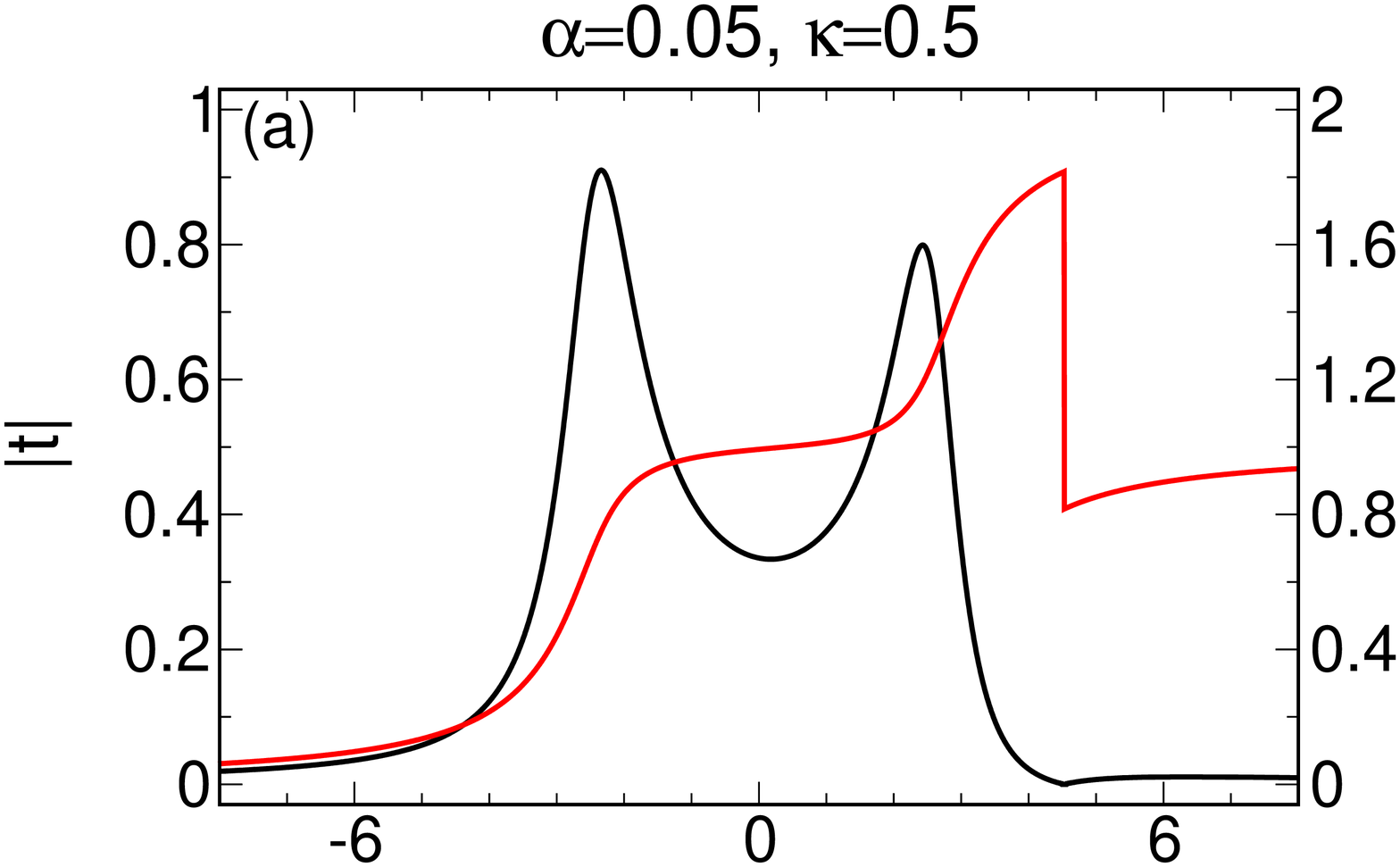}\hspace{0.015\textwidth}
        \includegraphics[width=0.435\textwidth,height=3.8cm,clip]{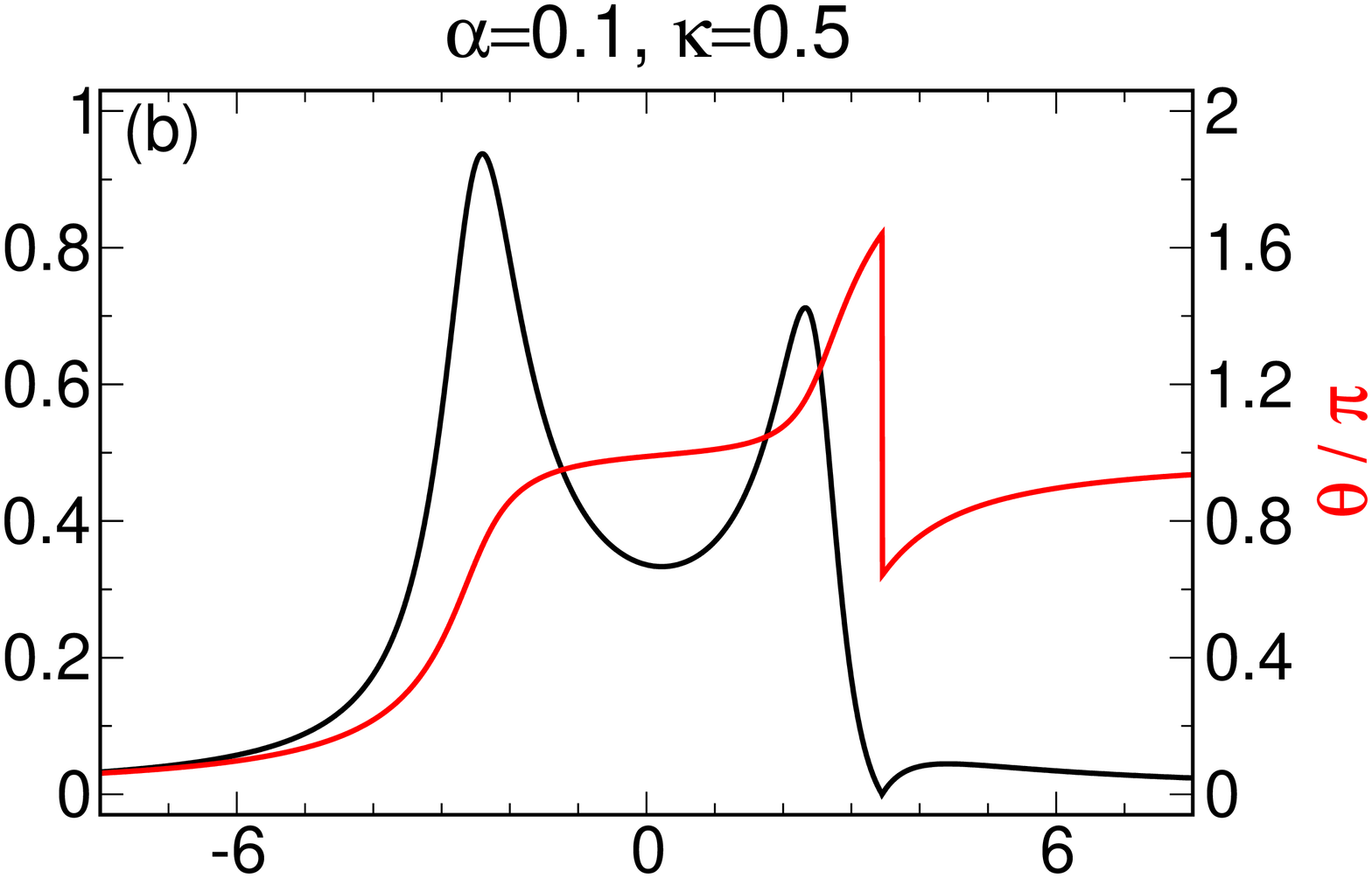}\vspace{0.3cm}
        \includegraphics[width=0.455\textwidth,height=4.6cm,clip]{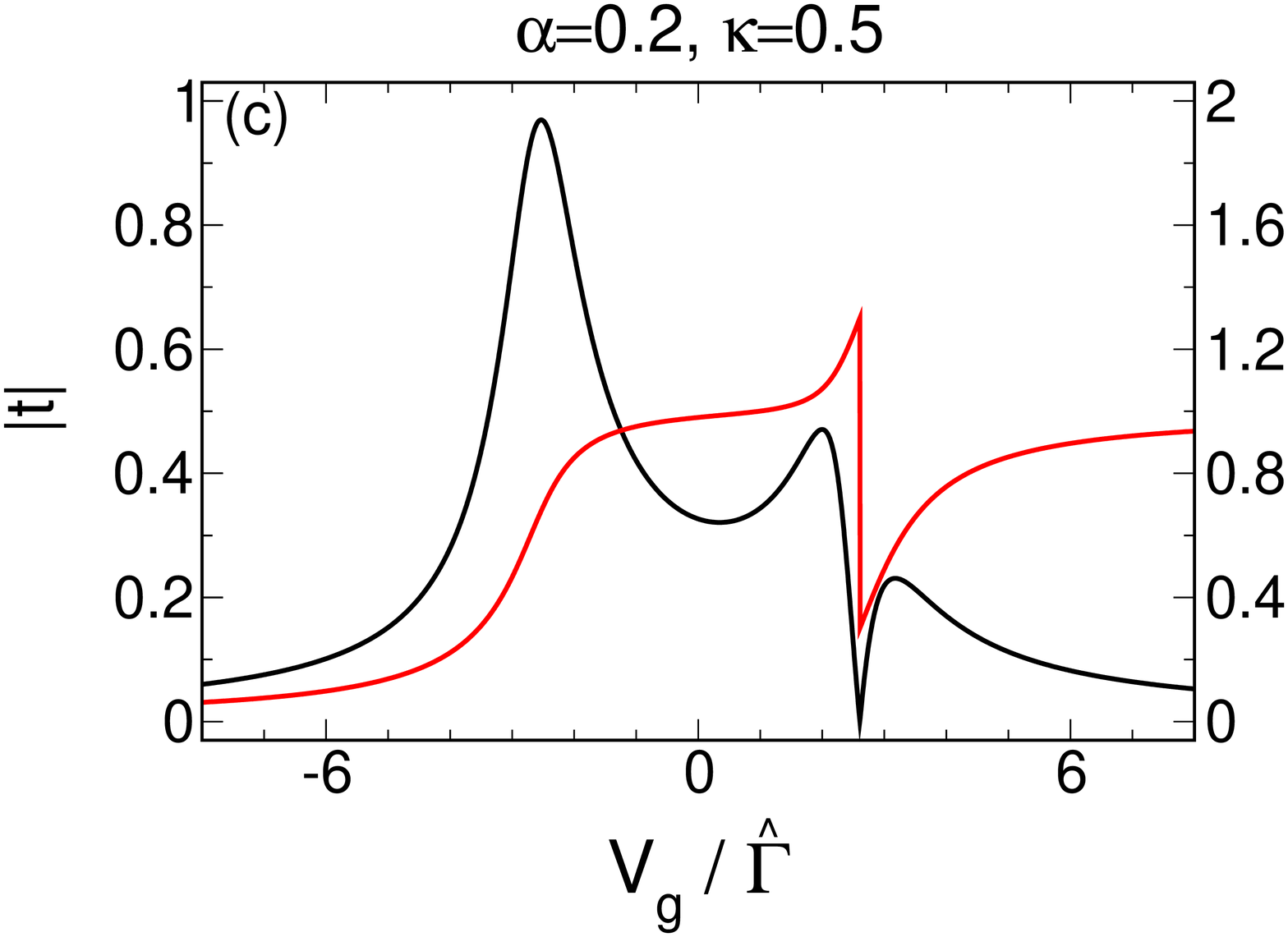}\hspace{0.015\textwidth}
        \includegraphics[width=0.435\textwidth,height=4.6cm,clip]{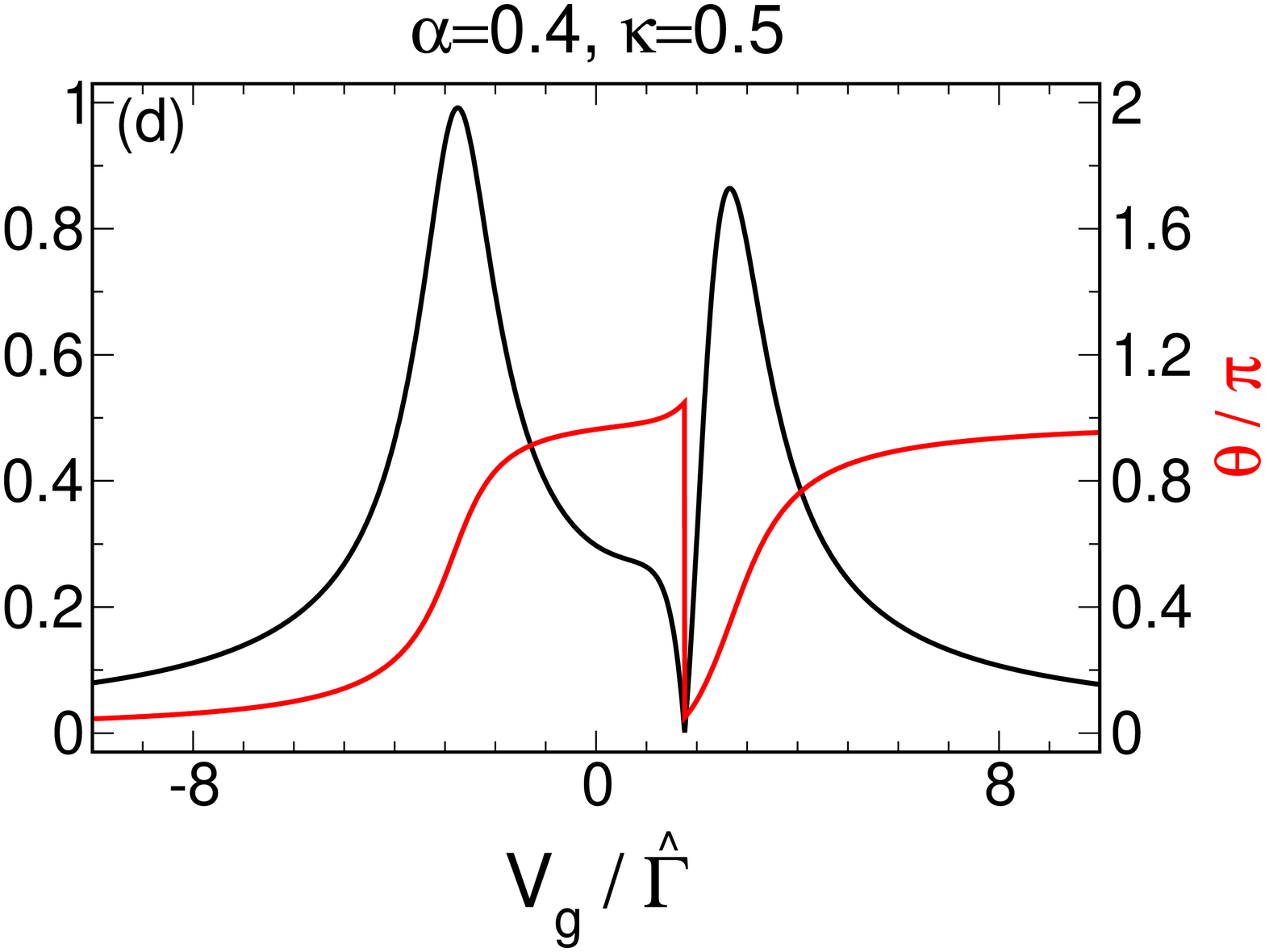}\vspace{0.5cm}
        \includegraphics[width=0.455\textwidth,height=3.8cm,clip]{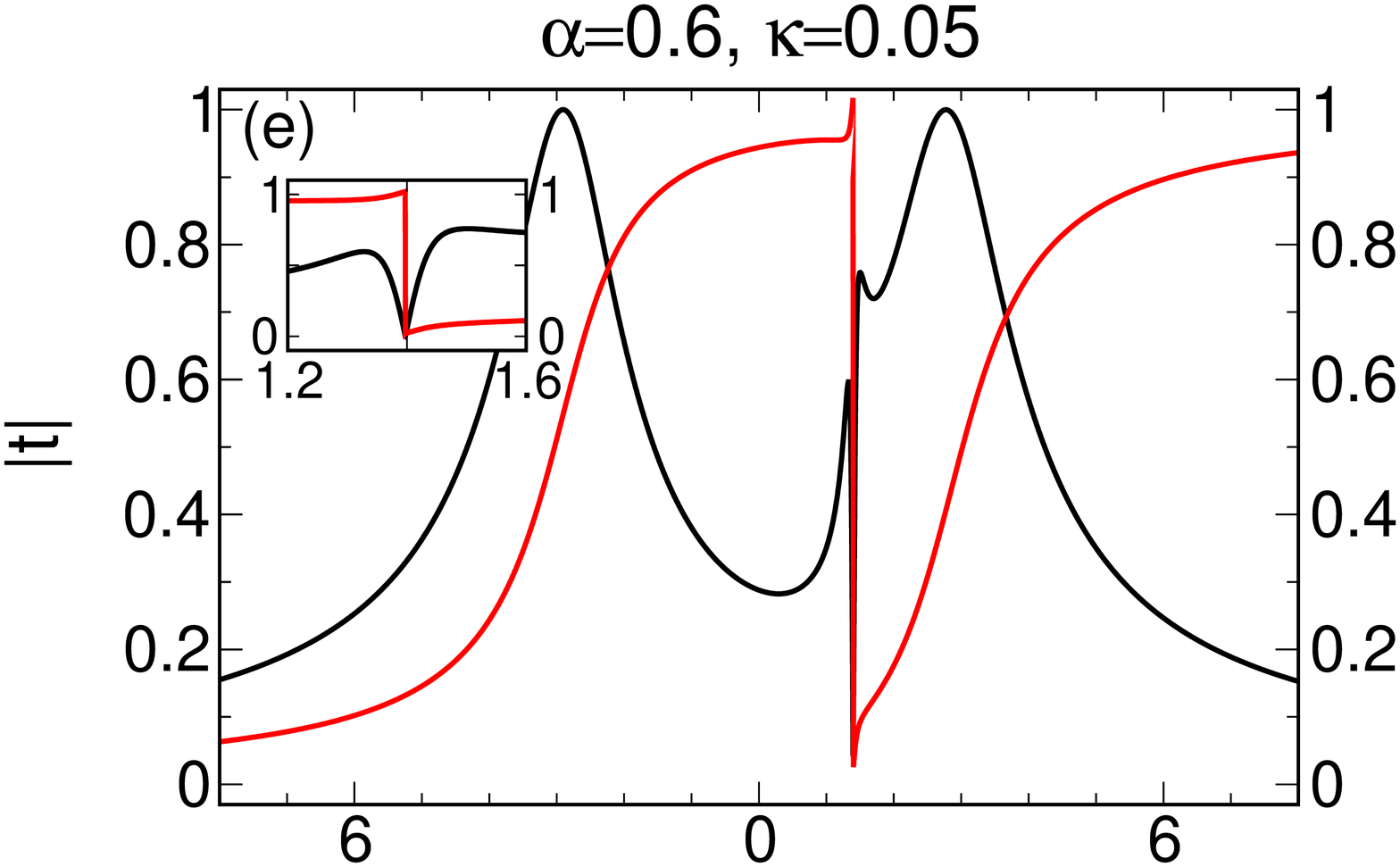}\hspace{0.015\textwidth}
        \includegraphics[width=0.435\textwidth,height=3.8cm,clip]{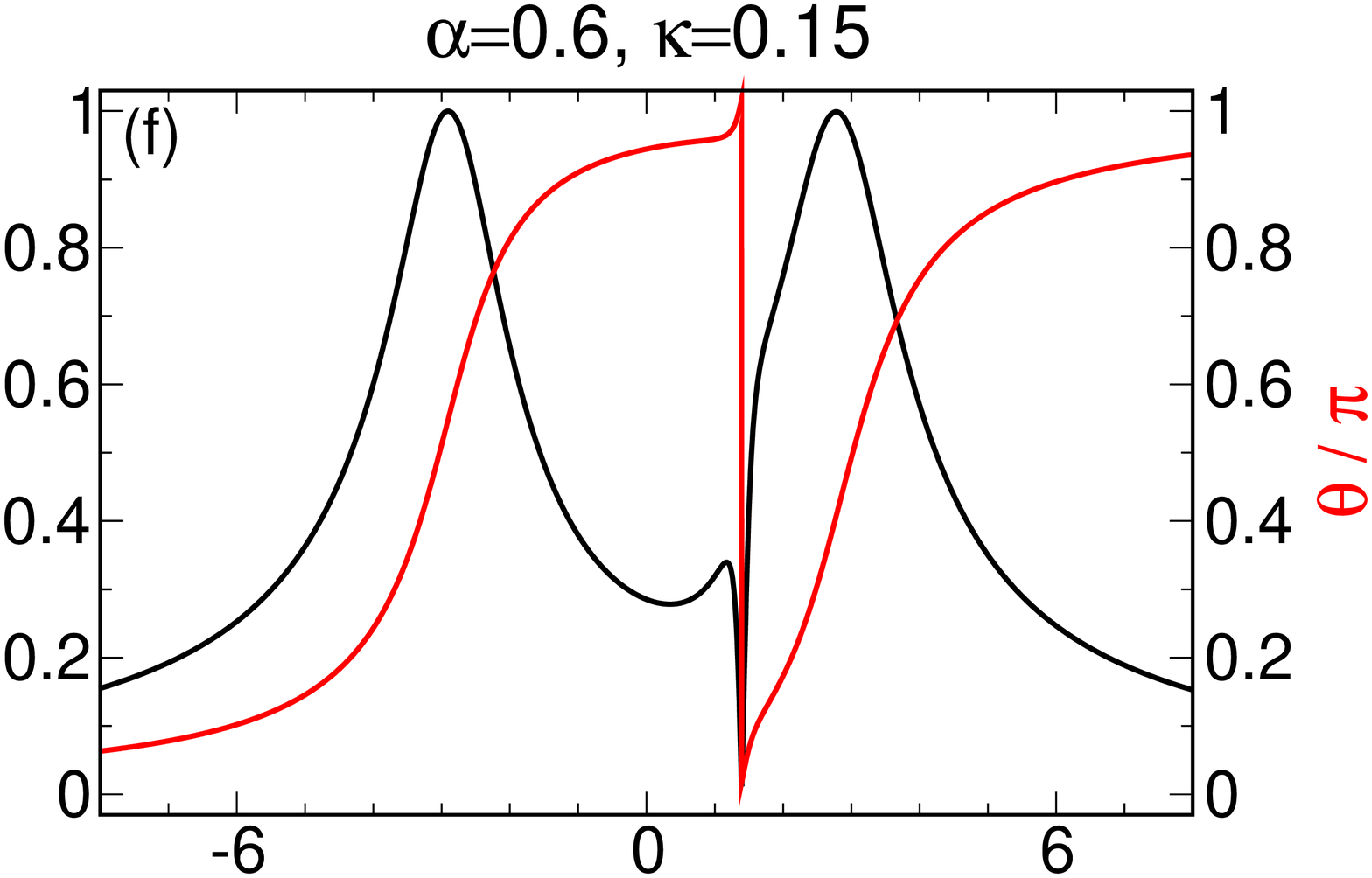}\vspace{0.3cm}
        \includegraphics[width=0.455\textwidth,height=4.6cm,clip]{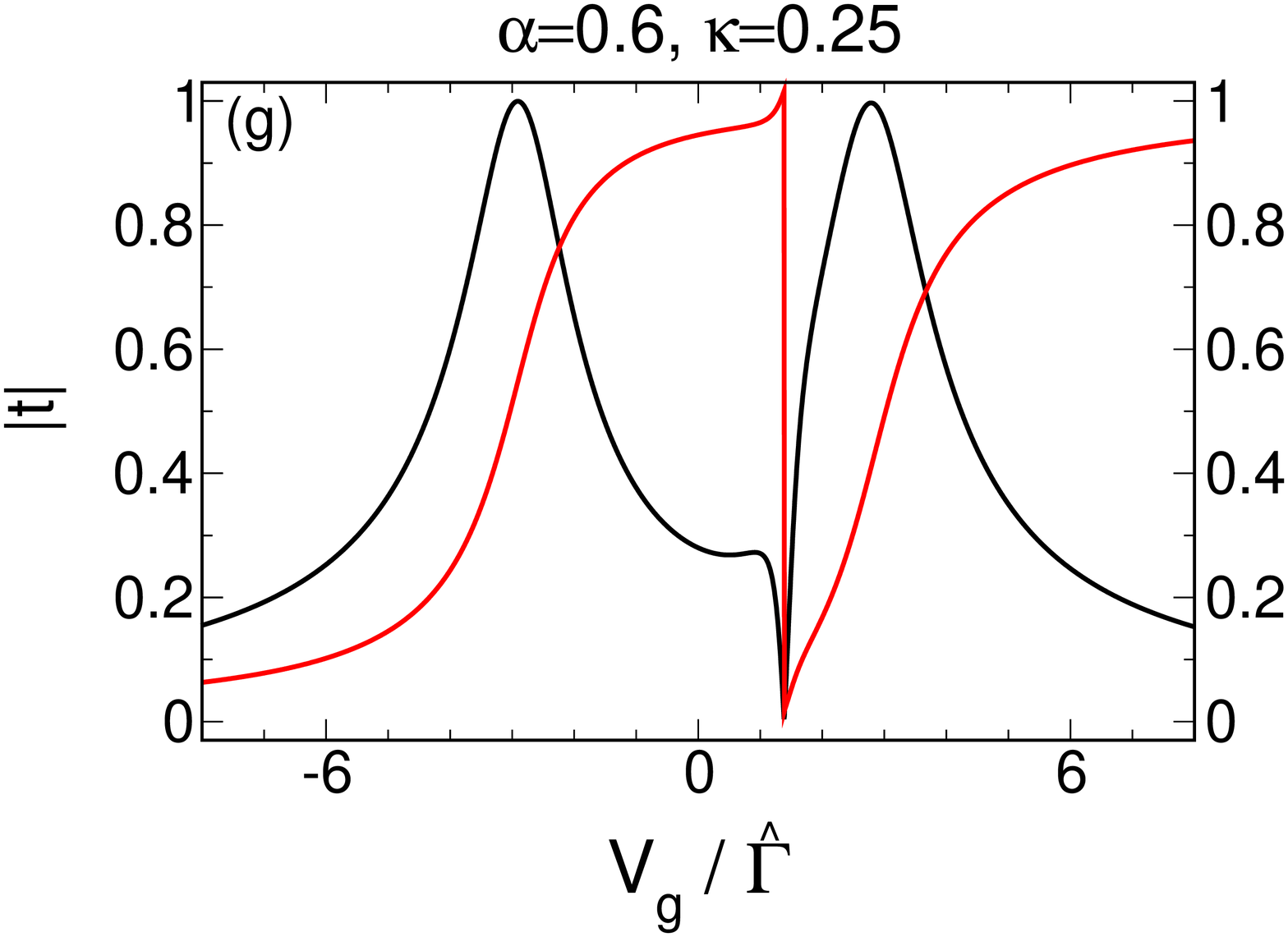}\hspace{0.015\textwidth}
        \includegraphics[width=0.435\textwidth,height=4.6cm,clip]{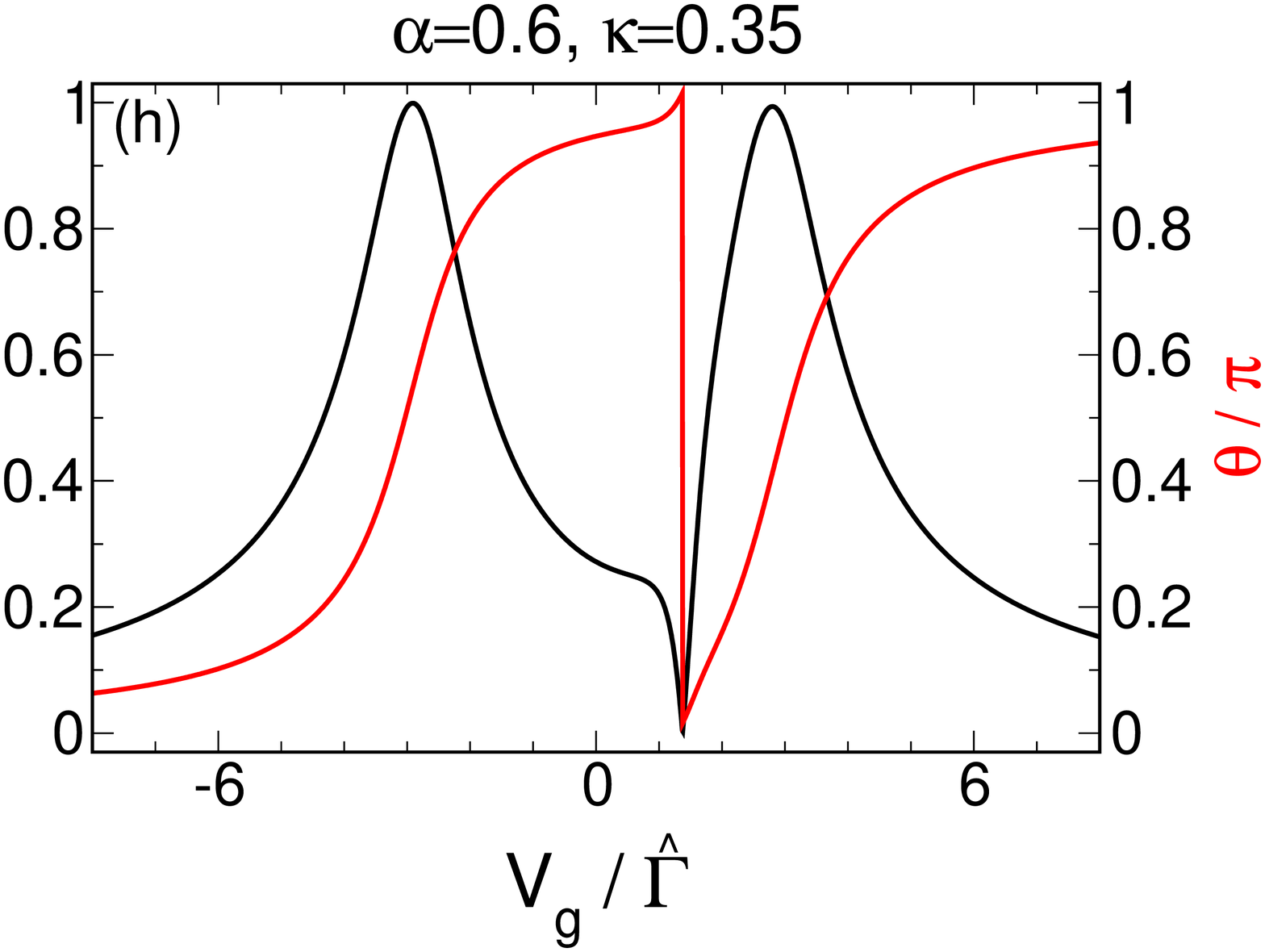}
        \caption{Two traces through the mean-field ``phase diagram''
          of GG. Figures (a)-(d) are for constant $\kappa$ and
          different $\alpha$ moving from GGs ``phase'' 2 into
          ``phase'' 1. Figures (e)-(h) are for constant $\alpha$
          and different $\kappa$ moving from GGs ``phase 3'' into
          ``phase 1''. The inset in Fig.\ (e) shows a zoom in of the
          gate voltage region around the phase lapse. 
          For a detailed comparison to the mean-field results see
          the text. The results were obtained at $T=0$ using the truncated fRG.}
\label{fig3}
\end{figure}

GG consider the subspace of level-lead
couplings defined by $\Gamma_1^L - \Gamma_1^R = \Gamma_2^R -
\Gamma_2^L$. Performing a unitary transformation on the dot states
the part of the Hamiltonian Eq.\ (\ref{generalham}) containing dot
operators can be transformed to (see Refs.\ \cite{Gefen} and
\cite{Gefenagain}) 
\begin{eqnarray}
\label{Gefenham}
\mbox{} \hspace{-1.0cm}  H_{\rm dot} + H_{\rm lead-dot} & = &\sum_{j=1,2} 
\hat \varepsilon_{j} \hat d^\dag_{j,\sigma} \hat d_{j,\sigma} + 
U  \hat d^\dag_{1} \hat d_{1}  \hat d^\dag_{2} \hat d_{2}
- \hat t (\hat d_1^\dag d_2 + \mbox{H.c.}) 
\nonumber \\*
\mbox{} \hspace{-1.0cm}  &&- \left[ c_{0,L} \left(\hat t_1 \hat d_1^\dag + \hat t_2
    \hat d_2^\dag \right) + c_{0,R} \left(\hat t_1 \hat d_1^\dag - \hat t_2
    \hat d_2^\dag \right) +  \mbox{H.c.} \right] \;,
\end{eqnarray}
with the transformed operators and parameters indicated by a
hat. The change of basis leads to a direct hopping 
$\hat t$ between the transformed levels. We here focus on the 
relative sign $\hat s=-$. 
GG then introduce the two new dimensionless parameters $\kappa$ and 
$\alpha$ as $\hat t = -\kappa \hat \delta /(2 \sqrt{1-\kappa^2})$ 
and $\alpha=(|\hat t_1|-|\hat t_2|)/\sqrt{\hat t_1^2 + \hat t_2^2}$. 
Varying $\kappa$ and $\alpha$ GG investigate the phase lapse behavior for fixed 
level spacing (in the new basis) $\hat \varepsilon_{2} - 
\hat \varepsilon_{1} = \hat \delta = 0.256\hat \Gamma$ and 
interaction $U/\hat \Gamma=6.4$. As the parameters in the new basis 
are rather complicated combinations of the original ones, the variation of
$\alpha$ and $\kappa$ corresponds to the variation 
of the $\Gamma_j^l$ (within the above  specified subspace), $\delta$ 
and even $s$.  Our simple picture that 
increasing the level spacing $\delta$ of the untransformed model 
leads from  the universal to the mesoscopic regime cannot 
easily made explicit using the parameters of GG. 
To make the comparison of our results to the mean-field study 
definite we nevertheless follow the steps of GG.

Varying $\alpha,\kappa \in[0,1[$ at $\hat s=-$ we move around in the
right part of what is called the ``phase diagram'' by GG
(Fig.\ 4 of Ref.\ \cite{Gefenagain}). In Fig.\ \ref{fig3} we show the
behavior of $|t|$ and $\theta$ varying $\alpha$ at fixed $\kappa$
[Figs.\ \ref{fig3} (a)-(d)] and $\kappa$ at fixed $\alpha$ [Figs.\
\ref{fig3} (e)-(h)], respectively.  In the first case, upon increasing
$\alpha$ at constant $\kappa=0.5$, we move from GGs ``phase 2'', (red
in Fig.\ 4 of Ref.\ \cite{Gefenagain}), with the transmission zero and phase lapse outside the two
CB peaks, into ``phase 1'' (blue in Fig.\ 4 of Ref.\ \cite{Gefenagain}),
with the transmission zero and phase lapse between the peaks. 
The mean-field approximation correctly captures the presence of these
two regimes. We find a 
smooth crossover between them (which is why we prefer the notion of 
different ``regimes'' rather than ``phases''): the $V_g$ value at 
which the transmission zero and phase lapse occur \emph{smoothly} crosses from lying outside 
the right CB peak to lying between the two CB peaks. This is similar 
to the smooth crossover we observe in the $s=-$ case of the 
untransformed model when moving from the mesoscopic to the 
universal regime (see columns of Fig.\ \ref{fig2}).
  
For fixed $\alpha=0.6$ and increasing $\kappa$ we move from ``phase
3'' (green in Fig.\ 4 of Ref.\ \cite{Gefenagain}) into ``phase 1''. 
In contrast to the mean-field approximation where an abrupt
transition from ``phase 3'' to  ``phase 1'' occurs, Figs.\ \ref{fig3} (e)-(h)
show a rather smooth evolution.
The mean-field "phase 3" is characterized by discontinuous
population switching and a phase lapse between the CB peaks which,
surprisingly, is smaller than $\pi$ \cite{Gefen,Berkovits}. 
Furthermore, in this parameter regime the mean-field results show 
no transmission zero, as discussed in Ref.\ \cite{Berkovits}. However, from 
Figs.\ \ref{fig3} (e)-(h) it is
apparent that ``phase 3'', is an artifact of the mean-field
approximation: upon taking fluctuations into account via fRG, the
$\pi$ phase lapse and the transmission zero are found to remain in the CB valley.  The
evolution with increasing $\kappa$ is similar to the $s=+$ case of the
untransformed model when the level spacing is increased at fixed
$\Gamma_j^l$ (see the second and third columns of Fig.\ \ref{fig1}). 
``Phase 3'' then corresponds to
the parameter regime with small level spacing and a sizable $U/\Gamma$
in which correlations are of particular importance leading e.g.\ to
the correlation-induced resonances [see Figs.\ \ref{fig3} (e)-(h)].  That the mean-field
approximation fails to properly describe this strongly correlated
regime is not surprising and has been recognized earlier
\cite{VF}. In Ref.\ \cite{Slava} a connection between the small
$\delta$ regime of the present model and the local moment (Kondo) regime of
the single-impurity Anderson model was established. 
Thus, the artifacts of the mean-field approximation for the spinless 
two-level dot are reincarnations of the the well-known artifacts it 
produces when applied to the Anderson model in the local moment regime.

Upon taking fluctuations into account, the
discontinuities of the $n_{1/2}$ in ``phase 3'' are washed 
out. The only choice of parameters for which we find discontinuous 
behavior is the one with l-r symmetric $\Gamma_j^l$, $s=+$ and $\delta=0$, 
a case which was already identified as being nongeneric \cite{VF}.    
Any arbitrarily small deviation from these conditions leads to 
a continuous gate voltage dependence of $n_{1/2}$.  
For parameters close to the nongeneric point the change in $n_j$ 
at first sight appears to be rather sharp
and an extremely high resolution in $V_g$ is required to identify
the behavior as continuous.   

The fact that the mean-field treatment at $U>0$ and small level
  spacings incorrectly causes the transmission zero to disappear and the
  corresponding phase lapse to become smaller than $\pi$ is its
  most consequential problem.  By generating such features,  the 
  mean-field treatment masks the very simple scenario that emerges upon properly including
fluctuations: for increasing $U$, the $\pi$ phase lapse and transmission zero found for small
$\delta/\Gamma$ at $U=0$ remain in the CB valley, while the
transmission peaks become well separated, Lorentzian-like and the
phase acquires an s-shape across the resonances.

\subsection{Finite temperatures}

\begin{figure}[t]	
        \centering
        \includegraphics[width=0.495\textwidth,height=4.4cm,clip]{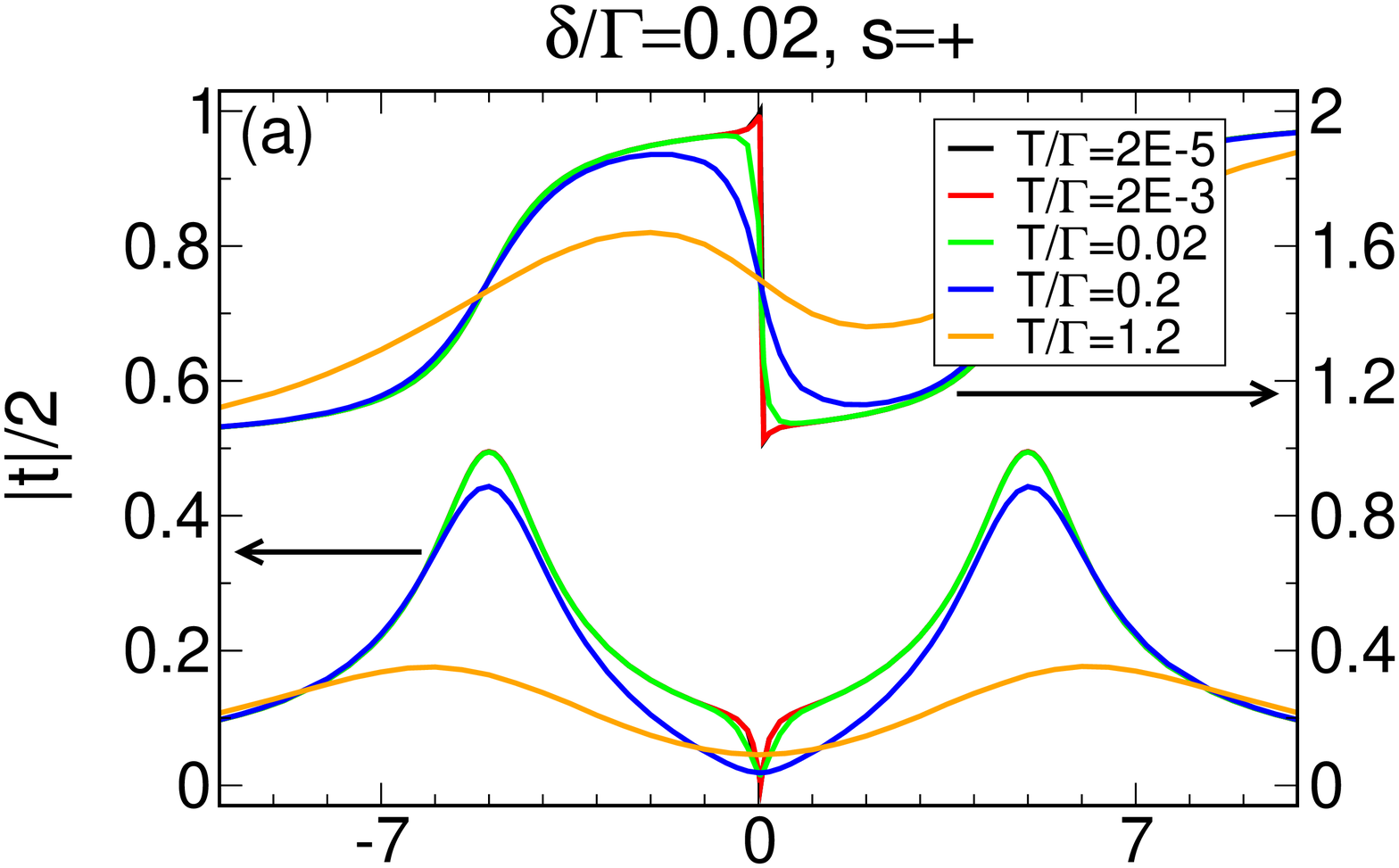}\hspace{0.015\textwidth}
        \includegraphics[width=0.475\textwidth,height=4.4cm,clip]{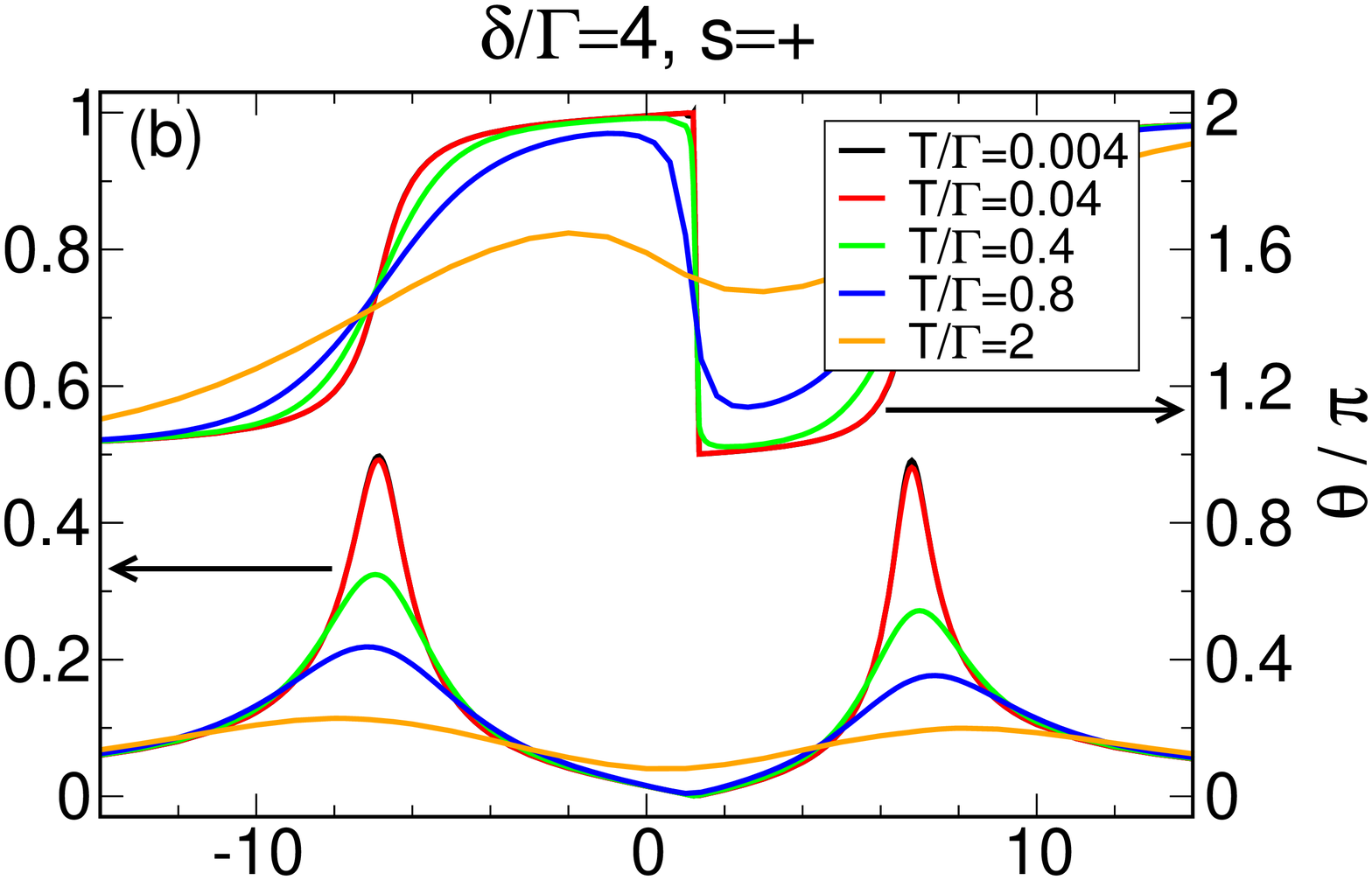}\vspace{0.3cm}
        \includegraphics[width=0.495\textwidth,height=5.2cm,clip]{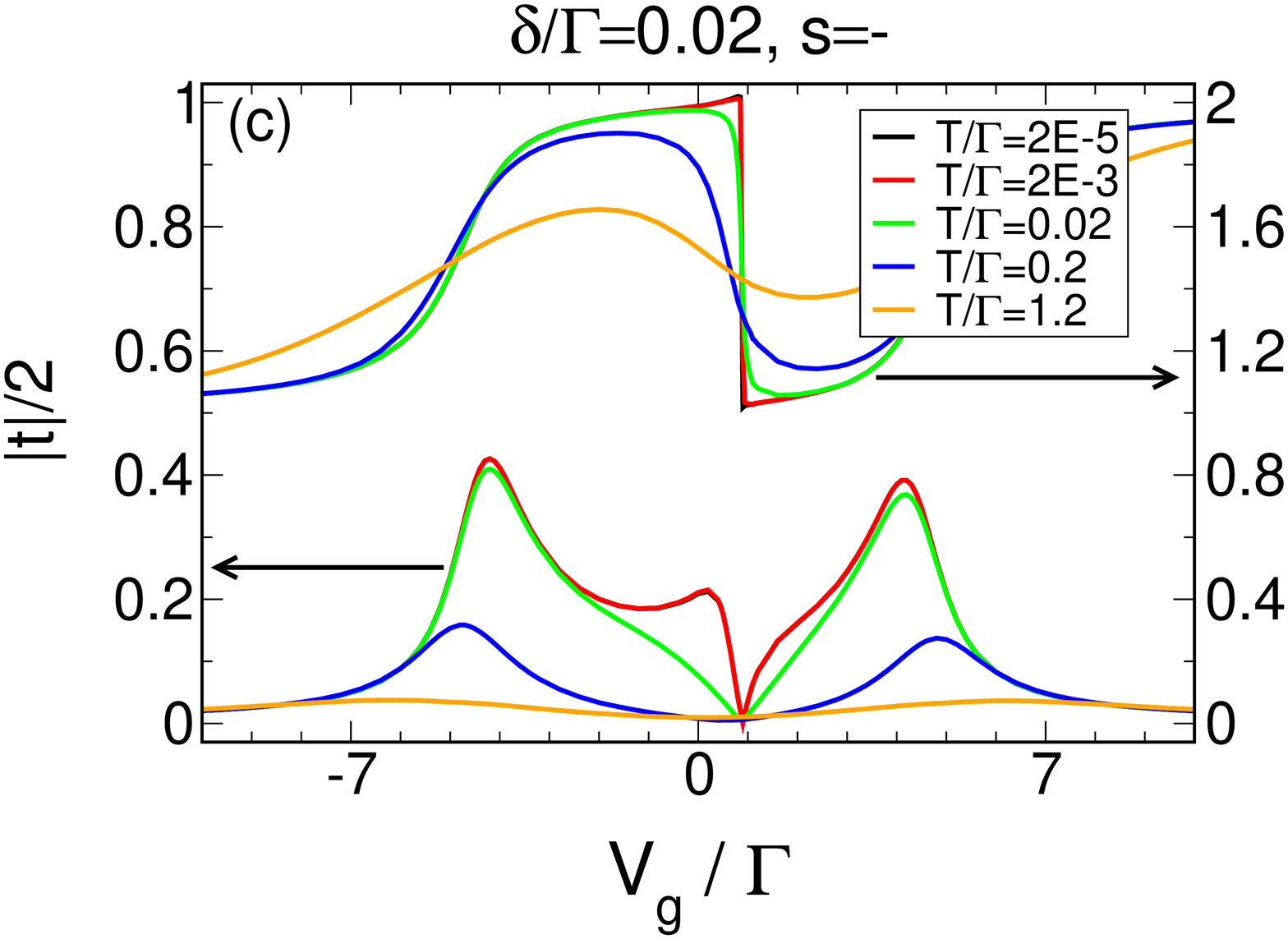}\hspace{0.015\textwidth}
        \includegraphics[width=0.475\textwidth,height=5.2cm,clip]{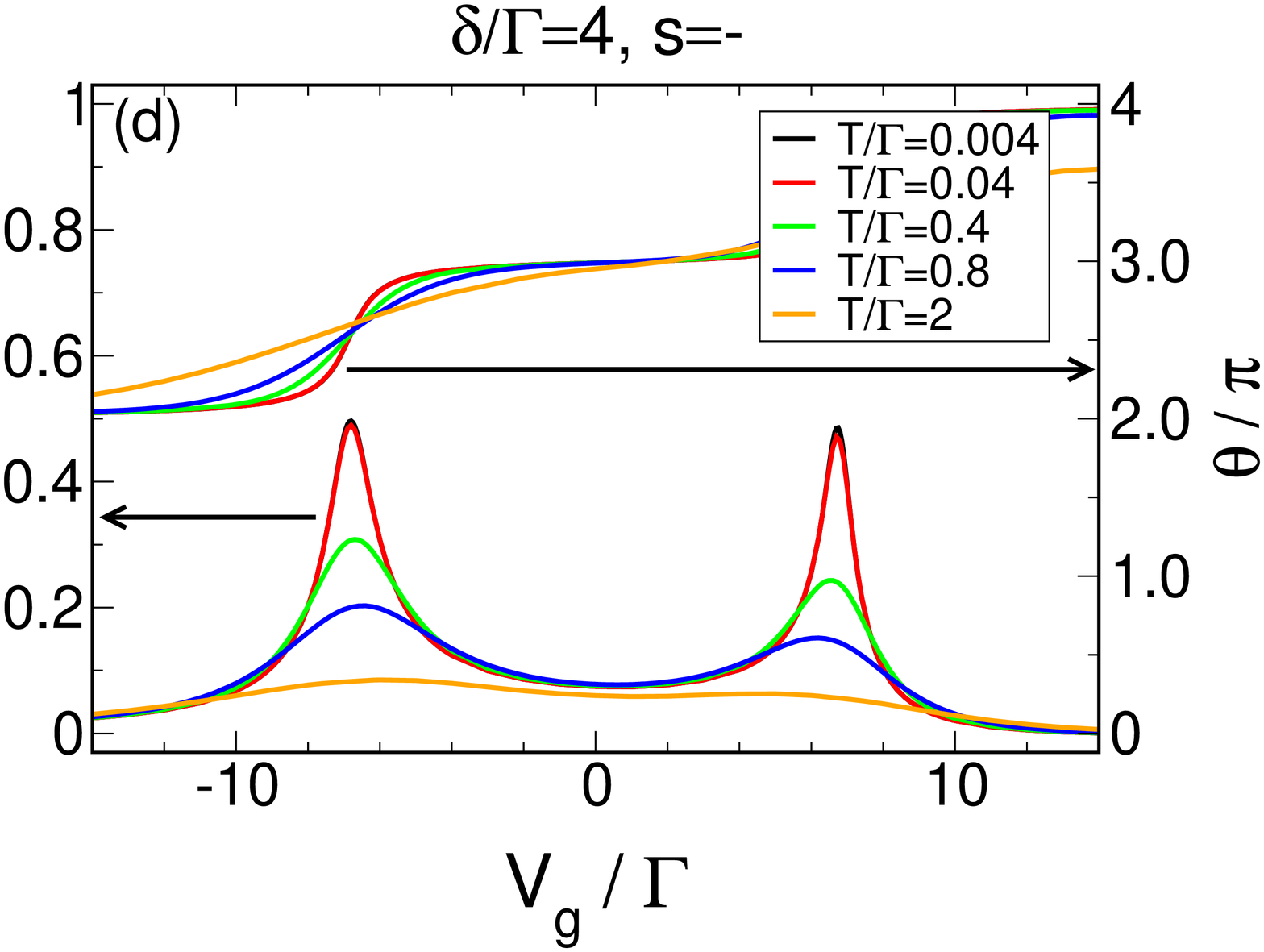}
        \caption{Temperature dependence of $|t(V_g)|$ and
          $\theta(V_g)$ obtained by NRG for $\gamma=\{0.27, 0.33,
          0.16, 0.24\}$, $U/\Gamma=10$ for $s=\pm$ and in the
          universal ($\delta/\Gamma=0.02$) and the mesoscopic
          ($\delta/\Gamma=4$) regime. We used the NRG parameters 
      $\Lambda_{\rm NRG}=2.3$ and $N_{\rm kept}\thicksim 512$.}
\label{fig4}
\end{figure}

We next investigate how the phase and magnitude of the transmission
are affected by finite temperatures. To this end, we use the new
  FDM-NRG algorithm recently proposed in Ref.\ \cite{Weichsel}. 
  In Fig.\ \ref{fig4} we show NRG data for $|t(V_g)|$ and $\theta(V_g)$ at
different $T$. We consider generic level-lead couplings
$\gamma=\{0.27, 0.33, 0.16, 0.24\}$, $U/\Gamma=10$, $s=\pm$ and
$\delta/\Gamma=0.02$ (universal regime) as well as $\delta/\Gamma=4$
(mesoscopic regime).  As expected, with increasing temperature the
sharp $\pi$ phase lapses are gradually smeared out, the transmission zero vanishes and the
change of $\theta$ at the phase lapse becomes smaller than $\pi$. Furthermore,
the CB peaks decrease and broaden. For $\delta/\Gamma=4$ and $s=-$
[Fig.\ \ref{fig4} (d)] the phase lapse lies outside the CB peaks and outside
the window of gate voltages shown.

In the mesoscopic regime [Figs.\ \ref{fig4} (b) and (d)]
the explicit temperature dependence of the
Green function entering Eq.\ (\ref{transamp}) via Eq.\
(\ref{transformula}) is rather
weak and the temperature dependence of the CB peaks and the phase lapse 
can be understood from the behavior in the
noninteracting model, but with level spacing $U+\delta$. For small $T$ the
height $h_j(T)$ of the $j$'th CB peak scales as $1-h_j(T)/h_j(0) \sim 
T^2/\Gamma_j^2$ and the width $w$ of the phase lapse as 
$w \sim T^2/(\delta+U)^2$ \cite{Silva}. The relevant scale for
sizable temperature effects in the peak height is thus
$\Gamma_j$ while it is $U+\delta$ in the smearing of the phase
lapses. 
Since we have chosen 
$\Gamma_j \ll \delta+U$ a reduction of $h_j$ is
visible for temperatures at which the phase lapse is still fairly sharp [see
Fig.\ \ref{fig4} (b)].   

Due to the importance of correlation effects at small $\delta/\Gamma$,
the $T$ dependence of $|t(V_g)|$ and $\theta(V_g)$ in the universal
regime is different from the noninteracting case.  Here the explicit
temperature dependence of $\mathcal G$ is much stronger and cannot be
neglected. The resulting $T$ dependence of $|t(V_g)|$ and $\theta(V_g)$
is shown in Figs.\ \ref{fig4} (a) and (c). A comparison to Fig.\
\ref{fig4} (b) shows that in the universal regime the smearing of the
phase lapse sets in at a lower energy scale than in the mesoscopic regime. This
scale depends on the relative sign $s$ of the level-lead hopping
matrix elements [compare Figs.\ \ref{fig4} (a) and (c)]. Furthermore,
in contrast to the mesoscopic regime the scales on which the CB peaks
and the phase lapse are affected by temperature are comparable. A more detailed
investigation of the temperature dependence in the universal regime,
which also discusses the fate of the correlation-induced resonances, is beyond the scope of the
present work and is left as subject for future studies.

\begin{figure}[t]	
	\centering
	      \includegraphics[width=0.495\textwidth,height=4.4cm,clip]{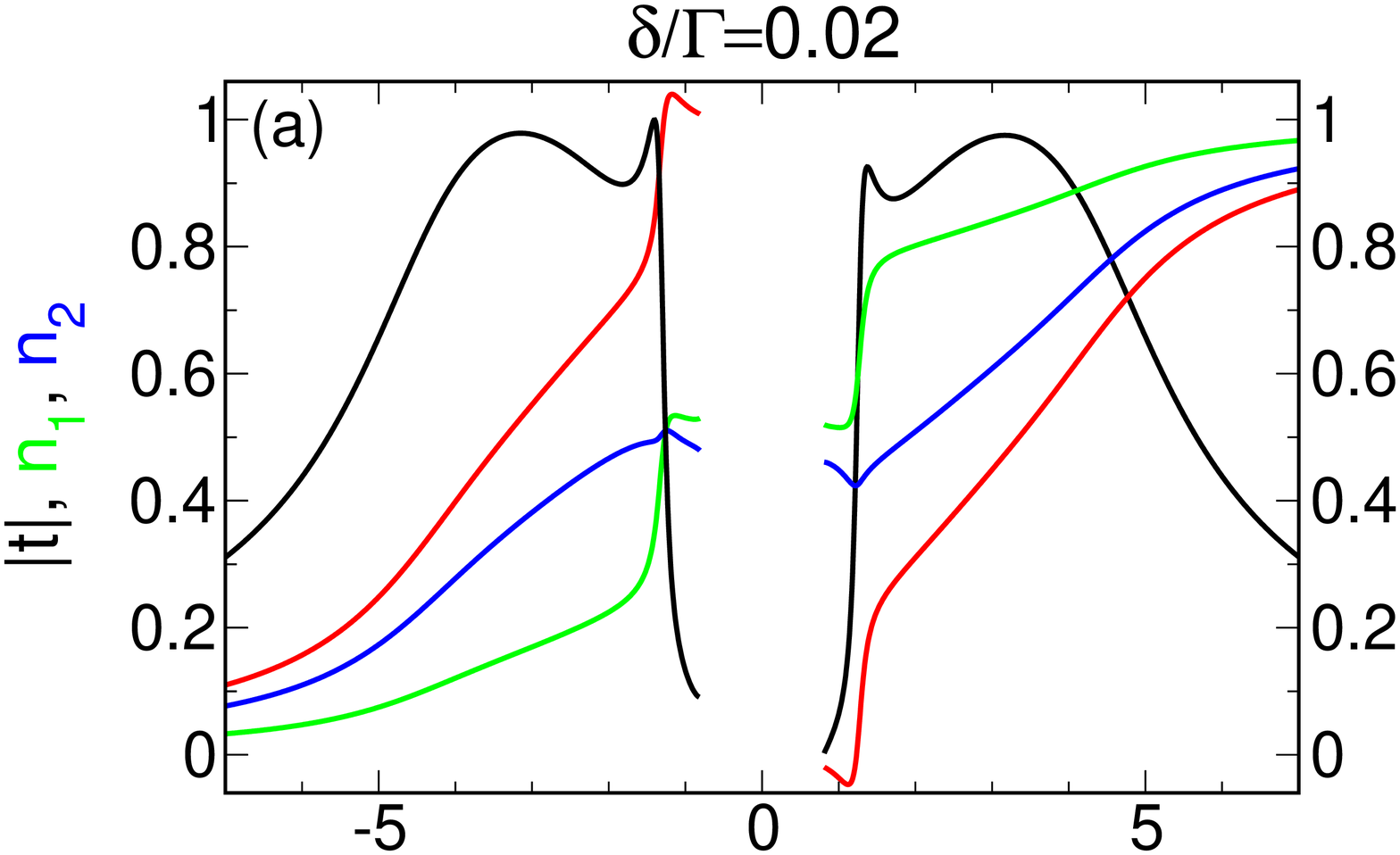}\hspace{0.015\textwidth}
        \includegraphics[width=0.475\textwidth,height=4.4cm,clip]{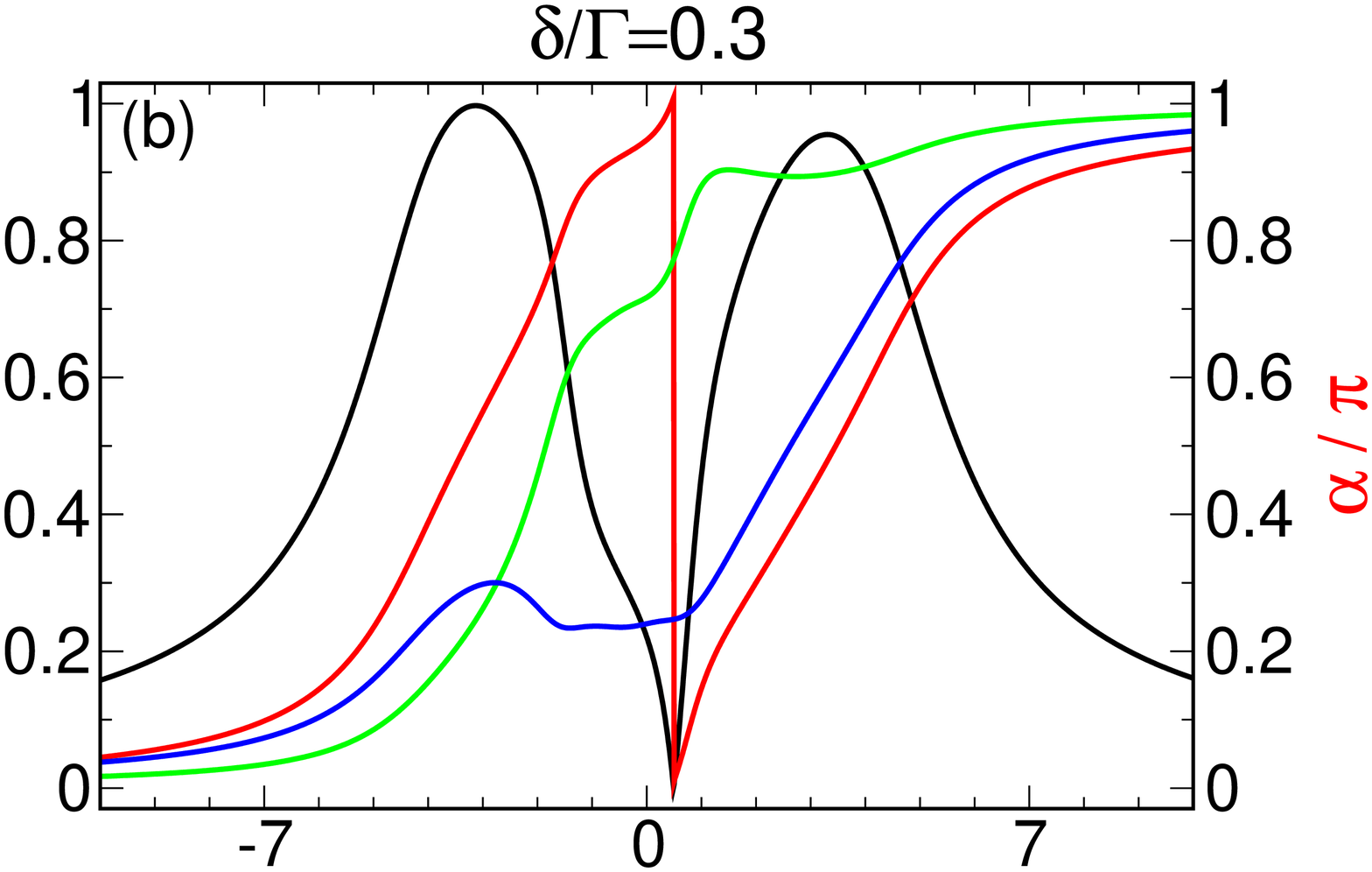}\vspace{0.3cm}
        \includegraphics[width=0.495\textwidth,height=5.2cm,clip]{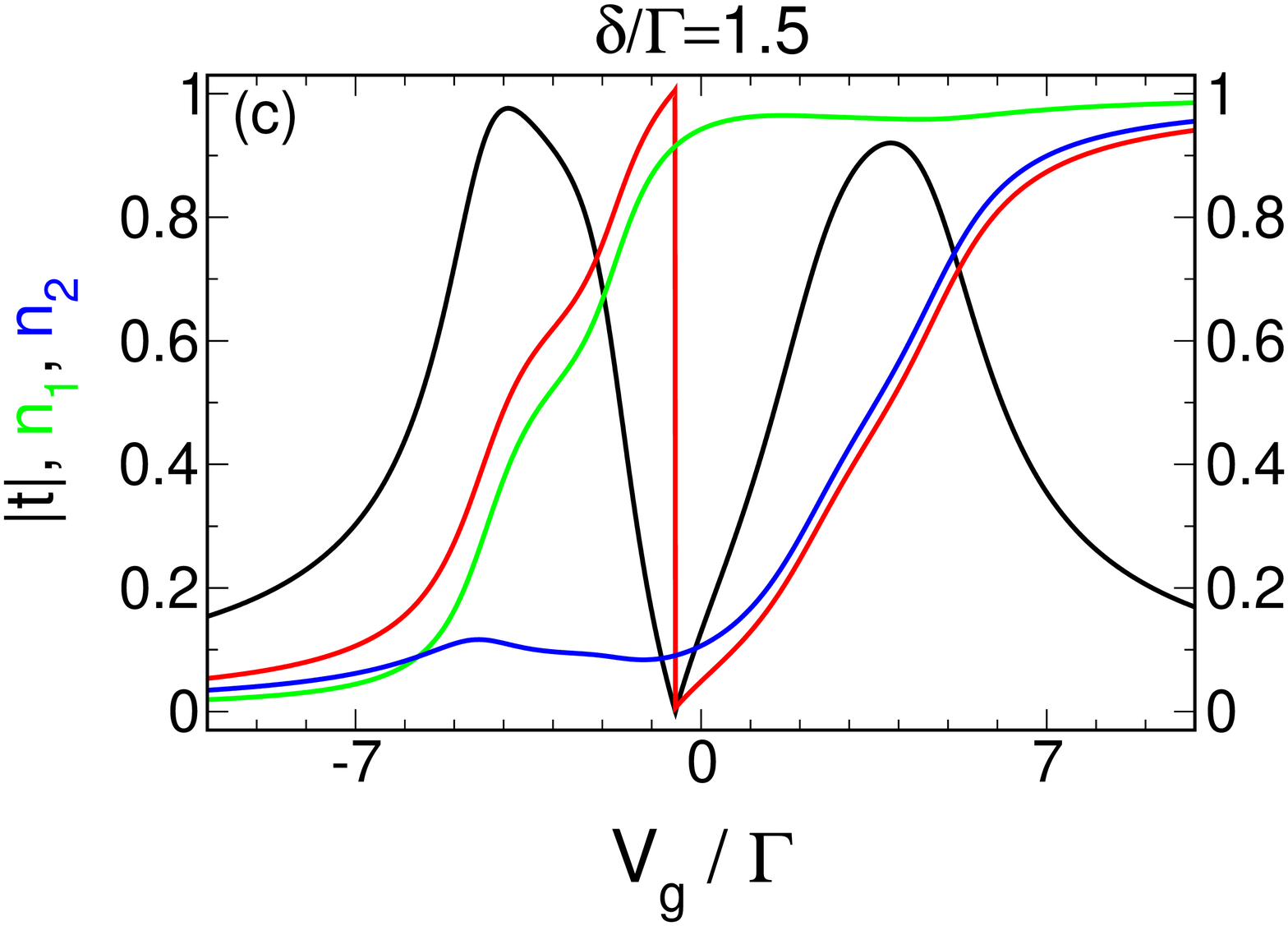}\hspace{0.015\textwidth}
        \includegraphics[width=0.475\textwidth,height=5.2cm,clip]{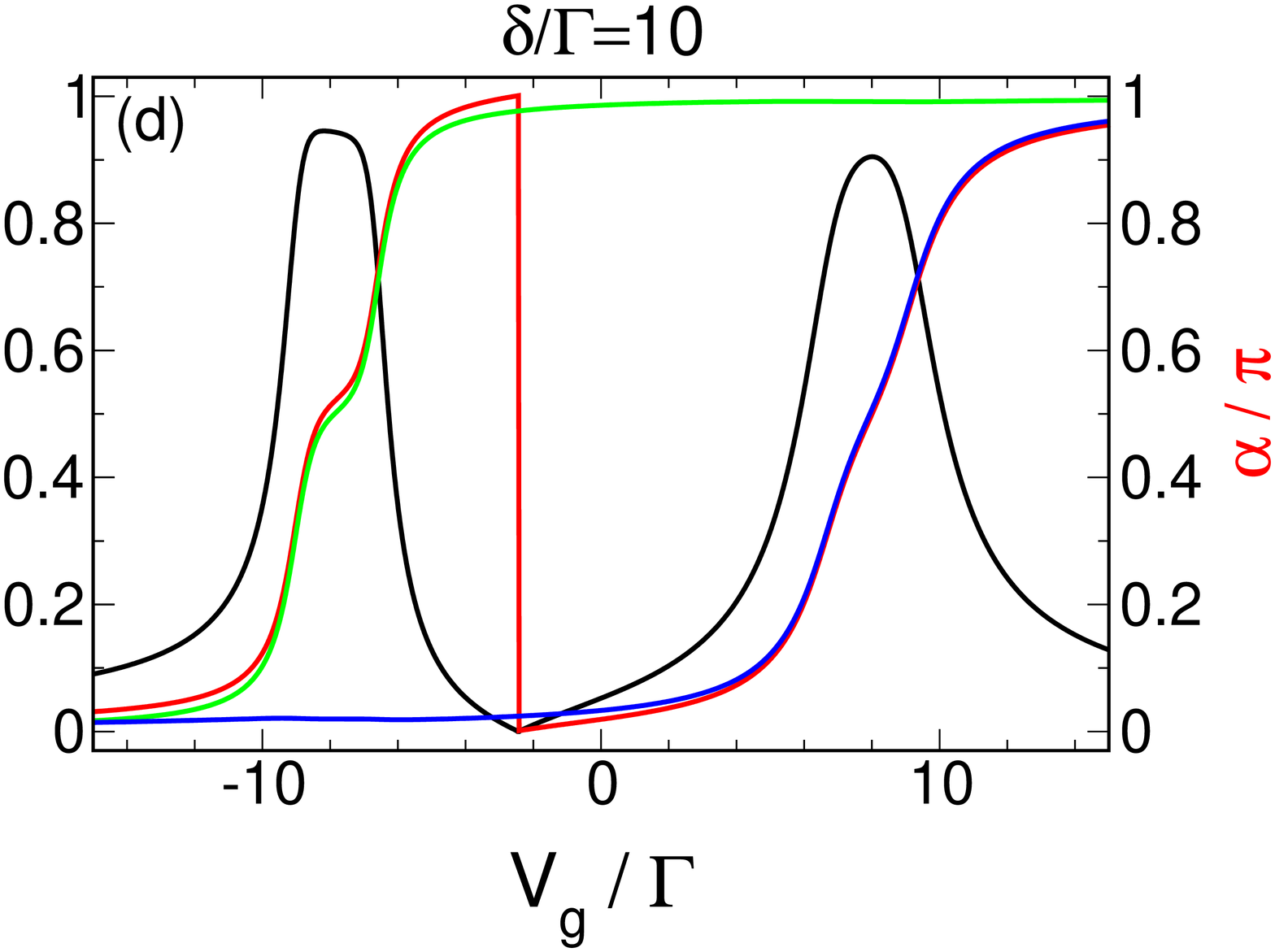}
        \caption{Gate voltage $V_g$ dependence of $|t|$ (black), 
$\theta$ (red) and the level occupancies per spin direction 
(green and blue) of a spinful two-level dot at $T=0$ for different $\delta$
obtained by fRG. 
The parameters are $U/\Gamma=3$, $\gamma=\{0.1, 0.2, 0.5, 0.2\}$ and
$s=+$.
In (a), no data are shown around $V_g = 0$ for reasons explained in the text.}
\label{fig5}
\vspace{0.6cm}
	      \includegraphics[width=0.495\textwidth,height=4.4cm,clip]{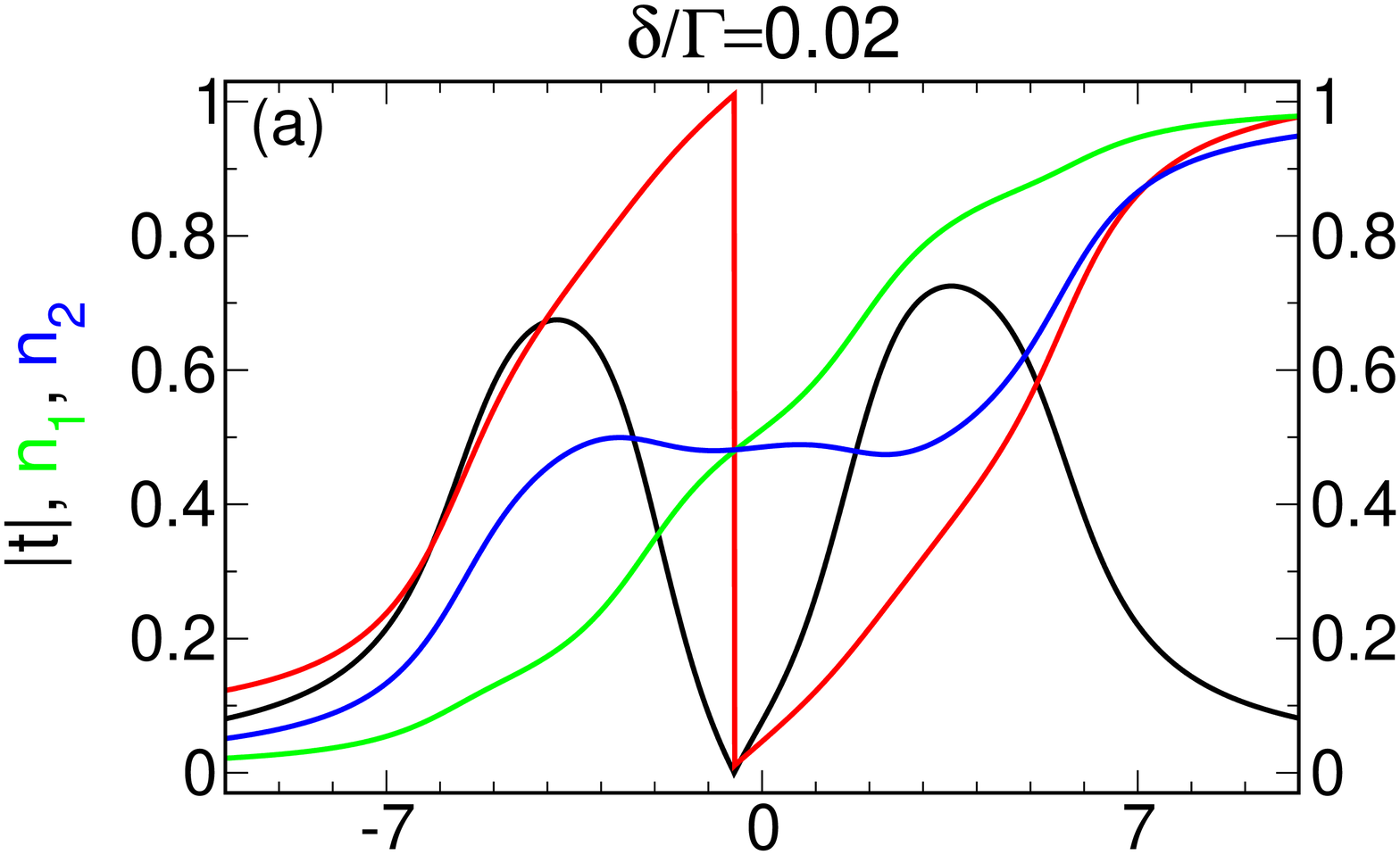}\hspace{0.015\textwidth}
        \includegraphics[width=0.475\textwidth,height=4.4cm,clip]{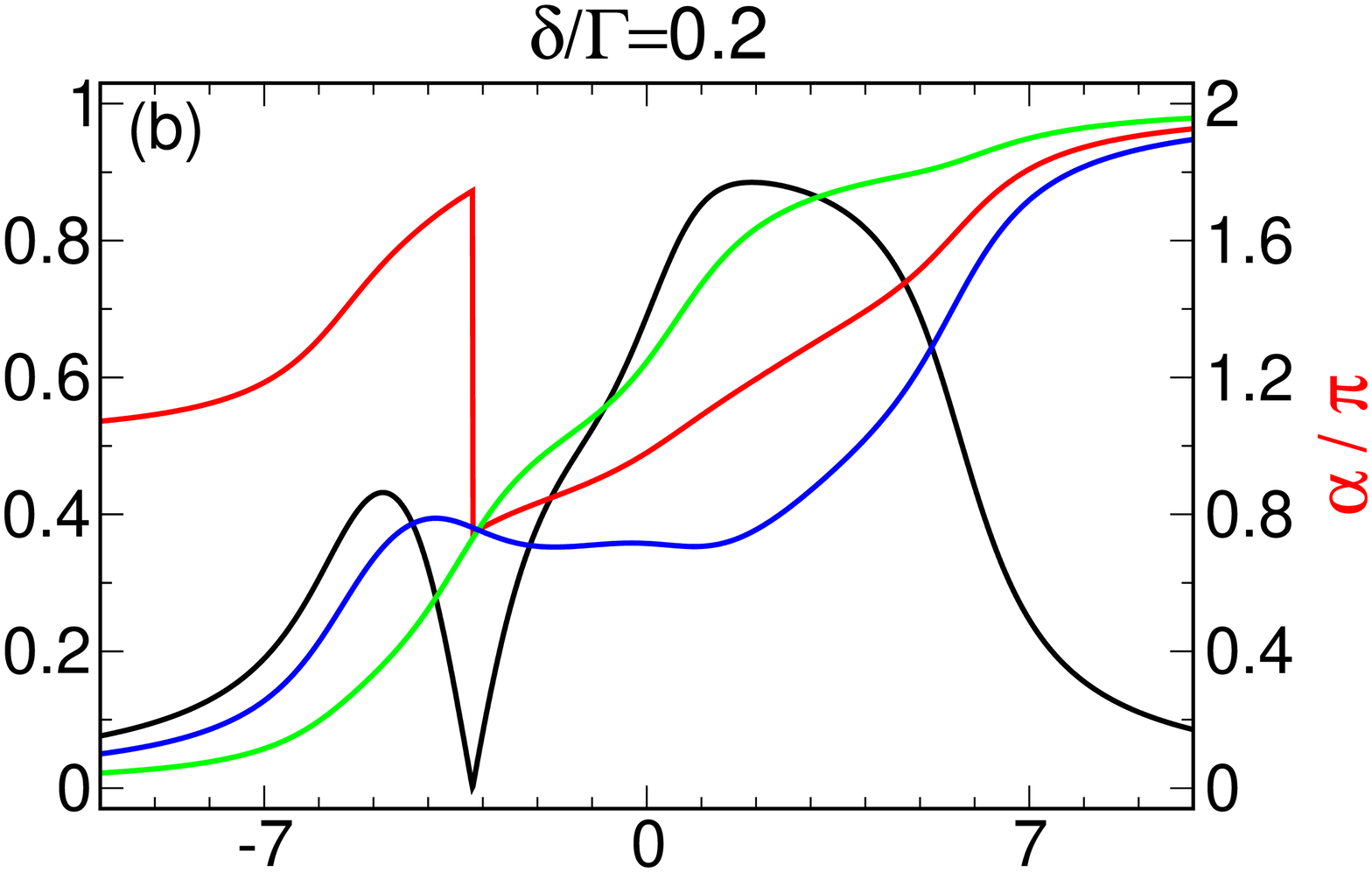}\vspace{0.3cm}
        \includegraphics[width=0.495\textwidth,height=5.2cm,clip]{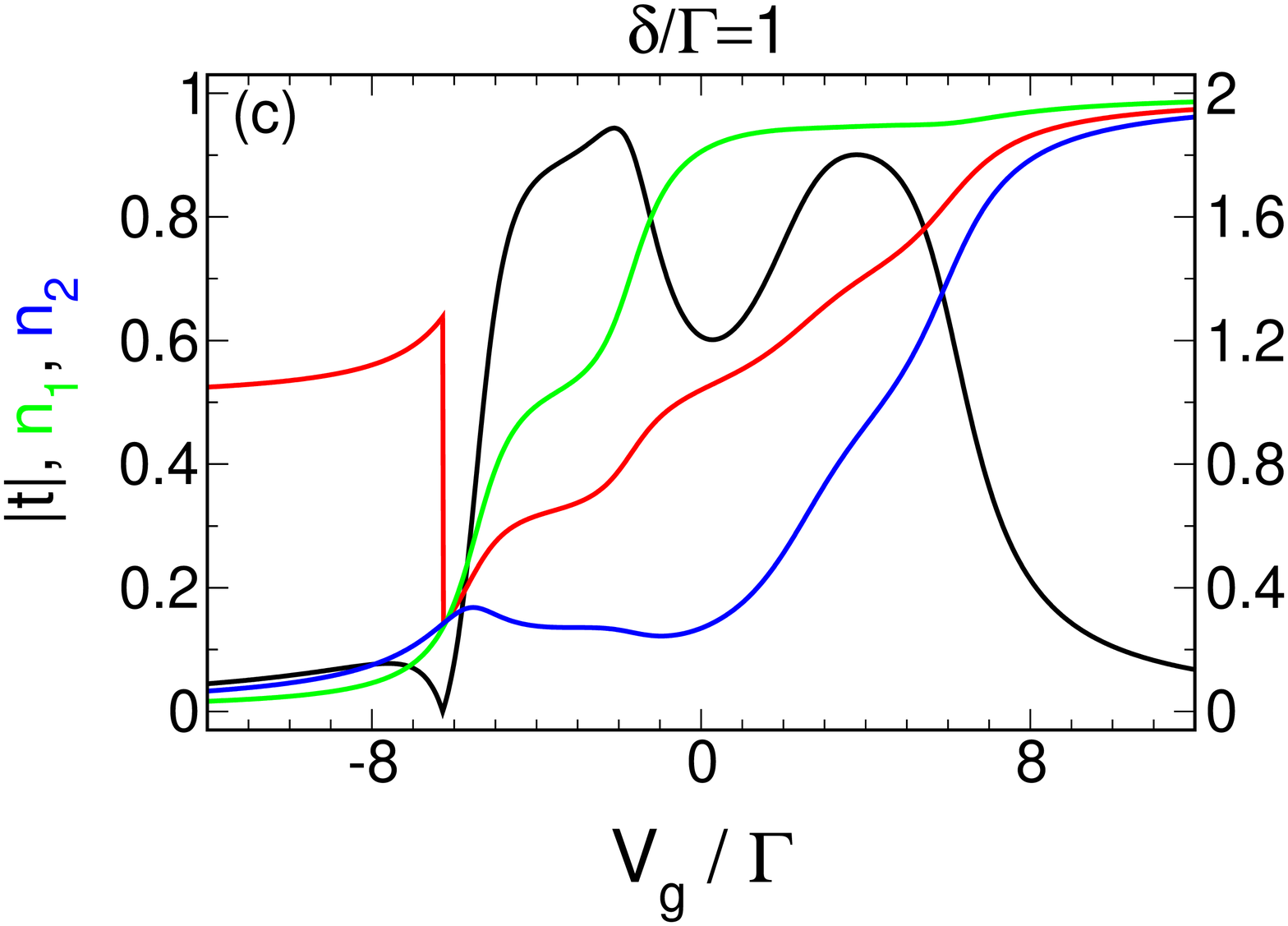}\hspace{0.015\textwidth}
        \includegraphics[width=0.475\textwidth,height=5.2cm,clip]{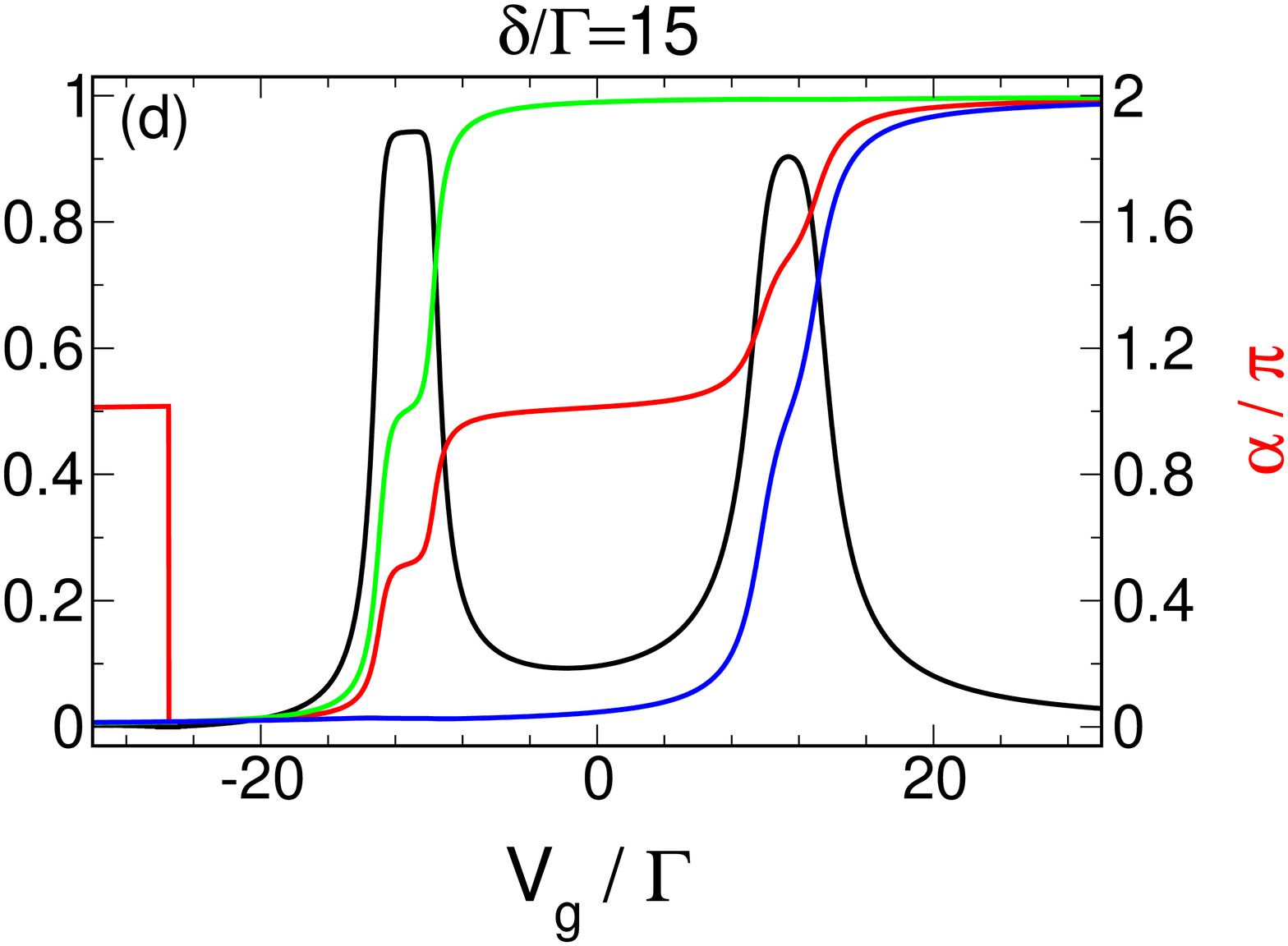}
        \caption{The same as in Fig.\ \ref{fig5}, but for $U/\Gamma=4$
        and $s=-$.}
\label{fig6}
\end{figure}
\afterpage{\clearpage}

\section{Results: Spin-degenerate dots}
\label{resultsspin}

We finally investigate the effect of the spin degree of freedom on the
discussed phase lapse scenario, at $T=0$.  
In Figs.\ \ref{fig5} ($s=+$, $U/\Gamma=3$) and \ref{fig6} 
($s=-$, $U/\Gamma=4$) 
we show fRG data for the evolution 
of $|t(V_g)|$, $\theta(V_g)$ and $n_j(V_g)$ (for a fixed 
spin direction) with increasing $\delta$ for a generic  
$\gamma=\{0.1, 0.2, 0.5, 0.2\}$. The overall dependence of
$\theta(V_g)$ on $\delta$ 
is similar to the one observed in the spinless case 
(compare to Figs.\ \ref{fig1} and \ref{fig2}).  
In particular, the behavior at small $\delta/\Gamma$ appears to be
almost unaffected by the presence of the spin degree of freedom 
[Figs.\ \ref{fig5} (a), (b) and \ref{fig6} (a), (b)].
For large $\delta/\Gamma$ [Figs.\ \ref{fig5} (d) and \ref{fig6} (d)] the
transmission resonances  
are located at odd average total filling of the
two-level dot indicated by shoulders in the $n_j$ [Figs.\ \ref{fig5}
(d) and \ref{fig6} (d)].  
At these fillings and for sufficiently large $U/\Gamma$ 
the Kondo effect is active and the  
resonances cannot be regarded as Lorentzian-like CB peaks. 
Instead they show a plateau-like shape known from the spinful
single-level dot (see Ref.\ \cite{Gerland} and references therein).      
Across the Kondo plateaus of $|t|$ the s-shaped increase of the phase 
is interrupted by a shoulder at $\theta \approx \pi/2$ as expected for
the Kondo effect \cite{Gerland}.  
It would be very interesting to
study how each of the Kondo plateaus of $|t|$ with increasing
temperature crosses over to two CB peaks and the phase behaves in the 
generated CB valley. This question is left for future investigations. 

The behavior of the phase in the presence of the Kondo effect was
experimentally investigated at temperatures comparable to the Kondo
temperature \cite{Yang}, and much below the Kondo temperature
\cite{Yang2}. As  we study the zero temperature case it is proper
to compare our calculations to the measurements at low 
temperatures. Indeed Fig.\ \ref{fig6} (c) for intermediate $\delta/\Gamma$ 
qualitatively reproduces the experimental results at low 
temperature as shown in Fig.\ 3 c of Ref.\
\cite{Yang2}. In particular the
increase of the phase by more then $\pi$ and the absence of clearly
developed Kondo plateaus are reproduced. 

\begin{figure}[t]	
	\centering
        \includegraphics[width=0.475\textwidth,height=4.4cm,clip]{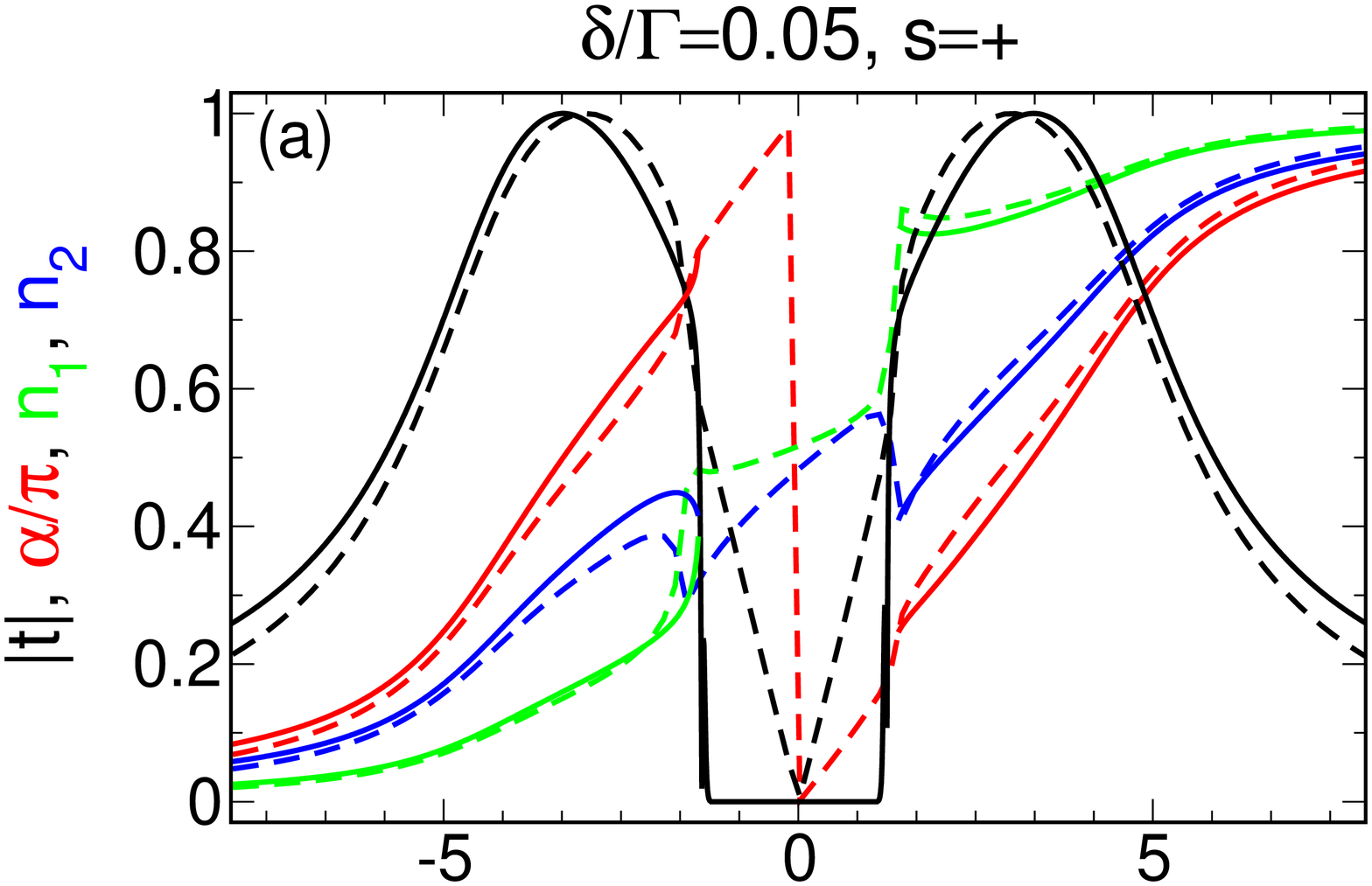}\hspace{0.035\textwidth}
        \includegraphics[width=0.475\textwidth,height=4.4cm,clip]{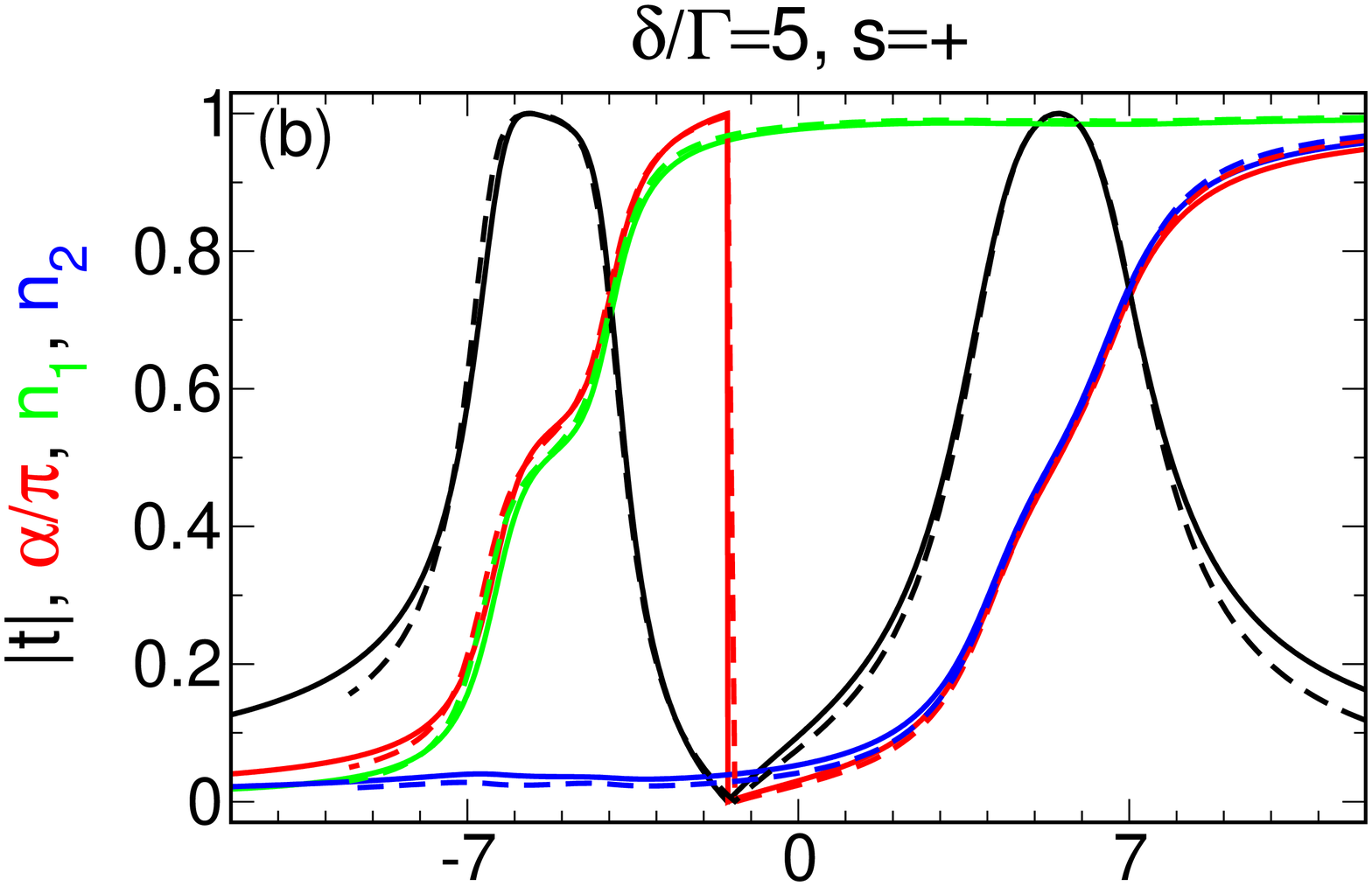}\vspace{0.3cm}
        \includegraphics[width=0.475\textwidth,height=5.2cm,clip]{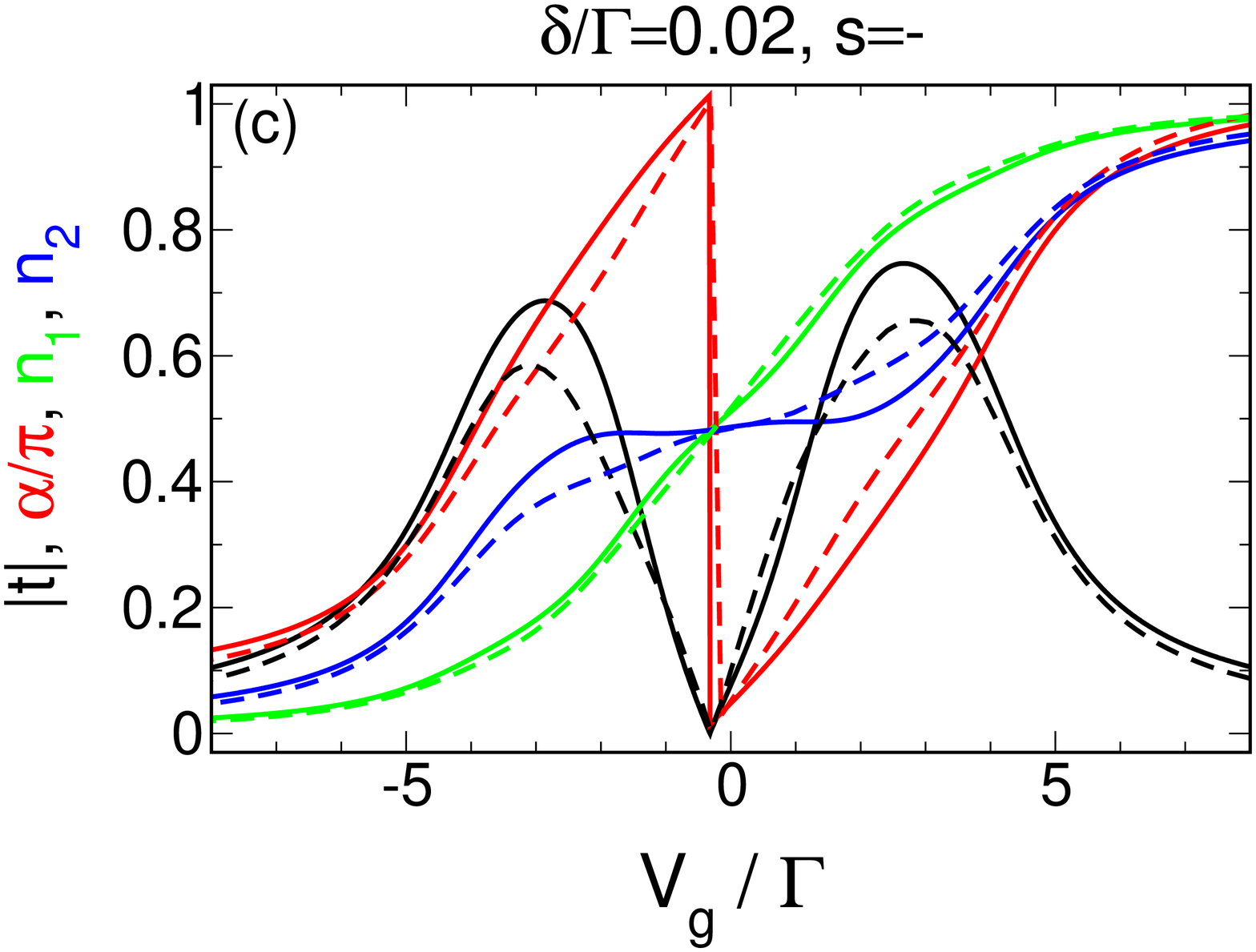}\hspace{0.035\textwidth}
        \includegraphics[width=0.475\textwidth,height=5.2cm,clip]{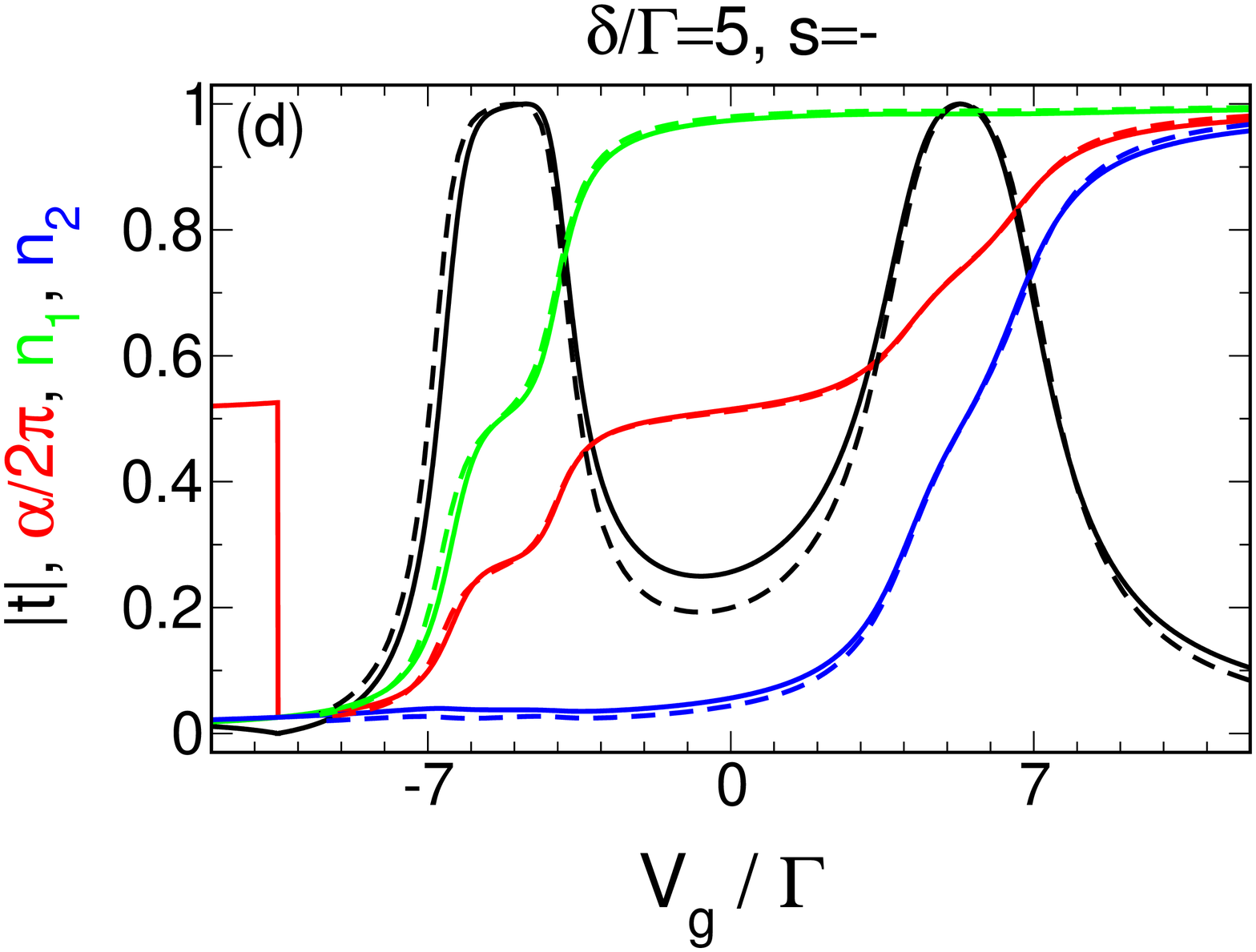}
        \caption{Comparison of fRG (solid) and NRG (dashed) data for
          $|t(V_g)|$, $\theta(V_g)$ and $n_j(V_g)$ (per spin
          direction) in the universal (small $\delta/\Gamma$) and 
          mesoscopic regime (large $\delta/\Gamma$) of a spinful 
          two-level dot at $T=0$. The parameters are $U/\Gamma=3$ and
          $\gamma=\{0.15,0.15,0.35,0.35\}$. In (a) the fRG data for
          $\theta(V_g)$ and $n_j(V_g)$ around $V_g=0$ are not shown. 
          Already the unsatisfactory comparison between fRG and NRG
          for $|t|$ indicate that the fRG becomes unreliable in
          this regime. For more details on this, see the text.
          For the NRG parameters we used $\Lambda_{\rm NRG}=2.5$,
          $N_{\rm kept}=1024$  for $s=+$ and  $N_{\rm kept}=2048$ for $s=-$.}
\label{fig7}
\end{figure}

In Fig.\ \ref{fig5} (a) we left out the fRG data around $V_g=0$ as for
these gate voltages some of the components of the flowing two-particle
vertex become large. This indicates the breakdown of our present truncation
scheme \cite{TCV,Christophdiplom} and the results for $|t|$, $\theta$
and $n_j$ become unreliable. For an explicit comparison to NRG data of
$|t|$ see Fig.\ \ref{fig7} (a). We note in passing that for $s=+$ 
correlation-induced resonances occur also in the model with spin [Fig.\ \ref{fig5} (a)]
\cite{TCV}.

In Fig.\ \ref{fig7} we compare fRG and NRG results for l-r symmetric
level-lead couplings. The computational resources required to obtain
NRG data away from l-r symmetry become large and such data are not 
required for the aim of the present paper. 
We can then use Eq.\ (\ref{t_LR}) and must only compute the
occupancies $n_j$, which is numerically less demanding.
For $\delta>0$ (as exclusively shown) it is only the absence of the 
correlation-induced resonances for 
$s=+$ [compare Figs.\ \ref{fig5} (a) and \ref{fig7} (a)] 
which is different from the results of
the generic $\gamma$ shown in Figs.\ \ref{fig5} and \ref{fig6}.  
With the exception of the  $V_g \approx 0$ regime in the case of small $\delta$ and
$s=+$ the fRG and NRG data compare quite well. In this case we only
show the fRG data for $|t|$ as the results for the phase and
occupancies become rather erratic. The reason
for the breakdown of the currently used truncated fRG is explained 
in Refs.\ \cite{TCV} and \cite{Christophdiplom} and is related 
to the fact that at small $\delta$ and small $V_g$ the correlations 
in effect become extremely large.

\section{Summary}
\label{summary}

In the present paper we studied the appearance of phase lapses in an
interacting two-level quantum dot considering the entire parameter 
space using NRG and a truncated fRG scheme. 

As a starting point we 
briefly discussed the noninteracting case at temperature $T=0$ and 
pointed out that for generic level-lead couplings, that is up to 
cases with increased symmetry, essential features of the universal phase lapse 
scenario are already established at $U=0$. 
For single-particle level spacings 
$\delta$ small compared to the  
level broadenings $\Gamma_j$ the transmission is characterized by two transmission
peaks of equal width with a transmission zero and an associated $\pi$
phase lapse between them (universal
regime). For large $\delta/\Gamma$ at $U=0$ the appearance or not of a transmission zero and
phase lapse between the two transmission peaks depends on the relative sign $s$ of the
level-lead couplings (mesoscopic regime). Within a spinless model we 
have shown that the separation of the two transmission peaks
increases linearly with the interaction $U$ while the $\pi$ phase lapse and transmission zero remain in the valley
between them. Furthermore, with increasing $U$ the increase of the phase across the peaks
takes an s-shape and the peaks become Lorentzian-like, 
thus assuming shapes resembling those observed experimentally \cite{ex1,ex2,ex3}. 
For $s=-$ and increasing $\delta$, a crossover occurs to a regime in which the
$\pi$ phase lapse and transmission zero lies outside the two CB peaks. The crossover scale
$\delta_c$ increases with increasing interaction and thus the Coulomb
repulsion stabilizes the universal phase lapse behavior. We have investigated
the relation between phase lapses and population inversions of the level occupancies $n_j$. 

We have shown that a mean-field treatment of the present problem
correctly reproduces certain features of the behavior discussed above,
but is not able to produce the universal phase lapse scenario at small
$\delta/\Gamma$  due to artifacts of the 
approximation such as a phase lapse by less than $\pi$ \cite{Gefen,Gefenagain} and a
vanishing of the transmission zero \cite{Berkovits}. Furthermore, the discontinuous
gate voltage dependence of the $n_j$ found in the mean-field
approximation turned out to be an artifact.  

Next, we studied how the phase lapse behavior is affect by temperatures
$T>0$. The universal phase lapse (at small $\delta/\Gamma$) is smeared out
but remains visible for not too large $T$. In the mesoscopic regime
with $\delta/\Gamma \gg 1$ the smearing of the phase lapse and
the decrease of the CB peaks can be understood in detail in analogy to
the noninteracting case. For  $\delta/\Gamma \ll 1$ correlations are
more important and a detailed understanding of the temperature
dependence of the transmission $t(V_g)$ requires further studies. 

The phase lapse behavior 
in both the universal and mesoscopic regimes
is also stable if the spin degree of freedom is
included. For sufficiently large $U/\Gamma$ in this case the Kondo effect 
is active at odd average dot
filling, leading to minor modifications of the scenario discussed
above. In particular, at large $\delta/\Gamma$ the $T=0$ transmission 
peaks are Kondo plateaus rather than Lorentzian-like CB
peaks. Across these Kondo plateaus the phase shows a shoulder at
$\theta \approx \pi/2$. In contrast, the behavior at
small $\delta/\Gamma$ appears to be almost unaffected by the spin
degree. A study of the combined effect of finite temperature and 
spin is left for future work.

\ack
We thank R.\ Berkovits, P.\ Brouwer, Y.\ Gefen, L. Glazman, 
D.\ Golosov, M.\ Heiblum,
J.\ Imry, V.\ Kashcheyevs, J.\ K\"onig, F.\ Marquardt, 
M.\ Pustilnik, H.\ Schoeller,
K.\ Sch\"onhammer, and A.\ Silva for valuable discussions.  This work
was supported in part, for VM, by the DFG through SFB602; 
for TH, AW and JvD by the DFG through SFB631 and De730/3-2, and also 
by Spintronics RTN (HPRN-CT-2002-00302), 
NSF (PHY99-07949) and DIP-H.2.1; and for YO by DIP-H.2.1, BSF and 
the Humboldt foundation.


\section*{References}
\begin{thebibliography}{99}

\bibitem{CBref} Sohn L L, Kouwenhoven L P and Sch\"on G (eds) 1997 
{\it Mesoscopic Electron Transport} (Kluwer, Dordrecht).

\bibitem{Kondo} Kouwenhoven L P and Glazman L 2001, {\it Phys. World} {\bf
    14}(1) 33   

\bibitem{Imry} Silvestrov P G and Imry Y 2000 {\it Phys.\ Rev.\ 
 Lett.} {\bf 85}
  2565; Silvestrov P G and Imry Y 2001 {\it Phys.\ Rev.} B {\bf 65} 035309

\bibitem{Koenig} K\"onig J and Gefen Y 2005 {\it Phys.\ Rev.} B {\bf 71}
  201308(R)

\bibitem{Sindel1} Sindel M, Silva A, Oreg Y and von Delft J (2005)
{\it Phys. Rev.} B {\bf 72} 125316

\bibitem{Berkovits} Goldstein M and Berkovits R (2006) cond-mat/0610810.

\bibitem{Hackenbroich} Hackenbroich G (2001) {\it Phys.\ Rep.} {\bf 343} 463

\bibitem{Gefen02}
Gefen Y. 2002 in {\it Quantum Interferometry with Electrons: Outstanding
Challenges} Lerner I V \emph{et al.} (eds) (Kluwer, Dordrecht) p. 13. 

\bibitem{Silva} Silva A, Oreg Y and Gefen Y (2002) 
{\it Phys.\ Rev.} B {\bf 66}, 195316 

\bibitem{Gefen} Golosov D I and Gefen Y (2006) {\it Phys.\ Rev.} B
  {\bf 74}, 205316

\bibitem{Gefenagain}   Golosov D I and Gefen Y (2007) {\it New
    J. Phys.} this issue, p.

\bibitem{VF}  Meden V and Marquardt F (2006) {\it Phys.\ Rev.\ Lett.}
    {\bf 96} 146801 

\bibitem{Slava} Kashcheyevs V, Schiller A, Aharony A and Entin-Wohlman
  O (2006) cond-mat/0610194 

\bibitem{ex1} Yacoby Y, Heiblum M, Mahalu D and Shtrikman H (1995) 
{\it Phys.\ Rev.\ Lett.} {\bf 74} 4047

\bibitem{ex2} Schuster R, Buks E, Heiblum M, Mahalu D, Umansky V and
  Shtrikman H (1997) {\it Nature} {\bf 385} 417

\bibitem{ex3} Avinun-Khalish M, Heiblum M, Zarchin O, Mahalu D and
  Umansky V  (2005) {\it Nature} {\bf 436} 529

\bibitem{Aharony}  Aharony A, Entin-Wohlman O, Halperin B and Imry Y
  (2002) {\it Phys.\ Rev.} B {\bf 66} 115311 

\bibitem{Felix} Berkovits R, von Oppen F and Kantelhardt J W (2004) 
{\it Euro.\ Phys.\ Lett.} {\bf 68} 699

\bibitem{PL1} Karrasch C, Hecht T, Oreg Y, von Delft J and Meden V
  (2006) cond-mat/0609191 

\bibitem{Fano} Fano U (1961) {\it Phys.\ Rev.} {\bf 124} 1866 

\bibitem{Yang} Ji Y, Heiblum M, Sprinzak D,   Mahalu D and
  Shtrikman H (2000) {\it Science} {\bf 290} 779  

\bibitem{Yang2} Ji Y, Heiblum M and
  Shtrikman H (2002) {\it Phys.\ Rev.\ Lett.} {\bf 88} 076601  

\bibitem{Krishna-murthy} Krishna-murthy H R, Wilkins J W and
  Wilson K G (1980) {\it Phys.\ Rev.} B {\bf 21} 1003

\bibitem{TCV} Karrasch C, Enss T and Meden V (2006) {\it Phys.\ Rev.\
  } B {\bf 73} 235337   

\bibitem{Christophdiplom} Karrasch C (2006) Diploma-thesis,
  Universit\"at G\"ottingen; cond-mat/0612329

\bibitem{Weichsel}
Weichselbaum A and von Delft J (2006) cond-mat/0607497

\bibitem{Gerland}  Gerland U, von Delft J, Costi T A and Oreg Y (2000)
 {\it Phys.\ Rev.\ Lett.} {\bf  84} 3710 

\bibitem{SalmhoferHonerkamp} Salmhofer M and Honerkamp C (2001) 
  {Prog.\ Theor.\ Phys.} {\bf 105} 1 

\bibitem{Ralf} Hedden R, Meden V, Pruschke Th and
  Sch\"on\-hammer K (2004) {\it J.~Phys.: Condens.\ Matter} 
{\bf 16} 5279 

\bibitem{lecturenotes} Meden V, lecture notes on the ``Functional
  renormalization group'',
  http://www.theorie.physik.uni-goettingen.de/$\sim$meden/funRG/

\bibitem{NegeleOrland} Negele J W and Orland H (1988) 
 {\it Quantum Many-Particle Physics} (Addison-Wesley, Reading)

\bibitem{KondoModel} Wilson K G and Kogut J
  (1974) {\it Phys. Rep.} {\bf 12} 75

\bibitem{Krishna2} Krishna-murthy H R, Wilkins J W and
  Wilson K G (1980) {\it Phys.\ Rev.} B {\bf 21} 1044

\bibitem{Costi} Costi T A, Hewson A C and Zlatic V
  (1994) {\it J.\ Phys.\ Cond.\  Matter} {\bf 6} 2519

\bibitem{BullaS} Bulla R, Hewson A C and Pruschke Th
  (1998) {\it J.\ Phys.\ Cond.\ Matter} {\bf 10} 8365 

\bibitem{Bulla2}  Bulla R, Costi T A and Vollhardt D
(2001){\it Phys.\ Rev.\ } B {\bf 64} 45103            

\bibitem{Hofstetter} Hofstetter W
  (2000) {\it Phys.\ Rev.\ Lett.\ } {\bf 85} 1508

\bibitem{Verstraete} Verstraete F, Weichselbaum A, Schollw\"ock U, Cirac
J I and von Delft J (2005) cond-mat/0504305

\bibitem{Anders}  Anders F B and Schiller A
  (2005) {\it Phys.\ Rev.\ Lett.\ } {\bf 95} 196801

\bibitem{KK} Pines D and Nozi\'eres
  (1966) {\it The Theory of Quantum Liquids, Vol I.}
  Benjamin W A, Inc.\ New York



\end{thebibliography}
\end{document}